\documentclass[seceq]{ptptex}

\usepackage{graphicx} 
\usepackage{dcolumn}  
\usepackage{bm}       
\usepackage{amsmath}
\usepackage{amssymb}
\begin{document}

\title{The Amplitude Equation for the Rosensweig Instability\\
       in Magnetic Fluids and Gels}

\author{Stefan Bohlius$^{1}$,
Harald Pleiner$^{1}$, and
Helmut R. Brand$^{2}$}

\inst{$^{1}$Max Planck Institute for Polymer Research, 55021 Mainz, Germany\\
$^{2}$Theoretical Physics, University of Bayreuth, 95440 Bayreuth, Germany}

\markboth{%
S. Bohlius, H. Pleiner, and H.R. Brand%
}{%
Amplitude Equation for the Rosensweig Instability%
}
\recdate{September 30, 2010; Revised November 12, 2010}
\pubinfo{Vol. 125, No. 1, January 2011}
\PTPindex{034, 054, 056}

\abst{
The Rosensweig instability has a special character among the frequently discussed instabilities. One distinct property is the necessary presence of a deformable surface, and another very important fact is, that the driving force acts purely via the surface and shows no bulk effect. These properties make it rather difficult to give a correct weakly nonlinear analysis. In this paper we give a detailed derivation of the appropriate amplitude equation based on the hydrodynamic 
equations emphasizing the conceptually new procedures necessary to deal with the distinct properties mentioned above. First the deformable surface requires a fully dynamic treatment of the instability and the observed stationary case can be interpreted as the limiting case of a frozen-in characteristic mode. Second, the fact that the driving force is manifest in the boundary conditions, only, requires a considerable change in the formalism of weakly nonlinear bifurcation theory. To obtain the amplitude equations a combination of solubility conditions and (normal stress) boundary conditions has to be invoked in all orders of the expansions.
}

\maketitle

\section{Introduction}

Since its discovery in 1967, the normal field or Rosensweig instability \cite{Cowley1967a} attracted the attention of experimentalists and theorists, alike. The phenomenon describes the transition of an initially flat ferrofluid surface to hexagonally ordered surface spikes as soon as an applied magnetic field exceeds a certain critical value. Ferrofluids are suspensions of magnetic nanoparticles in a suitable carrier liquid. They are coated by polymers or charged in order to prevent coagulation and show various distinct material properties \cite{Ferrohydrodynamics}. One of the most prominent examples of these properties is the superparamagnetic behavior in external magnetic fields, which accounts for the large magnetic susceptibility and the high saturation magnetization in rather low magnetic fields. If one starts the cross-linking process in a mixture of a ferrofluid and a polymer solution with cross-linking agents, a superparamagnetic elastic medium, called ferrogel, is obtained \cite{Zrinyi1996a}. As in usual ferrofluids, the initially flat surface of ferrogels becomes unstable beyond a critical magnetic field \cite{Bohlius2006a}.

With its discovery, a first theoretical description together with a linear stability analysis was given \cite{Cowley1967a}. At the free surface the stabilizing forces of gravity and surface tension compete with the destabilizing magnetic force. Although the applied magnetic field is homogeneous and therefore no net-force is acting on the medium, fluctuations of the surface lead to focusing effects rendering the local field at the surface inhomogeneous. With this model, the prediction of the critical magnetic field and the characteristic wavelength, which turned out to be the capillary wavelength, was possible. 
The linear growth behavior was discussed later on 
\cite{weilepp,Lange2000a,Lange2001a}. For magnetic gels the elastic force contributes as a stabilizing effect leading to a shift (as a function of the elastic shear modulus) to higher magnetic fields whereas the characteristic wavelength remains unchanged \cite{Bohlius2006a}. First experiments to confirm the threshold shift are performed using thermoreversible magnetic gels \cite{Lattermann2006a} as the magnetic medium.

A linear stability analysis provides us just with the threshold and the most unstable mode. No prediction of the arising pattern can be made, nor does it give the 
dynamic behavior beyond the threshold. A nonlinear analysis of the Rosensweig instability, however, turned out to be very complicated mainly due to the fact that the instability necessarily involves a deformable surface. In 1977 a very first approach to the nonlinear regime was given by A. Gailitis \cite{Gailitis1977a}. 
Since the pattern formed is static, Gailitis discussed the surface energy density, consisting of the gravitational energy, the energy contribution due the surface tension and the magnetic energy, as a function of the deflection of the surface from its flat ground state. Prescribing the regular surface patterns of stripes, squares and hexagons, he found upon minimizing the energy density, that at the linear threshold hexagons are the stable configuration that in turn transform into squares upon further increase of the magnetic field. Both transitions are accompanied by hysteretic regions. The stripe configuration instead is always unstable with respect to one of the other two patterns. A major drawback of this method is, that it is valid only in the asymptotic limit of vanishing magnetic susceptibility. Friedrichs and Engel \cite{Friedrichs2001a} extended Gailitis' method to systems with a finite depth and additionally gave an estimate of the maximal magnetic susceptibility up to which the method gives reasonable results. The energy method was extended to the Rosensweig instability in isotropic magnetic gels considering additionally the elastic surface energy density. \cite{Bohlius2006b} Another unsatisfying aspect of this method rests in the fact, that it cannot describe growth rates, since it ignores dissipative processes.

Based on the linear result of a static instability, an expansion of the basic static equations governing the ferrofluid behavior was discussed \cite{Twombly1983a,Silber1988a}. The analysis is, in contrast to the energy method, valid for any given magnetic susceptibility, however, for typical values no stable pattern could be found at the linear onset. Another approach \cite{Friedrichs2003a} considers a static regime, where only the normal stress boundary condition is considered for the nonlinear expansion and where a horizontal field component of the magnetic field was assumed to be strong enough to suppress two dimensional patterns \cite{Friedrichs2002a}. The dynamics of the system has first been taken into account by Kubstrup et al. \cite{Kubstrup1996a} who used a Swift-Hohenberg model to describe fronts between hexagons and squares. This approach, however, lacks the connection of the parameters introduced in the Swift-Hohenberg equation to the material properties of the medium. 
In addition, it is unclear whether the terms involving time derivatives in this model
are the appropriate ones.

What one would like to have is a systematic nonlinear expansion of the basic hydrodynamic equations in analogy to Ref.~\citen{Schlueter1965a}. To adapt this method in the case of the Rosensweig instability, the adjoint linear eigenvectors in the presence of a deformable surface are needed to satisfy Fredholm's theorem. To circumvent Fredholm's theorem, Malik and Singh \cite{Malik1985a,Malik1987a} restricted their discussions to potential flows only. However, as can be seen in the present paper and in Refs.~\citen{Bohlius2006a,Bohlius2006b}, the rotational flow contributions are needed to guarantee the free surface to be stress-free. Recently, the adjoint system for the Rosensweig instability in isotropic magnetic fluids and gels (and for the Marangoni instability) in presence of deformable surfaces was given by 
the present authors \cite{Bohlius2007a} as a prerequisite to access the nonlinear regime via a weakly nonlinear analysis. The latter, however, cannot be carried out straightforwardly, since the driving force acts on the surface, only. As a result, the (bulk) solvability conditions are not sufficient to give the amplitude equation, but have to be combined with the (normal stress) boundary conditions. This new procedure will be discussed in detail and executed explicitly in this manuscript, where we concentrate on the case of magnetic gels. 

Deformations of the free surface are crucial for the Rosensweig instability. The kinematic boundary condition relates the temporal changes of the surface deflection with the velocity of the bulk material at the surface. An a priori static description would, thus, completely miss this important boundary condition - one of the reasons, why previous attempts to deal systematically with the nonlinear instability regime have failed. For gels, in addition, the dynamic coupling between elastic deformations and flow also requires a fully dynamical treatment, even if the final instability is stationary. Only at the end the static limit can be performed. The pattern observed by the Rosensweig instability is thereby characterized as the limiting case of a frozen-in surface wave mode.

Our discussion is organized as follows. In \S\S2 and 3 we introduce the basic hydrodynamic equations and their general expansion into the nonlinear regime. Special emphasis is 
put on explaining the consequences of Fredholm's theorem for the present systems. 
In \S4 we solve the second perturbative order. The necessary solutions in the third order are derived in \S5, whereas in \S6 we give the amplitude equation for the Rosensweig instability in isotropic magnetic gels 
followed by a short discussion of the amplitude equation
for magnetic fluids in \S7.
Many of the detailed algebraic calculations are put into appendices.
Some of the results have recently been presented in a conference proceedings \cite{proc} without, however, laying out their derivation and the non-standard subtleties involved.

\section{Basic equations and general approach\label{BasicEquations}}

The basic equations we are concerned with when discussing the Rosensweig 
instability in magnetic gels,
are the hydrodynamic bulk equations derived for isotropic magnetic gels by Jarkova et al. \cite{Jarkova2003a} together with some approximations discussed below.
\begin{eqnarray}
\partial_{t}g_{i} + \partial_{j}T_{ij}
&=&
\rho G_{i}
 \label{NSgenerell} \\
\big(\partial_{t} + v_k\partial_k \big) \epsilon_{ij} - \frac{1}{2}\,\big(
\partial_iv_j + \partial_j v_i \big) &=& 0 \quad
\label{Elasticitygenerell}
\\
\partial_{i}v_{i } &=& 0 \quad
\label{Continuitygenerell}
\\[1ex]
\partial_i  B_i &=& 0
\label{NoMagneticMonopoles}
\\
\epsilon_{ijk} \partial_j H_k &=& 0
\label{RotH}
\end{eqnarray}
They account for the conservation of linear momentum (\ref{NSgenerell}), the conservation of mass (\ref{Continuitygenerell}) and the fact that the elastic network breaks continuous translational symmetry (\ref{Elasticitygenerell}). In our notation ${\bf g}$ is the momentum density, ${\bf v}$ the velocity, $p$ the pressure, ${\bf G}$ represents the acceleration due to gravity and ${\bf B}$ and ${\bf H}$ are the magnetic induction and the magnetic field, respectively. The second rank tensor $\epsilon_{ij}$ denotes the strain field, while the material parameters $\mu_2$ and $\nu_2$ stand for the shear elasticity and shear viscosity, respectively, and are contained in the stress tensor $\mathrm{T}$.

The underlying assumptions are as follows. Even though the magnetic field is considered a slowly relaxing variable in the hydrodynamic theory of Jarkova et al., we assume that it relaxes fast enough on the time scale considered in our discussion of the Rosensweig instability. This is justified by the fact, that the growth of surface spikes takes place on a time scale long compared to the temporal variations of the magnetic field. The magnetic field is then defined by the static Maxwell equations (\ref{NoMagneticMonopoles},\ref{RotH}) and the corresponding boundary conditions at the surface. We also assume, that the macroscopic material parameters like the shear modulus and the shear viscosity are independent of the magnetization in the medium. This also implies that we will neglect magnetostriction in our discussions. Furthermore we assume the magnetic gel to be incompressible ($\rho = \mathrm{const.}$, $\epsilon_{ii}=0$) and be described by linear elasticity theory. Although a realistic quantitative treatment of polymeric gels requires the use of nonlinear elasticity, there is no reason to expect the elastic nonlinearities to change the qualitative behavior of the Rosensweig instability.

The stress tensor of the magnetic medium is defined via the conservation equation for the momentum density (\ref{NSgenerell}) and is given in our notation by
\begin{eqnarray}
T_{ij} &=& g_iv_j + p\delta_{ij} - \big( B_{i}H_{j}
- \frac{1}{2}B_{k} H_{k}\,\delta_{ij} \big)
- \mu_2 (\epsilon_{jk}\epsilon_{ki} + \epsilon_{ik}\epsilon_{kj})
- 2\mu_2\epsilon_{ij}
\nonumber \\ &&
- \nu_2(\partial_{j}v_i + \partial_{i}v_{j}),
\label{StressTensorgenerell}
\end{eqnarray}
while the vacuum stresses are solely due to the magnetic field for what reason the stress tensor there reduces to the known vacuum Maxwell stress tensor $\mathrm{T}^{\mathrm{vac}}$ \cite{ClassicalEDynamics}.

The hydrodynamic and magnetic bulk equations are supplemented by boundary conditions at the deformable surface defined by $z=\xi$. Aside from the usual magnetic boundary conditions the tangential components of the mechanical stress between the magnetic medium and the vacuum above is required to vanish at $z=\xi$, while the normal stress difference is balanced by gravity and surface tension.
\begin{eqnarray}
{\bf n}\times\mathrm{T}\cdot{\bf n}
&=&
{\bf n}\times\mathrm{T}^{\mathrm{vac}}\cdot{\bf n}
\label{GeneralBCTangentialStress}
\\
{\bf n}\cdot\mathrm{T}\cdot{\bf n}
-
{\bf n}\cdot\mathrm{T}^{\mathrm{vac}}\cdot{\bf n}
&=&
 \sigma_T \mathrm{div}{\bf n} - \rho G \xi
\label{GeneralBCNormalStress}
\\
{\bf n}\times{\bf H} &=& {\bf n}\times{\bf H}^{\mathrm{vac}}
\label{GeneralBCTangentialHField}
\\
{\bf n}\cdot{\bf B} &=& {\bf n}\cdot{\bf B}^{\mathrm{vac}},
\label{GeneralBCNormalBField}
\end{eqnarray}
where ${\bf H}^{\mathrm{vac}}$ and ${\bf B}^{\mathrm{vac}}$ denote the magnetic field and the magnetic flux density in the vacuum, respectively, 
$\sigma_T$ is the surface tension,
and where we introduced the surface normal ${\bf n} = \bm{\partial} (z-\xi)/\mid\!\bm{\partial} (z-\xi)\!\mid$. Additionally, due to the deformable surface, we have to consider the kinematic boundary condition modeling the dynamics of the free surface at $z=\xi$
\begin{eqnarray} \label{GeneralKinematicBC}
d_{t}\xi &=& v_{z}.
\end{eqnarray}

\begin{figure}[t]
\begin{center}
\includegraphics[width=13.8cm]{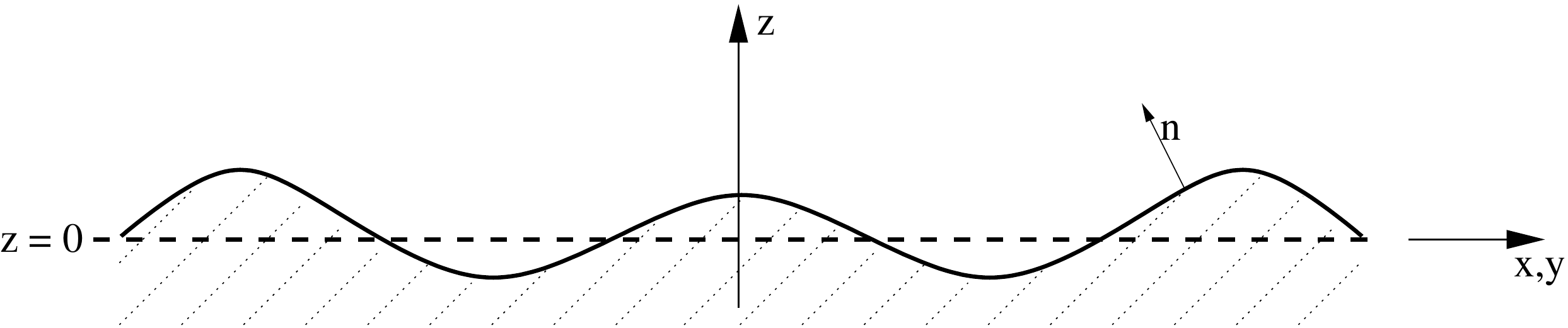}
\caption[Surface mode]{
\label{SurfaceModeCartoon}
Small periodic perturbations $\xi(x,y,t)$ of the initially flat surface $z=0$ between the ferrogel/ferrofluid of susceptibility $\chi$ in the lower half space and vacuum in the upper half space. The magnetic field is oriented parallel to the $z-$axis, whereas gravity is pointing in the opposite direction.
}
\end{center}
\end{figure}

Performing a weakly nonlinear analysis of the stationary state evolving slightly beyond the linear threshold $M_c$, we have to expand the macroscopic variables in terms of $\epsilon$, the normalized difference of the actual applied magnetic field to the critical one
\begin{eqnarray}
\{p,{\bf B},{\bf H},{\bf M}\}
&=&
\{p_{0},{\bf B}_{c},{\bf H}_{c},{\bf M}_{c}\}
+
\epsilon\{p^{(1)},{\bf B}^{(1)},{\bf H}^{(1)},{\bf M}^{(1)}\}
+ \dots
\label{ExpansionFields01}
\\
\{{\bf v},\epsilon_{ij},\xi\}
&=&
0 + \epsilon\{{\bf v}^{(1)},\epsilon_{ij}^{(1)},\xi^{(1)}\} + \dots
\label{ExpansionFields02}
\end{eqnarray}
The magnetic field, however, is an externally given parameter acting as the control parameter, the series expansion of ${\bf H}$ can therefore be reinterpreted as the definition of $\epsilon$.
The linear threshold is given \cite{Bohlius2006a} by 
\begin{eqnarray}
M_{c}^{2} = \frac{1+\mu}{\mu}\left( 2\sqrt{\sigma_T\rho\, G} + 2\mu_{2}\right) ,
\end{eqnarray}
which also determines the critical fields ${\bf B}_c, \,\,{\bf H}_c$ via the linear magnetic constitutive equation ${\bf B} \equiv {\bf H} + {\bf M} = \mu {\bf H}$ employed here.

\begin{figure}[t]
\begin{center}
\includegraphics[width=7cm]{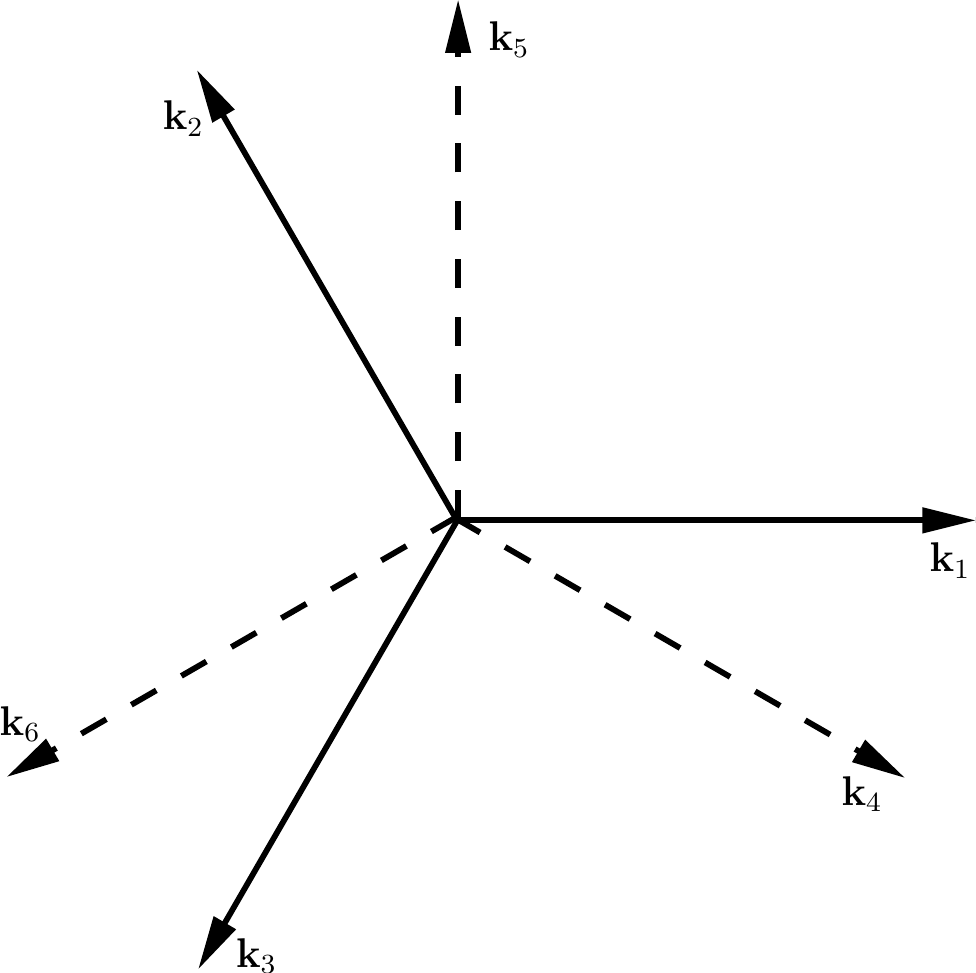}
\caption{
\label{GeometriesWaveVektors}
The sketch shows the relative orientation of the wave vectors under consideration in the amplitude equations (\ref{AmplitudeEquationNormalForm},\ref{AmplitudeEquationNormalFormSquares}). It allows to discuss the stability of hexagons and squares and their interaction.
}
\end{center}
\end{figure}

In our linear discussion \cite{Bohlius2006a}, the surface deflection $\xi(x,y,t)$ was modeled using plane waves $\xi(x,y,t) = \hat{\xi}e^{i\omega t - i{\bf k}\cdot{\bf r}}$. In a nonlinear discussion, this ansatz has to be expanded. The linear description can just provide the characteristic mode becoming unstable at the threshold. The most general ansatz as a starting point for a nonlinear discussion is to assume $N$ of these characteristic modes with different orientations. Each of these modes $i$ consists of a right and left traveling contribution (subscripts $R$ and $L$, respectively) 
\begin{eqnarray}
\xi^{(1)}
&=&
\sum_{i}^{N}\xi_{i} \equiv  \sum_{i}^{N} ( \xi_{iR} +  \xi_{iL} +  \xi_{iR}^* +  \xi_{iL}^* )   
\nonumber \\
&\equiv&
\sum_{i}^{N} (
\hat{\xi}_{iR}e^{i\omega_{i}t-i{\bf k}_{i}\cdot{\bf r}}
+
\hat{\xi}_{iL}e^{-i\omega_{i}t-i{\bf k}_{i}\cdot{\bf r}}
+
\hat{\xi}_{iR}^{\ast}e^{-i\omega_{i}t+i{\bf k}_{i}\cdot{\bf r}}
+
\hat{\xi}_{iL}^{\ast}e^{i\omega_{i}t+i{\bf k}_{i}\cdot{\bf r}}),\,\,\,\,\,\,\
\label{LinearSurfaceDeflection}
\end{eqnarray}
where the asterisk denotes the complex conjugate and ${\bf k}_i$ characterizes the direction  of the $i$-th mode. The wave number $k =|{\bf k}|= |{\bf k}_i|$ is the same for all modes. At the end it will turn out that basically three patterns are important, hexagons, squares and rolls (or stripes). They are described by six critical wave vectors, for which we choose the geometry of Fig.~\ref{GeometriesWaveVektors}. This geometry allows us to discuss hexagons ($\xi_{1}=\xi_{2}=\xi_{3}\not=0$ and $\xi_{4}=\xi_{5}=\xi_{6}=0$), squares ($\xi_{1}=\xi_{5}\not=0$ and $\xi_{i}=0$ for $i\in\{2,3,4,6\}$) and rolls ($\xi_{1}\not=0$ and $\xi_{i}=0$ for $i\in\{2,3,4,5,6\}$). As discussed already in the derivation of the adjoint system \cite{Bohlius2007a}, we have to treat the system dynamically and perform the limit towards a static system in the very end only. The eigenvectors in linear order are known to be modulated by $\xi^{(1)}$ \cite{Bohlius2006b} and therefore separate into left and right traveling contributions together with the complex conjugates, similarly as in Eq.~(\ref{LinearSurfaceDeflection}). The corresponding coefficients depend on the vertical direction $\sim\!e^{qz}$ and $\sim\!e^{kz}$ and are given in Ref.~\citen{Bohlius2006b}.

While performing a weakly nonlinear analysis, we have to specify the scales in space and time. In a first approach we will assume a surface pattern that arises homogeneously in space, which allows us not to rescale the spatial degrees of freedom. Time, however, will  be rescaled in the following manner
\begin{eqnarray}
t^{(1)} = \epsilon t \quad&\mathrm{and}&\quad t^{(2)} = \epsilon^2 t,
\label{AmplEqExpansionTimeDerivative}
\end{eqnarray}
which will lead to the substitution for the time derivative
\begin{eqnarray} \label{texpansion}
\partial_{t} &\longrightarrow&
\partial^{(0)}_{t} + \epsilon\partial^{(1)}_{t}
+ \epsilon^2\partial^{(2)}_{t} + \dots
\end{eqnarray}
We can interpret the scaling in time in the sense, that the dynamics of the amplitudes itself takes place on the slower time scales, $\xi_{i,\{R,L\}} \to \xi_{i,\{R,L\}}(t^{(1)},t^{(2)},\dots)$.

\section{Fredholm's theorem and the adjoint system\label{SecFredholm}}

With the scaling of time and the expansion of the macroscopic variables in terms of $\epsilon$ that we have introduced so far, the whole system of differential equations can be expanded in terms of $\epsilon$. Let $\mathcal{L}_{0}$ be the linear operator and $\mid\!\Psi\rangle = \mid\!\Psi^{(0)}\rangle + \epsilon\mid\!\Psi^{(1)}\rangle +\dots$ the macroscopic state vector. The different orders in $\epsilon$ are then given successively by 
\begin{eqnarray}
\mathcal{L}_{0}\mid\!\Psi^{(1)}\rangle &=& 0
\label{GeneralLinearEquation}
\\
\mathcal{L}_{0}\mid\!\Psi^{(2)}\rangle
&=&
\mid\!\mathcal{N}(\Psi^{(1)},\Psi^{(1)})\rangle
+
\mid\!\mathcal{T}(\partial^{(1)}_{t}\Psi^{(1)})\rangle
\label{GeneralSecondOrderEquation}
\\
\vdots &=& \vdots \nonumber
\end{eqnarray}
The first equation (\ref{GeneralLinearEquation}) represents the linearized set of equation used in the discussion regarding the linear stability \cite{Bohlius2006a}. Furthermore Eq.~(\ref{GeneralLinearEquation}) defines the kernel of the linear operator $\mathcal{L}_{0}$, given by the linear eigenvectors $\mid\!\Psi^{(1)}\rangle$. In the second perturbative order the set of equations (\ref{GeneralSecondOrderEquation}) becomes inhomogeneous due to the nonlinear nature of the basic set of equations (represented by $\mathcal{N}(\cdot,\cdot)$) and due to the rescaling of time (represented by $\mathcal{T}(\cdot)$). In the case that these inhomogeneities reproduce elements of the kernel of the linear operator $\mathcal{L}_{0}$, equation (\ref{GeneralSecondOrderEquation}) cannot be solved. The requirement that the inhomogeneities have to be orthogonal to the subspace spanned by the linear eigenvectors $\mid\!\Psi\rangle$ provides us with an additional solvability condition. It is named after Fredholm and reads in the second order
\begin{eqnarray}
\langle\Psi\!\mid
\mathcal{N}(\Psi^{(1)},\Psi^{(1)})\rangle
+
\langle\Psi\!\mid
\mathcal{T}(\partial^{(1)}_{t}\Psi^{(1)})\rangle
&=&
0,
\end{eqnarray}
%
%
where $\langle a\!\mid\!b\rangle$ denotes the suitable scalar product
\begin{eqnarray}
\langle a\!\mid\!b\rangle
&=&
\lim_{L\to\infty}\frac{1}{4L^2}
\int\limits_{-L}^{L} \!\!dx
\int\limits_{-L}^{L} \!\!dy
\int\limits_{-\infty}^{\xi} \!\!dz
\int\limits_{0}^{\tau} \!\!dt \,\,
a^* \, b
\label{AmplEqUsedScalarProduct}
\end{eqnarray}
taken over the whole range of the deformed sample. Application of this scalar product requires to explicitly expand all boundary values in terms of the surface deflection $\xi$.

The derivation of the required adjoint eigenvectors $\langle\Psi\!\mid$ for instabilities with a deformable surface is given in Ref.~\citen{Bohlius2007a}. Here we recall the results needed for the upcoming calculations. The components of the adjoint velocity field are given by
\begin{eqnarray}
\bar{v}_{x}
&=&
\bar{\omega}\frac{k_{i,x}}{k}\Big( e^{kz}
- \frac{2\bar{q}k}{\bar{q}^2+k^2}e^{\bar{q}z} \Big)
\frac{\bar{q}^2 + k^2}{\bar{q}^2 - k^2}\bar{\xi}_{i}
\label{AmplitudengleichungAdjointVelocityx}
\\
\bar{v}_{y}
&=&
\bar{\omega}\frac{k_{i,y}}{k}\Big( e^{kz}
- \frac{2\bar{q}k}{\bar{q}^2+k^2}e^{\bar{q}z} \Big)
\frac{\bar{q}^2 + k^2}{\bar{q}^2 - k^2}\bar{\xi}_{i}
\\
\bar{v}_z
&=&
i\bar{\omega}\Big( e^{kz} - \frac{2k^2}{\bar{q}^2 + k^2}e^{\bar{q}z} \Big)
\frac{\bar{q}^2 + k^2}{\bar{q}^2 - k^2}\bar{\xi}_{i}\,,
\label{AmplitudengleichungAdjointVelocityz}
\end{eqnarray}
where the adjoint frequency $\bar{\omega}$ is given by $-\omega$ and the adjoint inverse transverse decay length $\bar{q}$ is defined by $\bar{q}^2 = k^2 - \rho\bar{\omega}^2/(\mu_2 - i\bar{\omega}\nu_2)$ in the same way as the non-adjoint one, ${q}^2 = k^2 - \rho {\omega}^2/(\mu_2 + i {\omega}\nu_2)$. Here, $k_{i,x}$ and $k_{i,y}$ are the $x$ and $y$ component of the wave vector of mode $i$. The corresponding components of the adjoint strain field turn out to be
\begin{eqnarray}
\bar{\epsilon}_{zz} \label{AmplitudeEquationAdjointStrainzz}
&=&
2\mu_2 k \Big( e^{kz} - \frac{2\bar{q}k}{\bar{q}^2
+ k^2}e^{\bar{q}z} \Big)\frac{\bar{q}^2 + k^2}{\bar{q}^2 - k^2}\bar{\xi}_{i}
\\
\bar{\epsilon}_{xx}
&=&
2\mu_2 \frac{k_{i,x}^2}{k}\Big( e^{kz} - \frac{2\bar{q}k}{\bar{q}^2
+ k^2}e^{\bar{q}z} \Big)\frac{\bar{q}^2 + k^2}{\bar{q}^2 - k^2}\bar{\xi}_{i}
 \\ 
\bar{\epsilon}_{yy}
&=&
2\mu_2 \frac{k_{i,y}^2}{k}\Big( e^{kz} - \frac{2\bar{q}k}{\bar{q}^2
+ k^2}e^{\bar{q}z} \Big)\frac{\bar{q}^2 + k^2}{\bar{q}^2 - k^2}\bar{\xi}_{i}
 \\ 
\bar{\epsilon}_{xy}
&=&
-4\mu_2\frac{k_{i,x}k_{i,y}}{k}\Big( e^{kz} - \frac{2\bar{q}k}{\bar{q}^2
+ k^2}e^{\bar{q}z} \Big)
\frac{\bar{q}^2 + k^2}{\bar{q}^2 - k^2}\bar{\xi}_{i}
\quad \\
\bar{\epsilon}_{xz}
&=&
-4i\mu_2 k_{i,x}
\Big( e^{kz} - e^{\bar{q}z} \Big)\frac{\bar{q}^2 + k^2}{\bar{q}^2 - k^2}\bar{\xi}_{i}
\\
\bar{\epsilon}_{yz}
&=&
-4i\mu_2 k_{i,y}
\Big( e^{kz} - e^{\bar{q}z} \Big)\frac{\bar{q}^2 + k^2}{\bar{q}^2 - k^2}\bar{\xi}_{i}\, .
\label{AmplitudeEquationAdjointStrainyz}
\end{eqnarray}
%
%

\section{The second order}

The fact that within our assumptions the magnetic bulk equations completely decouple from the hydrodynamic bulk equations, has two important consequences. On the one hand this allows us to discuss and solve these two systems subsequently, i.e. we first solve the magnetic part in a given perturbative order for a given surface deflection $\xi$, and feed back this solution into the respective order of the hydrodynamic system. On the other hand, however, we have to face the problem, that the control parameter (the magnetization or the magnetic field in our case) does not occur in the hydrodynamic bulk equations and that the bulk equations for the magnetic system are homogeneous in all perturbative orders, which makes it impossible to obtain the control parameter in the next order by Fredholm's alternative, only. The coupling between these two systems is, however, mediated by the surface, and more precisely by the normal stress boundary condition. Satisfying the normal stress boundary condition provides us with an additional condition supplementing Fredholm's theorem.

\subsection{Magnetic contributions}

We start by solving the magnetic system of bulk equations together with the corresponding boundary conditions. The external magnetic field is the control parameter defining the expansion (\ref{ExpansionFields01}). Due to the deformations of the surface the actual magnetic field will be subject to perturbations and we separate the total magnetic field into the applied field ${\bf H}$ and the distortion field ${\bf h}$. These perturbations still obey the linear electromagnetic equations (${\bf b}=\mu {\bf h}$ and $\bm{\partial}\!\cdot\!{\bf b} = 0 = \bm{\partial}\!\times\!{\bf h}$), which allows for the introduction of a magnetic scalar potential ${\bf h}=-\bm{\partial}\Phi$. The perturbation potentials $\Phi$ (and $\Phi^{\mathrm{vac}}$ in the vacuum) are also expanded according to Eq.~(\ref{ExpansionFields01}) and fulfill the Laplace equation. In second order we obtain
\begin{eqnarray}
\Delta \Phi^{(2)} = 0
\quad & \mathrm{and} &\quad
\Delta \Phi^{(2)\mathrm{vac}} = 0
\label{LaplaceEquationsSecondOrder}
\end{eqnarray}
in the medium and in vacuum, respectively. In the second order of the $\epsilon-$expansion the magnetic boundary conditions for the tangential component of the total magnetic field ${\bf H} + {\bf h}$ can be simplified to (cf. App. \ref{AppMag}) 
\begin{eqnarray}
\partial_{x}\Phi^{(2)\mathrm{vac}}
-
\partial_{x}\Phi^{(2)}
&=&
\frac{2\mu}{1 + \mu}M_{c} \,\partial_{x}\xi^{(2)}
+  M^{(1)} \partial_{x} \xi^{(1)}
\label{TangBoundMag01}
\\
\partial_{y}\Phi^{(2)\mathrm{vac}}
-
\partial_{y}\Phi^{(2)}
&=&
\frac{2\mu}{1 + \mu}M_{c} \,\partial_{y}\xi^{(2)}
+  M^{(1)} \partial_{y} \xi^{(1)}\,,
\label{TangBoundMag02}
\end{eqnarray}
while the boundary condition for the normal component of the flux density ${\bf B}$ reads
\begin{eqnarray}
\partial_{z} \Phi^{(2)\mathrm{vac}}
-
\mu \partial_{z} \Phi^{(2)} 
&=&
-\frac{\mu}{1 + \mu} M_{c}
\sum_{i,j} (k_{1ij}^2\xi_{i}\xi_{j} + k_{2ij}^2\xi_{i}\xi^{\ast}_{j}
+ c.c. )\,,
\label{NormBoundMag01}
\end{eqnarray}
where we introduced abbreviations that depend on the angle $\theta_{ij}$ between the $i$-th and the $j$-th main characteristic mode
\begin{eqnarray} \label{Defk1ij}
k_{1ij} &=& k_{c}\sqrt{2 + 2\cos\theta_{ij}} \\
k_{2ij} &=& k_{c}\sqrt{2 - 2\cos\theta_{ij}} \label{Defk2ij}\,.
\end{eqnarray}

A convenient ansatz for the magnetic scalar potentials to solve this system of equations consists of two contributions. The first contribution $\Phi^{(2,1)}$ is proportional to the linear deflection $\xi^{(1)}$ to account for the contributions proportional to ${\bf M}^{(1)}$ in the boundary conditions (\ref{TangBoundMag01}) and (\ref{TangBoundMag02}). This automatically satisfies the Laplace equation (\ref{LaplaceEquationsSecondOrder}) for $\Phi^{(2,1)}$ \cite{Bohlius2006a}. The second contribution $\Phi^{(2,2)}$ accounts for the higher harmonic couplings of the linear characteristic modes proportional to $\xi^{(2)}$, which are modeled by the product of two characteristic modes
\begin{eqnarray}
\xi^{(2)}
&=&
k_{c} \sum_{i,j} (\xi_{i}\xi_{j} + \xi_{i}\xi^{\ast}_{j} + c.c.)\,.
\label{AnsatzSecondDeflection}
\end{eqnarray}
The characteristic wave vector $k_{c}$ in Eq.\@ (\ref{AnsatzSecondDeflection}) is just added to give $\xi^{(2)}$ the same unit as $\xi^{(1)}$.

The Laplace equation (\ref{LaplaceEquationsSecondOrder}) for $\Phi^{(2,2)}$  is satisfied by the ansatz
\begin{eqnarray}
\Phi^{(2,2)}
&=&
 k_c \sum_{i,j} (
\hat{\Phi}^{(2,2)R}_{ij}  \xi_{i}\xi_{j} e^{k_{1ij} z}
+ \hat{\Phi}^{(2,2)L}_{ij}  \xi_{i}\xi^{\ast}_{j} e^{k_{2ij} z}
+ c.c. )
\label{MagneticPotentialSecond}
\end{eqnarray}
and by a corresponding one for the magnetic potential in vacuum. 

The boundary conditions for the different Fourier modes decouple and can be satisfied separately. We obtain for the contributions proportional to $\xi^{(1)}$ 
\begin{eqnarray} \label{PhiMag2}
\Phi^{(2,1)}
&=&
-\frac{M^{(1)}}{1+\mu}\xi^{(1)} e^{k_{c} z}
\\ \label{PhiMag2vac}
\Phi^{(2,1)\mathrm{vac}}
&=&
\frac{\mu M^{(1)}}{1+\mu}\xi^{(1)} e^{-k_{c} z}\,,
\end{eqnarray}
which are of the same structure as in the linear case. The presence of $M^{(1)}$ guarantees $\Phi^{(2,1)}$ to be of second order.

The contributions due to the higher harmonics of the characteristic modes read
\begin{eqnarray}
\hat{\Phi}^{(2,2)R}_{ij}
&=&
\frac{\mu}{(1 + \mu)^2}M_{c}
  \Big( \frac{k_{1ij}}{k_{c}} - 2 \Big)
\label{MagPotHigherHarmonic}
\\
\hat{\Phi}^{(2,2)R\mathrm{vac}}_{ij}
&=&
\frac{\mu^2}{(1 + \mu)^2}M_{c}
  \Big( \frac{k_{1ij}}{\mu k_{c}} - 2 \Big)\,,
\label{MagPotHigherHarmonicVac}
\end{eqnarray}
while $\hat{\Phi}^{(2,2)L}_{ij}$ and $\hat{\Phi}^{(2,2)L\mathrm{vac}}_{ij}$ are obtained  replacing $k_{1ij}$ by $k_{2ij}$ in Eqs.\@ (\ref{MagPotHigherHarmonic}) and Eqs.\@ (\ref{MagPotHigherHarmonicVac}), respectively.

%
%

\subsection{Hydrodynamic contributions\label{HydroContribSecondOrder}}

According to the general expression (\ref{GeneralSecondOrderEquation}) the set of hydrodynamic bulk equations is given in the second perturbative order by
\begin{eqnarray}
\rho\partial_{t}^{(0)} v_{i}^{(2)}
+
\partial_{i} p^{(2)}
&-&
2\mu_2\partial_{j}\epsilon^{(2)}_{ij}
-
\nu_{2} \big(\partial_{j}\partial_{i}v^{(2)}_{j}
           + \partial_{j}\partial_{j}v^{(2)}_{i} \big)
\nonumber \\
&=&
-\rho\partial^{(1)}_{t} v^{(1)}_{i}
- \partial_{j} \big(\rho v^{(1)}_{i} v^{(1)}_{j}
                          - 2\mu_{2}\epsilon^{(1)}_{jk}\epsilon^{(1)}_{ki}
                     \big)
\label{SecondOrderNavierStokes}
\\
\partial^{(0)}_{t} \epsilon^{(2)}_{ij}
- \frac{1}{2} \big( \partial_{i} v^{(2)}_{j}
                      + \partial_{j} v^{(2)}_{i} \big)
&=& - \partial^{(1)}_{t} \epsilon^{(1)}_{ij}
- v^{(1)}_{k}\partial_{k}\epsilon^{(1)}_{ij}
\label{SecondOrderElasticity}
\\
\partial_{i} v^{(2)}_{i} &=& 0\,.
\label{SecondOrderContinuity}
\end{eqnarray}
The structure of these equations suggests two kind of solutions similar to the magnetic part. One contribution is proportional to the main characteristic modes $\xi^{(1)}$ and a second one proportional to the second harmonics $\xi^{(2)}$, Eq.~(\ref{AnsatzSecondDeflection}).

The corresponding boundary conditions at the surface $z=\xi$ are expanded in the same manner (for a detailed discussion cf. App. \ref{AnhangHydroBC}). For the tangential contributions we obtain
\begin{eqnarray}
2\mu_{2}\epsilon^{(2)}_{yz}
+
\nu_{2}
\big( \partial_{z}v^{(2)}_{y} + \partial_{y}v^{(2)}_{z} \big)
&=& \Omega_{yz}^{(2)}
\\
2\mu_{2}\epsilon^{(2)}_{xz}
+
\nu_{2}
\big( \partial_{z}v^{(2)}_{x} + \partial_{x}v^{(2)}_{z} \big)
&=& \Omega_{xz}^{(2)}\,, \label{SecondOrderBCtang2}
\end{eqnarray}
where the inhomogeneities are abbreviated by $\Omega^{(2)}_{ij}$ and are listed in App. \ref{AnhangHydroBC},  Eqs.\@ (\ref{AnhangHydroBCTangentialBC01SecondOrder}) and (\ref{AnhangHydroBCTangentialBC02SecondOrder}). In contrast to the gravitational and the magnetic force the surface tension always acts normal to the surface and therefore  enters only the normal stress boundary condition 
\begin{eqnarray}
2\mu_{2}\epsilon^{(2)}_{zz}
&+&
2\nu_{2}\partial_{z}v^{(2)}_{z}
- p^{(2)}
+ G\rho\xi^{(2)}
- 
       \mu H_{c}\partial_{z}\Phi^{(2)}
       + \mu_{0}H_{c}^{\mathrm{vac}}\partial_{z}\Phi^{(2)\mathrm{vac}}
\nonumber \\
&=& \Omega^{(2)}_{zz}
- \sigma_{T} \Delta \xi^{(2)}
+ \frac{\mu}{1+\mu}M^{(1)}M_{c}k_{c}\xi^{(1)}
\end{eqnarray}
with $\Omega^{(2)}_{zz}$ given in Eq.~(\ref{AnhangHydroBCNormalBCSecondOrder}).
Finally, the kinematic boundary condition describing explicitly the deformable surface reads in second order
\begin{eqnarray}
\partial^{(0)}_{t}\xi^{(2)}
+ \partial^{(1)}_{t}\xi^{(1)}
+ ({\bf v}^{(1)}\cdot\bm{\partial}) \,\xi^{(1)}
&=&
v^{(2)}_{z} + \xi^{(1)}\partial_{z}v^{(1)}_{z}\,.
\label{SecondOrderKinematicBC}
\end{eqnarray}
The last contribution in Eq.~(\ref{SecondOrderKinematicBC}) is due to the fact, that in second order the surface, at which the boundary conditions have to be evaluated, is already deflected.

\subsubsection{The solvability condition in second order}
\label{FredholmSecondOrder}

The general solvability condition discussed in \S{A}~\ref{SecFredholm} is applied to the set of second order equations (\ref{SecondOrderNavierStokes}-\ref{SecondOrderContinuity}) and explicitly reads
\begin{eqnarray}
\langle\bar{v}_{i}\!\mid\!
 -\rho\partial^{(1)}_{t} v^{(1)}_{i}
     -\partial_{j}
      (\rho v^{(1)}_{i}v^{(1)}_{j}
       - 2\mu_{2}\epsilon^{(1)}_{jk}\epsilon^{(1)}_{ki} )
 \rangle
\qquad\nonumber \\
+
\langle \bar{\epsilon}_{ij} \!\mid\!
- \partial^{(1)}_{t}\epsilon^{(1)}_{ij}
- v^{(1)}_{k}\partial_{k} \epsilon^{(1)}_{ij}
\rangle
&=& 0\,. \quad
\label{ExplFredholmSecondOrder}
\end{eqnarray}
At this point one might be tempted to use the fact that the Rosensweig instability is a stationary one (in linear approximation) and substitute $\omega^{(0)} = \sigma^{(0)} = 0$ as well as the stationary limits of the adjoint and original eigenvectors into condition (\ref{ExplFredholmSecondOrder}). The solvability condition would then reduce to
\begin{eqnarray}
\langle\bar{\epsilon}_{ij} \!\mid\!
- \partial^{(1)}_{t} \epsilon^{(1)}_{ij} \rangle
=
( \pm\, i \omega^{(1)} + \sigma^{(1)} )
\langle\bar{\epsilon}_{ij} \!\mid\! \epsilon_{ij} \rangle
&=&
0
\end{eqnarray}
corresponding to the solution $\omega^{(1)} = 0 = \sigma^{(1)}$.
Here, we have replaced $\partial_t^{(1)}$ by $\pm\,i \omega^{(1)} + \sigma^{(1)}$ (for right- and left-traveling waves, respectively) implying a normal mode ansatz for the time dependence of the amplitudes.
Of course, $\omega^{(0)}=0$ is the correct solution in the stationary limit. However, in that limit the connection between bulk equations and boundary conditions is lost  (cf. Eqs.~(\ref{Elasticitygenerell}) and (\ref{GeneralKinematicBC})) and an amplitude equation cannot be derived. Therefore, one must still treat the system as fully dynamic at least at those places related to the kinematic boundary condition and to the velocity/strain relation, and satisfy Fredholm's theorem with the derivative $\partial_{t}^{(0)}$ being finite. One can, however, at non-crucial instances simplify the calculations by the fact that $\omega^{(0)}$ is small, but only at the very end one can take $\omega^{(0)} \equiv 0$.

The solvability condition (\ref{ExplFredholmSecondOrder}) consists of two different parts. One containing spatial derivatives and the other the (scaled) time derivative $\partial^{(1)}_{t}$. We first discuss the latter part. The integration upon $x$ and $y$ is straightforwardly done and only retains contributions that are proportional to $\delta({\bf k}_{i} - {\bf k}_{j})$. After integration with respect to $z$ we end up with the following expression, 
\begin{eqnarray} 
\!\langle\bar{v}_{i} \!\!&\mid&\!
\rho\partial^{(1)}_{t} v^{(1)}_{i} \rangle
+
\langle\bar{\epsilon}_{ij}\!\mid\!
\partial^{(1)}_{t} \epsilon^{(1)}_{ij} \rangle
\nonumber \\
&=&
i\omega^{(1)}\Big(
\hat{\xi}^{\ast}_{iL}\hat{\xi}_{iR}e^{2i\omega t}
- \hat{\xi}_{iL}\hat{\xi}^{\ast}_{iR}e^{-2i\omega t}
\Big) e^{2\sigma t}
\Big\lbrace
8 \mu_2
\frac{k_c(k_c^2 + q^2)^2}
     {q(k_c+q)^3}
\nonumber \\ &&
-\rho([\omega^{(0)}]^2\!-\![\sigma^{(0)}]^2)
\frac{4 k_c^6 + 6 k_c^5 q+ 6 k_c^4q^2 + 6 k_c^3 q^3 + 2 k_c^2 q^4}
     {q k_c^3 (k_c + q)^3}
\Big\rbrace
\nonumber \\
&+&
\sigma^{(1)}\Big(
\hat{\xi}^{\ast}_{iL} \hat{\xi}_{iR} e^{2i\omega t}
+ \hat{\xi}_{iL} \hat{\xi}^{\ast}_{iR}e^{-2i\omega t}
+ \hat{\xi}_{iR}\hat{\xi}^{\ast}_{iR}
+ \hat{\xi}_{iL}\hat{\xi}^{\ast}_{iL}
\Big) e^{2\sigma t}
\Big\lbrace
8 \mu_2
\frac{k_c(k_c^2 + q^2)^2}
     {q(k_c+q)^3} 
\nonumber \\ &&
-\rho([\omega^{(0)}]^2\!-\![\sigma^{(0)}]^2)
\frac{4 k_c^6 + 6 k_c^5 q+ 6 k_c^4q^2 + 6 k_c^3 q^3 + 2 k_c^2 q^4}
     {q k_c^3 (k_c + q)^3}
\Big\rbrace \,.
\end{eqnarray}
For the second order contributions we finally get
\begin{eqnarray} \label{solv2dt}
\langle\bar{v}_{i}\!&\mid&\!
\partial^{(1)}_{t} ( \rho v^{(1)}_{i} ) \rangle
+
\langle\bar{\epsilon}_{ij}\!\mid\!
\partial^{(1)}_{t} \epsilon^{(1)}_{ij} \rangle
\nonumber \\
&=&
i\omega^{(1)}
4\mu_2k_c (\hat{\xi}^{\ast}_{iL}\hat{\xi}_{iR} - \hat{\xi}_{iL}\hat{\xi}^{\ast}_{iR})
\nonumber \\ &&
+
\sigma^{(1)}
4\mu_2k_c (\hat{\xi}^{\ast}_{iL}\hat{\xi}_{iR} + \hat{\xi}_{iL}\hat{\xi}^{\ast}_{iR}
+\hat{\xi}_{iR}\hat{\xi}^{\ast}_{iR}
+\hat{\xi}_{iL}\hat{\xi}^{\ast}_{iL}
)\,,
\end{eqnarray}
where the static limit has safely been performed.

Up to now it has been possible to do the calculations without specifying the actual number of modes contributing to the nonlinear pattern and the results are applicable for any value of $N$ and in particular for any angle between these modes. This is changed when the second part of Eq.~(\ref{ExplFredholmSecondOrder})), containing the spatial derivatives, is considered. Two of these terms turn out to be irrelevant for the second order solvability condition and are not discussed here. The only relevant term, $2\mu_{2}
\langle\bar{v}_{i} \!\mid\! \partial_{j} ( \epsilon^{(1)}_{jk}\epsilon^{(1)}_{ki} )
\rangle$, generally vanishes, except when three linear modes oriented at $\pi/3$ relative to each other are interacting. This hexagonal order is enforced by the integration upon $x$ and $y$. Integrating with respect to $z$ yields in lowest order of $\omega^{(0)}$ and $\sigma^{(0)}$
\begin{eqnarray} \label{solv2grad}
2\mu_{2}
\langle\bar{v}_{i}
\partial_{j} ( \epsilon^{(1)}_{jk}\epsilon^{(1)}_{ki} )
\rangle
&=&
-3 i\omega^{(0)} \mu_2 k_c^2
\big(
\hat{\xi}_{1R}\hat{\xi}_{2R}\hat{\xi}_{3R}
-\hat{\xi}_{1L}\hat{\xi}_{2L}\hat{\xi}_{3L}
+\hat{\xi}_{1R}\hat{\xi}_{2R}\hat{\xi}_{3L}
\nonumber \\ && \quad
+\hat{\xi}_{1R}\hat{\xi}_{2L}\hat{\xi}_{3R}
-\hat{\xi}_{1L}\hat{\xi}_{2R}\hat{\xi}_{3R}
-\hat{\xi}_{1L}\hat{\xi}_{2L}\hat{\xi}_{3R}
\nonumber \\ && \quad
-\hat{\xi}_{1L}\hat{\xi}_{2R}\hat{\xi}_{3L}
+\hat{\xi}_{1R}\hat{\xi}_{2L}\hat{\xi}_{3L}
- c.c.
\big)
\nonumber \\ &&
-3 \sigma^{(0)} \mu_2 k_c^2
\big(
\hat{\xi}_{1R}\hat{\xi}_{2R}\hat{\xi}_{3R}
+\hat{\xi}_{1L}\hat{\xi}_{2L}\hat{\xi}_{3L}
+\hat{\xi}_{1R}\hat{\xi}_{2R}\hat{\xi}_{3L}
\nonumber \\ && \quad
+\hat{\xi}_{1R}\hat{\xi}_{2L}\hat{\xi}_{3R}
+\hat{\xi}_{1L}\hat{\xi}_{2R}\hat{\xi}_{3R}
+\hat{\xi}_{1L}\hat{\xi}_{2L}\hat{\xi}_{3R}
\nonumber \\ && \quad
+\hat{\xi}_{1L}\hat{\xi}_{2R}\hat{\xi}_{3L}
+\hat{\xi}_{1R}\hat{\xi}_{2L}\hat{\xi}_{3L}
+ c.c.
\big)\,.
\end{eqnarray}
Eqs.~(\ref{solv2dt}) and (\ref{solv2grad}) are the two parts that enter the solvability condition Eq.~(\ref{ExplFredholmSecondOrder}), which we are now going to solve. The imaginary part yields the condition 
\begin{eqnarray}
4 i\omega^{(1)}
(\hat{\xi}^{\ast}_{iL}\hat{\xi}_{iR} \!-\! \hat{\xi}_{iL}\hat{\xi}^{\ast}_{iR})
&=&
- 3 i\omega^{(0)} k_c
(
\hat{\xi}_{1R}\hat{\xi}_{2R}\hat{\xi}_{3R}
\!-\!\hat{\xi}_{1L}\hat{\xi}_{2L}\hat{\xi}_{3L}
\!+\!\hat{\xi}_{1R}\hat{\xi}_{2R}\hat{\xi}_{3L}
\nonumber \\ && \quad
\!+\hat{\xi}_{1R}\hat{\xi}_{2L}\hat{\xi}_{3R}
\!-\!\hat{\xi}_{1L}\hat{\xi}_{2R}\hat{\xi}_{3R}
\!-\!\hat{\xi}_{1L}\hat{\xi}_{2L}\hat{\xi}_{3R}
\nonumber \\ && \quad
\!-\hat{\xi}_{1L}\hat{\xi}_{2R}\hat{\xi}_{3L}
\!+\!\hat{\xi}_{1R}\hat{\xi}_{2L}\hat{\xi}_{3L}
\!-\! c.c.
) \quad\,.
\end{eqnarray}
This condition is identically fulfilled by the ansatz
\begin{eqnarray}
\hat{\xi}_{iL} = \hat{\xi}_{iR} = \hat{\xi}_{i} \quad\quad \mathrm{and} \quad\quad
\label{StaticSolutionCondition}
\hat{\xi}_{iL}^{\ast} = \hat{\xi}_{iR}^{\ast} = \hat{\xi}_{i}^{\ast}\,,
\end{eqnarray}
which is the solution one expects for the stationary case, since in that limit one cannot distinguish right from left traveling waves.

Using this result for evaluating the real part, we obtain
\begin{eqnarray}
2 \sigma^{(1)}
 \sum\limits_{i} \hat{\xi}_{i}\hat{\xi}_{i}^{\ast} 
&=&
- 3 \sigma^{(0)}  k_c
(\hat{\xi}_{1}\hat{\xi}_{2}\hat{\xi}_{3}
+
\hat{\xi}_{1}^{\ast}\hat{\xi}_{2}^{\ast}\hat{\xi}_{3}^{\ast})\,,
\end{eqnarray}
which obviously is solved by
\begin{eqnarray}
\sigma^{(1)} \hat{\xi}_{1} = -\sigma^{(0)} k_c  \hat{\xi}_{2}^{\ast}\hat{\xi}_{3}^{\ast}  \quad\quad {\rm and} \quad\quad  | \hat{\xi}_{1}|^2 = | \hat{\xi}_{2}|^2= | \hat{\xi}_{3}|^2
\label{PrimitiveAmplEqSecondOrderReal}
\end{eqnarray}
and all its cyclic permutations $1\to 2 \to 3 \to 1$ and their complex conjugates. Equation (\ref{PrimitiveAmplEqSecondOrderReal}) tells us, that the slow variable $\sigma^{(1)}$ scales in the bulk with  $\sigma^{(0)}$, indicating that $\sigma^{(1)}/\sigma^{(0)}$ stays finite in the stationary limit. This behavior is mediated by the the kinematic boundary condition $d_{t}\xi=v_{z}$ (\ref{GeneralKinematicBC}). As a consequence, the velocity field as well as the adjoint velocity field are proportional to the time derivative as we realized in Ref.~\citen{Bohlius2006b} and in Eqs.~(\ref{AmplitudengleichungAdjointVelocityx})-(\ref{AmplitudengleichungAdjointVelocityz}). This is physically reasonable, since in the case of the Rosensweig instability the velocity field vanishes if the surface pattern has fully developed and the hydrodynamic bulk equations are trivially fulfilled by ${\bf v}\equiv 0$, the same solution as for the initial undeformed ground state. This singular behavior, unique for the Rosensweig instability, is scaled out by the choice of a dimensionless time derivative $\tilde{\partial}_{t}^{(1)} = \sigma^{(1)}/\sigma^{(0)}$ for the bulk hydrodynamic equations. Using this time derivative, Eq.~(\ref{PrimitiveAmplEqSecondOrderReal}) can be rewritten as
\begin{eqnarray} \label{PrimitiveAmplEqSecondOrderReal2}
\tilde{\partial}^{(1)}_{T}\hat{\xi}_{1}
&=&
-k_c \hat{\xi}_{2}^{\ast}\hat{\xi}_{3}^{\ast}\,.
\end{eqnarray}

Equation (\ref{PrimitiveAmplEqSecondOrderReal2}) gives the relation among the three amplitudes of the second order deflection, $\xi^{(1)}$, characteristic for hexagon patterns. 
For any other regular pattern the right hand side of Eq.~(\ref{solv2grad}) is zero
implying, that there is no nonlinear interaction between two different modes in the second order for those patterns.

What is missing in Eq.\@ (\ref{PrimitiveAmplEqSecondOrderReal2}), which in a sense can be viewed as a primitive form of an amplitude equation, is a contribution proportional to the control parameter ${\bf M}^{(1)}$. This is due to the fact, that the two bulk systems of magnetic and hydrodynamic equation decouple completely. The control parameter enters the amplitude equation via the normal stress boundary condition, the only way magnetic and hydrodynamic subsystems are interacting.

\subsubsection{Solutions proportional to $\xi^{(1)}$\label{HydroSolutSecondOrderMainModes}}

Before we can exploit the normal stress boundary condition in \S\ref{SecondOrderFredholmBoundary}, we have to determine the solution of the hydrodynamic contributions, Eqs.~(\ref{SecondOrderNavierStokes}) - (\ref{SecondOrderBCtang2}).  From Fredholm's theorem we learned, under what conditions we can find a solution to the system of equations in the second perturbative order. As in the magnetic part, we distinguish solutions of the system of equations that are either proportional to $\xi^{(1)}$ or proportional to $\xi^{(2)}$. In this subsection we concentrate on the part proportional to $\xi^{(1)}$. Inspired by the linear discussion, we use a scalar $\varphi^{(2,1)}$ and a vector potential ${\bf \Psi}^{(2,1)}$ for the potential and the vorticity flow, respectively. For the contributions proportional to the main characteristic modes $\xi^{(1)}$, the governing equations read
\begin{eqnarray}
\Delta\varphi^{(2,1)} &=& 0 \label{HydroSecondOrderIncompr}
\\
\rho\Delta\partial^{(0)}_{t}\varphi^{(2,1)} + \Delta p^{(2,1)}
&=&
-\rho\Delta\partial^{(1)}_{t}\varphi^{(1)}
\label{HydroSecondOrderPressure}
\\
\rho(\partial_{t}^{(0)})^3 \Psi^{(2,1)}_{i}
- \tilde{\mu}_{2}\Delta\partial^{(0)}_{t}\Psi^{(2,1)}_{i}
&=&
- \mu_{2}\Delta\partial^{(1)}_{t}\Psi^{(1)}_{i}
- \rho(\partial_{t}^{(0)})^2\partial^{(1)}_{t}\Psi^{(1)}_{i}
\label{HydroSecondOrderVectorPotential}
\end{eqnarray}
with the abbreviation $\tilde{\mu}_{2} = \mu_{2} + \nu_{2}\partial_{t}^{(0)}$. On the right hand side of these equations the first order (linear) potentials act as inhomogeneities. They are listed in Ref.~\citen{Bohlius2006a}.

The appropriate boundary conditions for the flow potentials are derived in App. \ref{AppSubSecSecondOrderBC} and read for tangential stress
\begin{eqnarray}
\tilde{\mu}_{2} \big( \partial^{2}_{z}-\partial^{2}_{y} \big) \Psi^{(2,1)}_{x}
+ \tilde{\mu}_{2} \partial_{y}\partial_{x} \Psi^{(2,1)}_{y}
+ 2 \tilde{\mu}_{2} \partial_{z}\partial_{y}\varphi^{(2,1)}
&=& 0
\label{Tangxi1yz}
\\
\tilde{\mu}_{2} \big( \partial^{2}_{z}-\partial^{2}_{x} \big) \Psi^{(2,1)}_{y}
+ \tilde{\mu}_{2} \partial_{x}\partial_{y} \Psi^{(2,1)}_{x}
- 2 \tilde{\mu}_{2} \partial_{z}\partial_{x}\varphi^{(2,1)}
&=& 0\,.
\label{Tangxi1xz}
\end{eqnarray}
The physical boundary conditions have to be taken at $z= \xi^{(1)}$ in the second order. This leads to additional contributions in $\xi^{(1)}$, which have already been taken into account in the effective boundary conditions Eqs.~(\ref{Tangxi1yz}) and (\ref{Tangxi1xz}). The latter therefore have to be taken at $z=0$.

The kinematic boundary condition now involves the slow timescale $t^{(1)}$ and reads 
\begin{eqnarray} \label{HydroSecondOrderKinemat}
v_{z}^{(2,1)} &=& \partial^{(1)}_{t}\xi^{(1)}\,.
\end{eqnarray}

We start with the particular inhomogeneous solutions of Eqs.~(\ref{HydroSecondOrderPressure}) and (\ref{HydroSecondOrderVectorPotential}) for the vector potential $\bf\Psi$ and the pressure $p$, respectively. It can be checked that the following fields satisfy the inhomogeneous bulk equations
\begin{eqnarray}
\Psi_{i}^{(2,1)}=\hat{\Psi}_{i}^{(2,1)\mathrm{inhom}}\xi^{(1)}ze^{qz}
\quad&\mathrm{and}&\quad
p^{(2,1)\mathrm{inhom}} = -\rho\partial_{t}^{(1)}\varphi^{(1)}
\label{SolutionsSecondOrderMainModesInhomo}
\end{eqnarray}
with the operators defined by
\begin{eqnarray}
\hat{\Psi}^{(2,1)\mathrm{inhom}}_{x}
=
- \frac{\mu_2+\tilde{\mu}_{2}}{\tilde{\mu}_{2} q}\partial^{(1)}_{t}\partial_{y}
\quad\mathrm{and}\quad
\hat{\Psi}^{(2,1)\mathrm{inhom}}_{y}
=
\frac{\mu_2+\tilde{\mu}_{2}}{\tilde{\mu}_{2} q}\partial^{(1)}_{t}\partial_{x}\,.
\end{eqnarray}
The inhomogeneous solutions do not yet satisfy the boundary conditions (\ref{Tangxi1yz}) and (\ref{Tangxi1xz}). Substituting ${\bf\Psi}^{\mathrm{inhom}}$ into Eq.~(\ref{Tangxi1yz}) results in an additional source of tangential stress at the boundary due to the inhomogeneous solutions, which can be balanced by the homogeneous ones
\begin{eqnarray}
\tilde{\mu}_{2} \big( \partial^{2}_{z}-\partial^{2}_{y} \big) \Psi^{(2,1)\mathrm{hom}}_{x}
+
\tilde{\mu}_{2} \partial_{y}\partial_{x} \Psi^{(2,1)\mathrm{hom}}_{y}
&+&
2 \tilde{\mu}_{2} \partial_{z}\partial_{y}\varphi^{(2,1)}
\nonumber \\ 
&=&
\partial_{y}
\left(
\tilde{\mu}_{2}
\frac{\mu_{2}\!+\!\tilde{\mu}_{2}}{\tilde{\mu}_{2}}\partial_{t}^{(1)}\xi^{(1)}
\right)\,.
\end{eqnarray}
If we use the following ansatz for the homogeneous solutions of the flow potentials ${\bf\Psi}^{(2,1)\mathrm{hom}}$ and $\varphi^{(2,1)}$
\begin{eqnarray}
\Psi_{x}^{(2,1)\mathrm{hom}}
=
-\partial_{y}\hat{\Psi}^{(2,1)}e^{qz}\xi^{(1)}
,\quad
\Psi_{y}^{(2,1)\mathrm{hom}}
=
\partial_{x}\hat{\Psi}^{(2,1)}e^{qz}\xi^{(1)}
\nonumber \\[0.2cm]
\quad\mathrm{and}\quad
\varphi^{(2,1)} = \hat{\varphi}^{(2,1)}e^{k_cz}\xi^{(1)} \,,\quad\quad\quad\quad\quad
\label{SolutionsSecondOrderMainModesHomoVecPot}
\end{eqnarray}
the amplitudes $\hat{\Psi}^{(2,1)}$ are given by 
\begin{eqnarray}
\hat{\Psi}^{(2,1)} &=& \frac{2k_c}{q^2+k_c^2}\hat{\varphi}^{(2,1)}
-
2\frac{\mu_2+\tilde{\mu}_2}{\tilde{\mu}_2(q^2+k_c^2)}\partial_{t}^{(1)}\,.
\label{SolutionsSecondOrderMainModesVectorPotential}
\end{eqnarray}
Note that $q$ is the inverse decay length of the linear transverse modes with $q^2 = k_c^2 + \rho [\partial_{t}^{(0)}]^2/ (\mu_2 + \nu_2\partial_{t}^{(0)})$ \cite{Bohlius2006a} and $\partial^{(1)}_{t}$ is a short hand notation for $\pm \, i \omega^{(1)} + \sigma^{(1)}$, as before.

The homogeneous solution of the pressure $p^{(2,1)\mathrm{hom}}$ is straightforwardly given by Eq.~(\ref{HydroSecondOrderPressure})
\begin{eqnarray}
p^{(2,1)\mathrm{hom}} &=& -\rho\partial_{t}^{(0)}\varphi^{(2,1)}
\end{eqnarray}
and if we exploit the kinematic boundary condition (\ref{HydroSecondOrderKinemat}), the solution of the scalar flow potential $\varphi^{(2,1)}$ can be determined as
\begin{eqnarray}
\hat{\varphi}^{(2,1)}
&=&
\frac{q^2 + k_c^2}{k_c (q^2 - k_c^2)}
\left(
\partial_{t}^{(1)}
-
2k_c^2\frac{\mu_2 + \tilde{\mu}_2}{\tilde{\mu}_{2}(q^2 + k_c^2)}\partial_{t}^{(1)}
\right)\,.
\label{SolutionsSecondOrderMainModesScalarPotential}
\end{eqnarray}
With the help of the flow potentials the velocity fields are determined
\begin{eqnarray}
\!\!\!\!\! v_{z}^{(2,1)}
= &&
\left\{\! \left[
q^2\!-\!\frac{2\mu_{2}\!+\!\tilde{\mu}_{2}}{\tilde \mu_{2}}k_c^2\right] e^{k_cz} \!+\!
2\frac{\mu_2}{\tilde \mu_{2}} k_c^2 e^{qz}
\!-\!
\frac{\mu_{2}\!+\!\tilde \mu_{2}}{\tilde \mu_{2}} k_c^2(q^2\!-\!k_c^2) \frac{ z e^{qz}}{q}
\!\right\} 
\frac{\partial_{t}^{(1)}\xi_{i}^{(1)}}{q^2\!-\!k_c^2}
\nonumber \\[0.2cm]
\label{AmplitudeEquationVelocityzSecondOrderMainWave}
\end{eqnarray}
\begin{equation}
v_{x}^{(2,1)} = \frac{ik_{i,x}}{\tilde \mu_{2}(q^2\!-\!k_c^2)} L(z) \partial_{t}^{(1)}\xi_{i}^{(1)} \quad {\rm and} \quad 
v_{y}^{(2,1)} = \frac{ik_{i,y}}{\tilde \mu_{2}(q^2\!-\!k_c^2)} L(z) \partial_{t}^{(1)}\xi_{i}^{(1)}
\end{equation}
with the abbreviation
\begin{eqnarray}
L(z) =&& \bigl[ \tilde \mu_{2}(q^2 - k_c^2) - 2 \mu_2 k_c^2 \bigr] \frac{e^{k_cz}}{k_c}
 \nonumber \\ &&+ \left[ 2 \mu q^2 -(\mu_2 - \tilde \mu_2) (q^2 - k_c^2) (1+ qz) \right] \frac{e^{qz}}{q}\,,
\end{eqnarray}
from which the strain fields follow
\begin{eqnarray}
\epsilon_{zz}^{(2,1)}
&=&
-  \frac{\mu_{2}+\tilde{\mu}_{2}}{\tilde{\mu}_{2}} k_c^2
L_{+}(z)
\frac{\partial_{t}^{(1)}}{\partial_{t}^{(0)}}\xi^{(1)}_{i}
\\
\epsilon_{ab}^{(2,1)}
&=& \frac{\mu_{2}+\tilde{\mu}_{2}}{\tilde{\mu}_{2}}
k_{i,a} k_{i,b} L_{-}(z) 
\frac{\partial_{t}^{(1)}}{\partial_{t}^{(0)}}\xi^{(1)}_{i}
\\
\epsilon_{az}^{(2,1)}
&=&
ik_{i,a} \frac{\mu_{2}+\tilde{\mu}_{2}}{2\tilde{\mu}_{2}}
\left\lbrace\!
\frac{2}{q^2-k_c^2}
\left[
2k_c^2e^{k_c z}-(q^2+k_c^2)e^{qz}
\right]
\right. \nonumber \\ && \left.
\quad\quad\quad\quad\quad
+
\Bigl(1+qz+\frac{k_c^2}{q}z\Bigr)e^{qz}
\!\right\rbrace
\frac{\partial_{t}^{(1)}}{\partial_{t}^{(0)}}\xi_{i}^{(1)}
\end{eqnarray}
for $\{a,b\} \in \{x,y\}$ with
\begin{eqnarray}
L_{\pm} =  
\frac{2}{q^2\!-\!k_c^2}
\left(
k_c e^{k_c z} \!-\! q e^{qz}
\right)
 \pm \frac{1+qz}{q}
e^{qz}\,.
\label{AmplitudeEquationShortCutStrainFieldSecondOrder}
\end{eqnarray}

This concludes the derivation of the second order eigenfunctions  that are proportional to $\xi^{(1)}$. These solutions satisfy every condition except the normal stress boundary condition. The latter will be used to determine the still unknown first order correction to the control parameter, $M^{(1)}$, which finally enters the amplitude equation as the linear contribution. We postpone the actual derivation of these contributions to \S\ref{SecondOrderFredholmBoundary}.

\subsubsection{Solutions proportional to $\xi^{(2)}$\label{HydroSolutSecondOrderHarmonicModes}}

We are left with solving the system of hydrodynamic equations in the second perturbative order, Eqs.~(\ref{SecondOrderNavierStokes})-(\ref{SecondOrderContinuity}), for the higher harmonic contributions proportional to $\xi^{(2)}$. The appropriate set of bulk equations reads, if we use again the representation with a scalar potential and a vector potential,
\begin{eqnarray}
\Delta
\bigl[
\rho (\partial^{(0)}_{t})^2 \varphi^{(2,2)} + \partial^{(0)}_{t}p^{(2,2)}
\bigr]
&=&
\partial_{i}
\bigl[
- 2\mu_{2}\partial_{j}(v^{(1)}_{k}\partial_{k}\epsilon^{(1)}_{ij})
\label{VelPotentialSecO2}
 \nonumber \\
&& 
\,\,\,\,\,\,\, - \partial^{(0)}_{t}\partial_{j}
(
\rho v^{(1)}_{i}v^{(1)}_{j} - 2\mu_{2}\epsilon^{(1)}_{jk}\epsilon^{(1)}_{ki}
)
\bigr] 
\\ \nonumber
\bigl[ \rho(\partial^{(0)}_{t})^2
- \tilde{\mu}_{2}\Delta \bigr] \bigl[\partial_{i}\partial_{m}
\Psi^{(2,2)}_{m} 
&-&
\Delta \Psi^{(2,2)}_{i} \bigr] 
\label{VorticitySecondOrderHigherHarmonics}
\\ \nonumber
&=& 
\epsilon_{ijk}
\partial_{j}
\big[
-
2\mu_{2}\partial_{m}(v^{(1)}_{l}\partial_{l}\epsilon^{(1)}_{km})
\nonumber \\ && \qquad\quad
-
\partial^{(0)}_{t}\partial_{l}
(\rho v^{(1)}_{k}v^{(1)}_{l}
- 2\mu_{2}\epsilon^{(1)}_{lm}\epsilon^{(1)}_{km}
)
\big]
\\
\Delta\varphi^{(2,2)}
&=&
0\,.
\label{ContinuitySecondOrderHigherHarmonics}
\end{eqnarray}
The first equation determines the pressure contribution $p^{(2,2)}$. Since the pressure appears only in the normal stress boundary condition, this is dealt with in the subsequent section. 
Next we construct a particular inhomogeneous solution of Eq.~(\ref{ContinuitySecondOrderHigherHarmonics}) for the vector potential ${\bf \Psi}$. The most general ansatz necessary reads 
\begin{eqnarray}
\Psi_{k}^{(2,2)\mathrm{inhom}}
&=& -\epsilon_{zkl} \sum\limits_{N,M}
\sum\limits_{i,j} \partial_l  \Bigl(
{\Psi}_{NMij}^{\mathrm{inhom}}(z)\,\xi_{iN}\xi_{jM} +  \tilde{\Psi}_{NMij}^{\mathrm{inhom}}(z)\,
\xi_{iN}^*\xi_{jM} 
+ c.c. \Bigr)\,.
\nonumber \\
\label{InhomSolSecOrderPsi}
\end{eqnarray}
Here, summation over all relevant modes $i,j$ is implied (e.g. $\{i,j\} \in \{1,2,3\}$, $\{i,j\} \in \{1,5\}$, and $i=j=1$ for hexagons, squares, and rolls, respectively, Fig.\ref{GeometriesWaveVektors}) as well as over right and left traveling waves $\{N,M\} \in \{R,L\}$, cf. Eq.~(\ref{LinearSurfaceDeflection}).
Substituting this ansatz into the dynamic equations and matching the coefficients with the inhomogeneous contributions of the vorticity equation (\ref{VorticitySecondOrderHigherHarmonics}) yields the functions ${\Psi}_{NMij}^{\mathrm{inhom}}(z)$ and $ \tilde{\Psi}_{NMij}^{\mathrm{inhom}}(z)$. Since their general form is extremely bulky, we list in App.\ref{AppendixEigenvectors} only ${\Psi}_{NMij}^{\mathrm{inhom}}(z)$
and $\tilde{\Psi}_{NMij}^{\mathrm{inhom}}(z)$
for  hexagonal ($ij=ji=13=23=31$) and square patterns ($ij=ji=15$) as well as for stripe solutions ($ij=11$).

The general solution is the sum of the particular inhomogeneous and a general homogeneous solution, $\Psi_{k}^{(2,2)} = \Psi_{k}^{(2,2)\mathrm{inhom}} + \Psi_{k}^{(2,2)\mathrm{hom}}$. It has to satisfy the effective tangential boundary conditions (cf. App.\ref{AppSubSecSecondOrderBC})
\begin{eqnarray}
\tilde{\mu}_{2}
(\partial_{z}^{2} - \partial_{y}^{2})\Psi^{(2,2)}_{x}
+
\tilde{\mu}_{2}
\partial_{y}\partial_{x}\Psi^{(2,2)}_{y}
+
2\tilde{\mu}_{2}\partial_{z}\partial_{y}\varphi^{(2,2)}
&=& \label{TangBCzyPotentials} \nonumber
 \\
\partial_{y} \sum\limits_{N,M} \sum\limits_{i,j}
(
\hat{F}_{NMij}^{\prime} \xi_{iN}\xi_{jM} + \tilde{\hat {F}}_{NMij}^{\prime} \xi_{iN} \xi_{jm}^* &+&\, c.c.)
\end{eqnarray}
with a suitably abbreviated amplitudes $\hat F_{NMij}^{\prime}$. The special form of the right hand side is obtained, if in Eq.~(\ref{AnhangHydroBCTangentialBCPotentialsSecondOrder}) the first order expressions for the variables are explicitly put in.
Substituting the inhomogeneous solutions $\Psi_{i}^{(2,2)\mathrm{inhom}}$ into Eq.~(\ref{TangBCzyPotentials}) a modified boundary condition for the homogeneous solution results
\begin{eqnarray} \label{modBC-hom22}
\tilde{\mu}_{2}
(\partial_{z}^{2} - \partial_{y}^{2})\Psi^{(2,2)\mathrm{hom}}_{x}
+
\tilde{\mu}_{2}
\partial_{y}\partial_{x}\Psi^{(2,2)\mathrm{hom}}_{y}
+
2\tilde{\mu}_{2}\partial_{z}\partial_{y}\varphi^{(2,2)}
&=&  \nonumber \\
\partial_{y} \sum\limits_{N,M} \sum\limits_{i,j}
(
\hat{F}_{NMij} \xi_{iN}\xi_{jM} + \tilde{\hat {F}}_{NMij} \xi_{iN} \xi_{jm}^* &+&\, c.c.)\,,
\end{eqnarray}
since the inhomogeneous solution does not satisfy the boundary condition. In particular, on the right hand side the inhomogeneous part of the boundary conditions at $z=0$ is modified 
\begin{eqnarray}
\hat F_{NMij} &=& \hat{F}_{NMij}^{\prime} + \hat F_{NMij}^{\mathrm{inhom}}
\\
{\mathrm{with}} \quad\quad \hat F_{NMij}^{\mathrm{inhom}} \xi_{iN} \xi_{jM}
&=&
-\tilde{\mu}_{2}
(
2\partial_{x}^{2} - \partial_{z}^2
)\Psi_{NMij}^{\mathrm{inhom}}(z) \mid_{z=0}  \xi_{iN} \xi_{jM}\,.
\end{eqnarray}
Similarly one obtains the $y$ component of the tangential boundary condition starting from Eq.~(\ref{AnhangHydroBCTangentialBCPotentialsSecondOrder2}).

Now the general solution of $\Psi_{k}^{(2,2)\mathrm{hom}}$, Eq.~(\ref{modBC-hom22}),  can be obtained using an ansatz similar to that used for the solution of the magnetic potential, Eq.~(\ref{MagneticPotentialSecond}) 
\begin{eqnarray}
\varphi^{(2,2)}
&=&
\sum\limits_{N,M} \sum\limits_{i,j} (
\hat{\varphi}_{NMij}
\,e^{k_{1ij}z}
k_{c}
\xi_{iN}\xi_{jM}
+ \tilde{\hat{\varphi}}_{NMij}
\,e^{k_{2ij}z}
k_{c}
\xi_{iN}^* \xi_{jM} + c.c.) 
\\
\!\!\!\! \Psi^{(2,2)\mathrm{hom}}_{k}
&=&
 -\epsilon_{zkl} \sum\limits_{N,M} \sum\limits_{i,j} \partial_l ( 
\hat{\Psi}_{NMij}^{\mathrm{hom}}\,e^{q_{1ij}z}
k_{c}
\xi_{iN}\xi_{jM}
\nonumber \\ && \qquad\qquad\qquad\quad
+ \tilde{\hat{\Psi}}_{NMij}^{\mathrm{hom}}\,e^{q_{2ij}z}
k_{c}
\xi_{iN}^* \xi_{jM} + c.c. )\,,
\end{eqnarray}
where, again, the first summation is over right and left traveling waves and the second one over the fundamental modes involved. The inverse decay length for the rotational flow contributions, $q$, depends on the angle between the $i$-th and the $j$-th mode
\begin{eqnarray} \label{Defq1ij}
q_{1ij}^{2} &=& k_{1ij}^{2} + \frac{\rho[D_{t}^{(0)}]^2}{\mu_{2} + \nu_{2}\,D^{(0)}_{t}}
\end{eqnarray}
and accordingly $q_{2ij}$ by substituting $k_{2ij}^{2}$ for $k_{1ij}^{2}$ in Eq.~(\ref{Defq1ij}), where $k_{1ij}$ is defined in Eq.~(\ref{Defk1ij}).
Here, $D_{t}^{(0)}$ is an abbreviation for the Fourier transformed time derivative and takes the values $i\omega^{(0)} + \sigma^{(0)}$,  $ \sigma^{(0)}$, and  $-i\omega^{(0)} + \sigma^{(0)}$ when applied to RR, RL or LR, and LL modes, respectively. The bulk equations and boundary conditions are fulfilled for the amplitudes
\begin{eqnarray}
\hat{\Psi}_{RRij}^{\mathrm{hom}}
&=&
\frac{q_{1ij}^{2}}{\tilde{\mu}_{2} k_{c}(q_{1ij}^{4}+q_{1ij}^2k_{1ij}^2)}
\big(
\hat F_{RRij} - 2\tilde{\mu}_{2} k_{1ij}k_{c}\hat{\varphi}_{RRij}
\big)
\label{VectorPotentialHomSecondOrderHigherHarmonics}
\end{eqnarray}
and
\begin{eqnarray}
\hat{\varphi}_{RRij} \xi_{iR}\xi_{jR}
&=&
\frac{q_{1ij}^2+k_{1ij}^2}{k_{c}k_{1ij}(q_{1ij}^2-k_{1ij}^2)}
\Big\lbrace
k_{c}D_{t}^{(0)}\xi_{iR}\xi_{jR}
\label{ScalarPotentialSecondOrderHigherHarmonics}
\\ && \qquad
-
k_{1ij}^2\frac{\hat{F}_{RRij}}{\tilde{\mu}_{2}(q_{1ij}^2+k_{1ij}^2)}\xi_{iR}\xi_{jR}
-
2\xi^{(1)}\partial_{z}v^{(1)}_{z}
\nonumber \\
&&\qquad
+
\frac{1}{q_{1ij}^2+k_{1ij}^2}
\Big\lbrack
(k_{1ij}^2\partial^{2}_{z}
+
q_{1ij}^2
[\partial^{2}_{x} + \partial^{2}_{y}])
\hat{\Psi}_{RRij}^{\mathrm{inhom}}\,
\xi_{iR} \xi_{jR} 
\Big\rbrack_{z=0}
\Big\rbrace
\nonumber
\end{eqnarray}
For the last expression we explicitly used the kinematic boundary condition for the second perturbative order, Eq.~(\ref{SecondOrderKinematicBC}). In App.\ref{AppendixEigenvectors} these solutions for the flow potentials are specified for hexagons, Eqs.~(\ref{Phihex}) and (\ref{PsiHomHex}), and squares, Eqs.~(\ref{Phisquare}) and (\ref{PsiHomSquare}). The amplitudes with a tilde are obtained from those without one by replacing $k_{1ij}$ or $q_{1ij}$ by $k_{2ij}$ or $q_{2ij}$, respectively. For $\tilde{\hat{\varphi}}_{RRij} \xi_{iR}\xi_{jR}$ this leads to a denominator $\sim k_{2ij}$, which vanishes for $i=j$ according to Eq.~(\ref{Defk2ij}). Nevertheless, all physical quantities derived from that potential, like velocities and strain components, stay finite. The amplitudes in Eqs.~(\ref{VectorPotentialHomSecondOrderHigherHarmonics}) and (\ref{ScalarPotentialSecondOrderHigherHarmonics}) for the RL and LL (instead of RR) components are obtained by choosing the appropriate expressions for $q_{1ij}$ and $D_{t}^{(0)}$, according to the rules given above. The only remaining condition not yet satisfied is the normal stress boundary condition, which we will discuss in the next section.

\subsection{The normal stress boundary condition \label{SecondOrderFredholmBoundary}}

To find the solutions to the hydrodynamic bulk equations (\ref{SecondOrderNavierStokes}-\ref{SecondOrderContinuity}), it was not necessary to use the normal stress boundary condition. The same situation appears in the derivation of the linear eigenvectors. There,  substituting the eigenvectors into the normals stress boundary condition yields the dispersion relation restricting the linear solution to those with a specific $\omega(k)$ relation. The second order normal stress boundary condition, as will be shown below, leads to the determination of $M^{(1)}$, the first correction to the control parameter entering the final amplitude equation in linear order. 

The second order normal stress boundary condition has been derived in App.~\ref{AppSubSecSecondOrderBC} and is given as Eq.~(\ref{AnhangHydroBCNormalBCSecondOrder}). It consists of two parts, one is proportional to $\xi^{(1)}$, Eq.~(\ref{BCnormXi1})
and the other to $\xi^{(2)}$. 
The latter equation can easily be fulfilled by splitting the pressure $p^{(2,2)} = p^{(2,2)B} + p^{(2,2)S}$ into one part, $p^{(2,2)B}$, that is determined by the bulk equation Eq.\@ (\ref{VelPotentialSecO2}) and the other, $p^{(2,2)S}$, by the $\xi^{(2)}$-boundary condition. This ansatz works, if $\Delta p^{(2,2)S}= 0$ in the bulk. Indeed, $p^{(2,2)S}\sim \xi_{i}\xi_{j} e^{k_{1ij} z}$ or $\sim \xi_{i}\xi_{j}^* e^{k_{2ij} z}$ leads to the required result. This additional pressure contribution is due to the inhomogeneities arising in the normal stress boundary condition, in particular the one due to surface tension. Since the surface tension always acts normally to the surface, this is the only point, where it can enter the nonlinear dynamics. It just contributes to the Laplace pressure, which is proportional to the curvature of the surface, a quite intuitive result.

However, this additional pressure contribution is of no importance because of two reasons. First, the pressure always enters linearly the hydrodynamic bulk equations and therefore it will never give rise to inhomogeneous contributions, which have to be accounted for by Fredholm's theorem. Second, the pressure enters only the normal stress boundary condition, which is actually the governing equation for the appropriate pressure contribution in the next order. In addition, also $p^{(2,2)B}$ is not needed in the following and we refrain from showing it here. 

The situation is different for the first part of the normal boundary condition
\begin{eqnarray}
2\mu_{2}\epsilon^{(2,1)}_{zz} + 2\nu_{2}\partial_{z}v_{z}^{(2,1)} - p^{(2,1)}
- \mu H_{c}\partial_{z}\Phi^{(2,1)} &+& H_{c}^{\mathrm{vac}}\partial_{z}\Phi^{(2,1)\mathrm{vac}}
\nonumber \\
&=&
\frac{k_c\mu}{1\!+\!\mu}M^{(1)}M_{c}\xi^{(1)}\,.
\label{BCnormXi1}
\end{eqnarray}
It serves to determine the yet unknown control parameter $M^{(1)}$, which defines the expansion parameter $\epsilon$, on which the amplitude equation concept is based on. In contrast to bulk instabilities, where $M^{(1)}$ follows directly from Fredholm's alternative, here we have to employ the normal boundary condition, since the Rosensweig instability basically is a surface instability. The same is true for the Marangoni instability, where again the driving force of the instability is not contained in the bulk equations, but acts purely at the surface. In some previous discussions this problem was circumvented by using a scalar product artificially implementing the driving force into Fredholm's theorem. This special scalar product made use of the fact, that the free boundary was treated as undeformable. In the presence of a deformable surface, however, this specific scalar product seems to fail.

Satisfying the normal stress boundary condition (\ref{BCnormXi1}) provides us with the necessary relation between the control parameter $M^{(1)}$ and the scaled growth rate $\sigma^{(1)}$. The actual calculations to get these quantities are displayed in App.\ref{AppendixNormalBC2nd} and here the results are given
\begin{eqnarray} \label{nBC2nd}
\sigma^{(1)}
=
\frac{\mu M^{(1)}M_{c}}{\nu_{2}(1\!+\!\mu)} \quad \mathrm{and} \quad 
\omega^{(1)}
=
0\,.
\end{eqnarray}
The fact that $\omega^{(1)}$ vanishes states, that the instability remains stationary and excludes possible soft mode oscillatory branches beyond the linear threshold. For the slow growth rate $\sigma^{(1)}$ we obtain the physical result that the growth is the faster the farther one is beyond the  linear threshold and it is the slower the more viscous the medium under consideration is. We also observe that the elastic contributions in Eq.~(\ref{BCnormXi1}) cancel upon substituting the solutions of the eigenvectors. This, on the one hand, states that Eq.~(\ref{nBC2nd}) applies to ferrofluids and ferrogels, alike, and on the other hand it states that the growth process is solely given by the dissipative mechanisms in the system under consideration. Eq~(\ref{nBC2nd}) additionally tells us, that the boundary behaves qualitatively different with respect to the temporal properties when compared to the bulk (cf. Eq.~(\ref{PrimitiveAmplEqSecondOrderReal})), since it does not scale with $\sigma^{(0)}$. This qualitative difference is manifest in the kinematic boundary condition which always connects the velocity field to the temporal change of the amplitude, as we discussed already. It is therefore reasonable to compare the scaled time derivative from the bulk with the time derivative in Eq.~(\ref{nBC2nd}).

In order to combine the results from the surface with the solvability condition of the bulk equations (\ref{PrimitiveAmplEqSecondOrderReal}), we need to rewrite the growth rate $\sigma^{(1)}$ in dimensionless form. By multiplying Eq.~(\ref{nBC2nd}) with the typical (linear) time scale $\tau_{0}=\nu_{2}k_c(\rho G + \mu_{2} k_c)^{-1}$ and by defining $\tau_{0}\sigma^{(1)}$ as $\tilde{\partial}_{T}^{(1)}$ we obtain
\begin{eqnarray}
\tilde{\partial}_{T}^{(1)}\hat{\xi}_{i}
&=&
\frac{k_c \mu M^{(1)}M_c}{2(1\!+\!\mu)(\rho G + \mu_{2}k_c)}
\hat{\xi}_{i}
\label{PrimAmplEqBoundarySecondOrder}
\end{eqnarray}
and by adding the solvability condition from the bulk equation (\ref{PrimitiveAmplEqSecondOrderReal2}) with the one from the surface (\ref{PrimAmplEqBoundarySecondOrder}), we finally end up with a rudimentary form of an amplitude equation for the second order
\begin{eqnarray}
\tilde{\partial}_{T}^{(1)}\hat{\xi}_{i}
&=&
\frac{k_c \mu M^{(1)}M_c}{2(1\!+\!\mu)(\rho G+\mu_{2}k_c)}
\hat{\xi}_{i}
-
\frac{k_c}{4}
\sum\limits_{j,k}^{i\not=j\not=k}
\hat{\xi}_{j}^{\ast}\hat{\xi}_{k}^{\ast}
\label{AmplEqSecondOrder}
\end{eqnarray}
In the last step we explicitly assumed that the dimensionless time derivatives at the surface and in the bulk are of the same order. By adding the two subsystems we therefore accounted for the singular behavior of the kinematic boundary condition.

By now we have solved the second order problem completely, with the amplitudes of the critical modes satisfying Eq.~(\ref{AmplEqSecondOrder}).

\section{Third order}

With the complete solution of the second order problem at hand we can now discuss the third order, in order to obtain the desired amplitude equation. As in the second order, the solvability condition consists of two parts. One due to Fredholm's theorem and one that guarantees the normal stress to be compensated at the boundary. However, we will have to find solutions of the third order problem proportional to the main characteristic modes, only.
\subsection{Magnetic contributions proportional to $\xi^{(1)}$}

We again start with the magnetic contributions and restrict our discussion to the parts proportional to the main characteristic mode $\xi^{(1)}$. The differential equations for the scalar potentials of the distortions to the magnetic fields read
\begin{eqnarray}
\Delta\Phi^{(3,1)} = 0
\quad&\mathrm{and}&\quad
\Delta\Phi^{(3,1)\mathrm{vac}} = 0
\end{eqnarray}
with the corresponding boundary conditions (at $z=0$) given by
\begin{eqnarray}
\partial_{y}\Phi^{(3,1)\mathrm{vac}}
-
\partial_{y}\Phi^{(3,1)}
&=&
M^{(2)}\partial_{y}\xi^{(1)}
\\
\partial_{x}\Phi^{(3,1)\mathrm{vac}}
-
\partial_{x}\Phi^{(3,1)}
&=&
M^{(2)}\partial_{x}\xi^{(1)}
\\
\partial_{z}\Phi^{(3,1)\mathrm{vac}}
-
\mu\partial_{z}\Phi^{(3,1)}
&=&
0\,.
\end{eqnarray}
The solutions of this set of equations is obtained following the lines of the second order calculations, Eqs.~(\ref{PhiMag2}) and (\ref{PhiMag2vac}), leading to
\begin{eqnarray}
\Phi^{(3,1)} = -\frac{M^{(2)}}{1 + \mu}\xi^{(1)}e^{k_c z}
\quad&\mathrm{and}&\quad
\Phi^{(3,1)\mathrm{vac}}
=
\frac{\mu M^{(2)}}{1 + \mu}\xi^{(1)}e^{-k_c z}\,.
\end{eqnarray}

The contributions due to the higher harmonic modes could in principle be calculated in the same way as in the second order. However, these contributions again contribute only to the pressure offset and are therefore of no importance for the amplitude equation.

\subsection{Hydrodynamic contributions proportional to $\xi^{(1)}$\label{AmplitudeEquationThirdOrderDiscussion}}

The complete set of hydrodynamic bulk equations for the hydrodynamic variables reads in third perturbative order 
\begin{eqnarray}
\rho\partial_{t}^{(0)}v_{i}^{(3)}
&+&
\partial_{i} p^{(3)}
- 2\mu_2\partial_{j}\epsilon^{(3)}_{ij}
- \nu_{2} \big(\partial_{j}\partial_{i}v^{(3)}_{j}
          + \partial_{j}\partial_{j}v^{(3)}_{i} \big)
\label{AmplEqThirdOrderNavierStokes}
\\ 
&=&
\rho\,\partial^{(2)}_{t} v^{(1)}_{i} - \rho\,\partial^{(1)}_{t} v^{(2)}_{i}
\nonumber \\
&& -
 \partial_{j} \big(\rho\, v^{(1)}_{i} v^{(2)}_{j} + \rho \, v^{(2)}_{i} v^{(1)}_{j}
                          - 2\mu_{2}\epsilon^{(1)}_{jk}\epsilon^{(2)}_{ki}
                          - 2\mu_{2}\epsilon^{(2)}_{jk}\epsilon^{(1)}_{ki}
                     \big) \quad
\nonumber
\\
\partial^{(0)}_{t} \epsilon^{(3)}_{ij}
&-& \frac{1}{2} \big( \partial_{i} v^{(3)}_{j}
                      + \partial_{j} v^{(3)}_{i} \big)
\nonumber \\
&=& - \partial^{(2)}_{t} \epsilon^{(1)}_{ij}
- \partial^{(1)}_{t} \epsilon^{(2)}_{ij}
- v^{(1)}_{k}\partial_{k}\epsilon^{(2)}_{ij}
- v^{(2)}_{k}\partial_{k}\epsilon^{(1)}_{ij}
\\
\partial_{i} v^{(3)}_{i} &=& 0\,.
\label{AmplEqThirdOrderContinuity}
\end{eqnarray}

We restrict our attention now to the contributions proportional to $\xi^{(1)}$. The hydrodynamic equations, written in terms of the flow potentials, then reduce to
\begin{eqnarray}
\Delta\varphi^{(3)} &=& 0
\\
\rho\,\Delta\partial_{t}^{(0)}\varphi^{(3)} + \Delta p^{(3)}
&=&
-\partial_{t}^{(1)}\rho\,\Delta\varphi^{(2,1)}
-\partial_{t}^{(2)}\rho\,\Delta\varphi^{(1)}
\label{ThirdOrderxi1Pressure}
\\
\rho\,[\partial_{t}^{(0)}]^3\Psi_{m}^{(3)} - \tilde{\mu}_{2}\Delta \partial_{t}^{(0)}\Psi_{m}^{(3)}
&=&
-\mu_{2}\Delta\partial_{t}^{(1)}\Psi_{m}^{(2,1)}
-\rho\,[\partial_{t}^{(0)}]^2\partial_{t}^{(1)}\Psi_{m}^{(2,1)}
\nonumber \\&& -\mu_{2}\Delta\partial_{t}^{(2)}\Psi_{m}^{(1)}
-\rho\,[\partial_{t}^{(0)}]^2\partial_{t}^{(2)}\Psi_{m}^{(1)}\,.
\end{eqnarray}

To find the solutions, we follow the same lines as in the previous order. The particular inhomogeneous solutions for the vector potential read
\begin{eqnarray}
\!\!\!\!\!\!\! \Psi^{(3,1)\mathrm{inhom}}_{a} 
&=& \epsilon_{zba}
\frac{\mu_{2}\!+\!\tilde{\mu}_{2}}{q\tilde{\mu}_{2}}
\left[\!
\partial_{t}^{(2)}\!-\!\frac{[\partial_{t}^{(1)}]^2}{\partial_{t}^{(0)}}
\!-\!(1\!-\!qz)\rho
\frac{\mu_{2}\!+\!\tilde{\mu}_{2}}{4q^2\tilde{\mu}_{2}^2}
\partial_{t}^{(0)}[\partial_{t}^{(1)}]^2
\!\right]\!
ze^{qz}\partial_{b}\xi^{(1)}
\nonumber \\
\end{eqnarray}
for $\{a,b\} \in \{x,y\}$
while the inhomogeneities in (\ref{ThirdOrderxi1Pressure}) are compensated by
\begin{eqnarray}
p^{(3,1)\mathrm{inhom}}
&=&
-\rho\partial_{t}^{(1)}\varphi^{(2,1)} - \rho\partial_{t}^{(2)}\varphi^{(1)}\,.
\end{eqnarray}

The general homogeneous solutions take the form
\begin{eqnarray}
\Psi_{x}^{(3,1)\mathrm{hom}}
=
-\partial_{y}\hat{\Psi}^{(3,1)}e^{qz}\xi^{(1)}\,,
\qquad
\Psi_{y}^{(3,1)\mathrm{hom}}
=
\partial_{x}\hat{\Psi}^{(3,1)}e^{qz}\xi^{(1)}
\nonumber \\[0.2cm]
\mathrm{and}\quad
\varphi^{(3,1)} = \hat{\varphi}^{(3,1)}e^{k_cz}\xi^{(1)}\,,
\quad\quad\quad\quad\quad\quad
\end{eqnarray}
where the amplitude for the vector potential is given by
\begin{eqnarray}
\hat{\Psi}^{(3,1)} &=&
\frac{2k_c}{q^2+k_c^2}\hat{\varphi}^{(3,1)}
-
2\frac{\mu_{2}+\tilde{\mu}_{2}}{\tilde{\mu}_{2}(q^2+k_c^2)}
\left(
\partial_{t}^{(2)}
-
\frac{[\partial_{t}^{(1)}]^2}{\partial_{t}^{(0)}}
\right)\,.
\end{eqnarray}
The homogeneous solution for the pressure reads
\begin{eqnarray}
p^{(3,1)\mathrm{hom}}
&=&
-\rho\partial_{t}^{(0)}\varphi^{(3,1)}
\end{eqnarray}
and upon exploiting the kinematic boundary condition we obtain the amplitude for the scalar potential
\begin{eqnarray}
\hat{\varphi}^{(3,1)}
&=&
\frac{q^2+k_c^2}{k_c(q^2-k_c^2)}
\left(
\partial_{t}^{(2)} - 2k_c^2\frac{\mu_{2}+\tilde{\mu}_{2}}{\tilde{\mu}_{2}(q^2+k_c^2)}
\left(
\partial_{t}^{(2)} + \frac{[\partial_{t}^{(1)}]^2}{\partial_{t}^{(0)}}
\right)
\right)\,.
\end{eqnarray}

As in the second order, the normal stress boundary condition is not used when deriving the solutions. Again, it allows to calculate the linear contributions to the amplitude equations. This is done in App. \ref{AppE2} with the result
\begin{eqnarray}
\tilde{\partial}_{T}^{(2)}\xi^{(1)}
+
\frac{\mu_{2}k_c}{\rho G + \mu_{2}k_c}[\tilde{\partial}_{T}^{(1)}]^2\xi^{(1)}
&=&
\frac{k_c \mu (2M^{(2)}M_{c} + [M^{(1)}]^2)}{2(1\!+\!\mu) (\rho G+\mu_{2}k_c)}\xi^{(1)}\,,
\label{ThirdOrderNormalStressSolvabilityCondition}
\end{eqnarray}
where $\tilde{\partial}_{T}^{(2)} \equiv \tau_0 \sigma^{(2)}$. Note that the second term is absent in a ferrofluid without elasticity.

\section{Amplitude equation \label{FinalAmplitudeEquation}}

We are finally left with satisfying Fredholm's theorem for the third order bulk hydrodynamic equations. The general solvability condition for the equations (\ref{AmplEqThirdOrderNavierStokes})-(\ref{AmplEqThirdOrderContinuity}) reads
\begin{eqnarray}
\langle\bar{v}_{i}\!\mid\!&\rho&\!\partial_{t}^{(2)}v_{i}^{(1)}\rangle
+
\langle\bar{\epsilon}_{ij}\!\mid\!\partial_{t}^{(2)}\epsilon_{ij}^{(1)}\rangle
+
\langle\bar{v}_{i}\!\mid\!\rho\partial_{t}^{(1)}v_{i}^{(2,1)}\rangle
+
\langle\bar{\epsilon}_{ij}\!\mid\!\partial_{t}^{(1)}\epsilon_{ij}^{(2,1)}\rangle
\label{AmplitudeEquationFredholmsTheoremThirdOrder}
\\
&=&
- \langle\bar{v}_{i}\!\mid\!\rho\partial_{t}^{(1)}v_{i}^{(2,2)}\rangle
- \langle\bar{\epsilon}_{ij}\!\mid\!\partial_{t}^{(1)}\epsilon_{ij}^{(2,2)}\rangle
\nonumber \\ &&
+ 2\mu_{2} \langle\bar{v}_{i}\!\mid\!\partial_{j}
(
\epsilon_{jk}^{(1)}\epsilon_{ki}^{(2,1)} + \epsilon_{jk}^{(2,1)}\epsilon_{ki}^{(1)}
)\rangle
- \rho \langle\bar{v}_{i}\!\mid\!\partial_{j}(
v_{i}^{(1)}v_{j}^{(2,1)} + v_{i}^{(2,1)}v_{j}^{(1)}
)\rangle
\nonumber \\ &&
-
\langle\bar{\epsilon}_{ij}\!\mid\!(v_{k}^{(1)}\partial_{k})\epsilon_{ij}^{(2,1)}
+ (v_{k}^{(2,1)}\partial_{k})\epsilon_{ij}^{(1)} \rangle
- \rho\langle\bar{v}_{i}\!\mid\!\partial_{j}(v_{i}^{(1)}v_{j}^{(2,2)} + v_{i}^{(2,2)}v_{j}^{(1)})\rangle
\nonumber \\ &&
+ 2\mu_{2} \langle\bar{v}_{i}\!\mid\!\partial_{j} (
\epsilon_{jk}^{(1)}\epsilon_{ki}^{(2,2)} + \epsilon_{jk}^{(2,2)}\epsilon_{ki}^{(1)}
)\rangle
- \langle\bar{\epsilon}_{ij}\!\mid\!
(v_{k}^{(1)}\partial_{k})\epsilon_{ij}^{(2,2)}
+ (v_{k}^{(2,2)}\partial_{k})\epsilon_{ij}^{(1)} \rangle\,,
\nonumber
\end{eqnarray}
where we already separated the contributions from the second order eigenvectors that are proportional to $\xi^{(1)}$ from those proportional to $\xi^{(2)}$. The first two contributions on the left hand side of Eq.~(\ref{AmplitudeEquationFredholmsTheoremThirdOrder}) can be discussed in the same way as the equivalent terms in the second perturbative order by replacing in Eq.~(\ref{solv2dt}) $\sigma^{(1)}$ and $\omega^{(1)}$ with $\sigma^{(2)}$ and $\omega^{(2)}$, respectively. Thus these contributions yield the scaled dimensionless time derivative $\tilde{\partial}_{T}^{(2)}=\sigma^{(2)}/\sigma^{(0)}$ for the bulk part. The third and the fourth contribution on the left hand side of Eq.~(\ref{AmplitudeEquationFredholmsTheoremThirdOrder}) can in principle contribute to the second time derivative, since the second order eigenvectors $v_{i}^{(2,1)}$ and $\epsilon_{ij}^{(2,1)}$ are proportional to $\partial_{t}^{(1)}$ (cf. Eqs.~(\ref{AmplitudeEquationVelocityzSecondOrderMainWave})-(\ref{AmplitudeEquationShortCutStrainFieldSecondOrder})). Discussing the last contribution first, we obtain upon exploiting the result of the second order, $\omega^{(1)}=0$,
\begin{eqnarray}
\langle\bar{\epsilon}_{ij}\!\mid\! \partial_{t}^{(1)}\epsilon_{ij}^{(2,1)}\rangle
&=&
4 \frac{\rho\mu_{2}}{\nu_{2}k_c}[\sigma^{(1)}]^2
\sum_{i=1}^{N}\hat{\xi}_{i}\hat{\xi}_{i}^{*}
+ \mathcal{O}([\omega^{(0)}]^{5})\,.
\end{eqnarray}
This contribution is at least of the order $[\sigma^{(0)}]^{2}$ and therefore vanishes in the limit of a static instability. Similarly, the contribution due to $\langle\bar{v}_{i}\!\mid\!\partial_{t}^{(1)}v_{i}^{(2,1)}\rangle$ is at least of the order $[\sigma^{(0)}]^{3}$ and can also be neglected. Let us now focus on the right hand side of Eq.~(\ref{AmplitudeEquationFredholmsTheoremThirdOrder}) and discuss those contributions first that are due to the eigenvectors $\epsilon_{ij}^{(2,1)}$ and $v_{i}^{(2,1)}$ of the second perturbative order, which are proportional to the main characteristic modes $\xi^{(1)}$. These contributions involve the combinations of three amplitudes $\xi^{(1)}$ and due to the lateral integration they therefore remain finite only in the case of hexagons. If we use Eq.~(\ref{PrimitiveAmplEqSecondOrderReal}) to substitute e.g. $\hat{\xi}_{1}\hat{\xi}_{2}\sigma^{(1)}\hat{\xi}_{3}$ by $-k_c\sigma^{(0)}\\ \mid\!\hat{\xi}_{1}\!\mid^2\mid\!\hat{\xi}_{2}\!\mid^2$, we finally obtain
\begin{eqnarray}
\langle\bar{\epsilon}_{ij}\!\mid\!
(v_{k}^{(1)}\partial_{k})\epsilon_{ij}^{(2,1)}\rangle
&=&
\frac{64}{9}\mu_{2}k_c^3\sigma^{(0)}
\left(
\mid\!\hat{\xi}_{1}\!\mid^2\mid\!\hat{\xi}_{2}\!\mid^2
+
\mid\!\hat{\xi}_{1}\!\mid^2\mid\!\hat{\xi}_{3}\!\mid^2
+
\mid\!\hat{\xi}_{2}\!\mid^2\mid\!\hat{\xi}_{3}\!\mid^2
\right)
\nonumber \\ &&
+ \mathcal{O}([\omega^{(0)}]^3)\,.
\label{AmplitudeEquationThirdOrderFredholmTerm5}
\end{eqnarray}
Note, that this term only contributes to the cubic coefficient for the hexagonal pattern and vanishes for any other pattern. All other contributions in (\ref{AmplitudeEquationFredholmsTheoremThirdOrder}) involving the eigenvectors $v_{i}^{(2,1)}$ or $\epsilon_{ij}^{(2,1)}$ are at least of the order $[\sigma^{(0)}]^{2}$ and vanish in the static limit. The remaining contributions involve the eigenvectors of the second perturbative order that are proportional to the higher harmonics $\xi^{(2)}$. Since their analytical expressions are bulky, the corresponding contributions to the cubic coefficients have been calculated with {\sl Mathematica}. For the term $\langle\bar{\epsilon}_{ij}\!\mid\!\partial_{t}^{(1)}\epsilon_{ij}^{(2,2)}\rangle$ one has to exploit Eq.~(\ref{PrimitiveAmplEqSecondOrderReal}) in the same manner as done for Eq.~(\ref{AmplitudeEquationThirdOrderFredholmTerm5}). The final results for the cubic coefficients $A'$ and $B'(\theta_{ij})$ are given, for the different regular surface patterns under consideration, by
\begin{eqnarray}
A' &=& 184\mu_{2}k_c^3
\label{AmplEqFerrogelAutoCorrelationCoefficient}\\
B'(\theta_{ij}=2\pi/3) &=&
(1256315969/10368-69828\sqrt{3})\mu_{2}k_c^3
\\
B'(\theta_{ij}=\pi/2) &=&
(31831/2 - 11072\sqrt{2})\mu_{2}k_c^3
\end{eqnarray}
and the solvability condition in the third perturbative order which is due to the bulk equations can be written for the hexagonal pattern as
\begin{eqnarray}
\tilde{\partial}_{T}^{(2)}\hat{\xi}_{1}
&=&
-\frac{A'}{16\mu_{2}k_c}\mid\!\hat{\xi}_{1}\!\mid^2\hat{\xi}_{1}
-\frac{B'(\theta_{ij}\!=\!2\pi/3)}{32\mu_{2}k_c}(\mid\!\hat{\xi}_{2}\!\mid^2 + \mid\!\hat{\xi}_{3}\!\mid^2)\hat{\xi}_{1}
\label{AmplitudeEquationsThirdOrderFredholm}
\end{eqnarray}
with all its cyclic permutations $1\to 2\to 3\to 1$. Correspondingly one finds in the case of the square pattern
\begin{eqnarray}
\tilde{\partial}_{T}^{(2)}\hat{\xi}_{1}
&=&
-\frac{A'}{16\mu_{2}k_c}\mid\!\hat{\xi}_{1}\!\mid^2\hat{\xi}_{1}
-\frac{B'(\theta_{ij}\!=\!\pi/2)}{32\mu_{2}k_c}\mid\!\hat{\xi}_{5}\!\mid^2\hat{\xi}_{1}\,.
\label{AmplitudeEquationsThirdOrderFredholmSquares}
\end{eqnarray}
From those equations (\ref{AmplEqFerrogelAutoCorrelationCoefficient})-(\ref{AmplitudeEquationsThirdOrderFredholmSquares}) it becomes clear that the dependence of the cubic coefficients on the material parameters is solely given by the characteristic wave vector $k_c$. Thus they are independent of the elastic shear modulus and the magnetic susceptibility. The same is true for the quadratic coefficient as observed in Eq.~(\ref{PrimitiveAmplEqSecondOrderReal2}). This behavior could have been anticipated by inspecting the general expressions for Fredholm's theorem (Eqs.~(\ref{ExplFredholmSecondOrder}) and (\ref{AmplitudeEquationFredholmsTheoremThirdOrder})). The lowest order in the expansion with respect to $\partial_{t}^{(0)}$ is always proportional to the shear modulus $\mu_{2}$ (since the adjoint strain field, Eqs.~(\ref{AmplitudeEquationAdjointStrainzz}-\ref{AmplitudeEquationAdjointStrainyz}), is proportional to the shear modulus) which therefore cancels in Eqs.~(\ref{solv2dt}) and (\ref{solv2grad}). This behavior is due to the assumption of linear elasticity. Similarly the assumption of a linearly magnetizable medium and neglecting magnetostrictive effects results in cubic coefficients that are independent of the magnetic susceptibility.

Adding Fredholm's theorem in the third order expansion (\ref{AmplitudeEquationsThirdOrderFredholm}) to the corresponding solvability condition from the normal stress at the boundary (\ref{ThirdOrderNormalStressSolvabilityCondition}), we obtain for the hexagonal pattern
\begin{eqnarray}
\tilde{\partial}_{T}^{(2)}\hat{\xi}_{1}
+
\frac{\mu_{2} k_c}{2(\rho G \!+\!\mu_{2}k_{c})}[\!\!\!&\tilde{\partial}\!\!&_{T}^{(1)}]^2\hat{\xi}_{1}
\label{BCThirdOrder}  \nonumber \\ 
&=&
\frac{k_c \mu\left(
2 M^{(2)}M_{c} \!+\! [M^{(1)}]^2
\right)}{4(1\!+\!\mu)(\rho G \!+\! \mu_{2}k_{c})}
\hat{\xi}_{1}
-\frac{A'}{32\mu_{2}k_c}\mid\!\hat{\xi}_{1}\!\mid^2\hat{\xi}_{1}
\nonumber \\ &&
- \,\frac{B'(\theta_{ij}\!=\!2\pi/3)}{64\mu_{2}k_c}(\mid\!\hat{\xi}_{2}\!\mid^2 \!+\! \mid\!\hat{\xi}_{3}\!\mid^2
)\hat{\xi}_{1}\,,
\end{eqnarray}
where we assume, as done in the second order, that the scaled time derivatives at the surface and in the bulk are the same, because of the kinematic boundary condition.

Recall now the results for the hexagonal pattern that we obtained from the solvability condition in the second order, Eq.~(\ref{AmplEqSecondOrder})
\begin{eqnarray}
\tilde{\partial}_{T}^{(1)}\hat{\xi}_1
&=&
\frac{k_c \mu M_cM^{(1)}}{2 (1+\mu)(\rho G\!+\!\mu_{2}k_{c})} \hat{\xi}_1
-
\frac{k_c}{2} \hat{\xi}_2^{\ast}\hat{\xi}_3^{\ast}\,.
\label{BCSecondOrder}
\end{eqnarray}
If we follow the standard methods and multiply the third order equation (\ref{BCThirdOrder}) by $\epsilon^3$ and the second order equation (\ref{BCSecondOrder}) by $ \epsilon^2$, we obtain
\begin{eqnarray}
(\epsilon^2\tilde{\partial}_{T}^{(2)} &+& \epsilon\tilde{\partial}_{T}^{(1)})\epsilon\hat{\xi}_{1}
+ \frac{\mu_{2}k_c}{2(\rho G +\mu_{2}k_{c})}\epsilon^2[\tilde{\partial}_{T}^{(1)}]^2\epsilon\hat{\xi}_{1}
\nonumber \\
&=&
\frac{k_c \mu(2\epsilon^2 M_c M^{(2)} + \epsilon^{2}[M^{(1)}]^2 + 2 \epsilon M_cM^{(1)})}{4(1+\mu)(\rho G+\mu_{2}k_{c})}\epsilon\hat{\xi}_{1}
- \frac{k_c}{2} \epsilon^2 \hat{\xi}_2^{\ast}\hat{\xi}_{3}^{\ast}
\nonumber \\ && 
- \frac{A'}{32 \mu_2 k_c}\epsilon^3\mid\!\hat{\xi}_1\!\mid^2\hat{\xi}_{1}
- \frac{B'(\theta_{ij}\!=\!2\pi/3)}{64 \mu_2 k_c}\epsilon^3
(\mid\!\hat{\xi}_2\!\mid^2 + \mid\!\hat{\xi}_3\!\mid^2)\hat{\xi}_1\,.
\end{eqnarray}
 By the series expansion of the magnetization, Eq.~(\ref {ExpansionFields01}), we can write 
\begin{eqnarray}
M^2 - M_c^2
&=&
\left(
M_c + \epsilon M^{(1)} + \epsilon^2 M^{(2)} + \dots
\right)^2 - M_c^2
\nonumber \\ &=&
2\epsilon M_c M^{(1)} + 2\epsilon^2 M_c M^{(2)} 
+ \epsilon^2 [M^{(1)}]^{2} + \dots
\end{eqnarray}
and define the control parameter $\tilde{\epsilon}$ in the usual way as the relative quadratic deviation from the critical value (of the magnetization) 
\begin{eqnarray}
(M^2 - M_c^2) = M_c^2 \tilde{\epsilon}\,.
\end{eqnarray}
Substituting the series expansion of the time derivative in terms of $\epsilon$ (cf. Eq.~\ref{AmplEqExpansionTimeDerivative})
\begin{eqnarray}
\epsilon\tilde{\partial}_{T}^{(1)}+\epsilon^2\tilde{\partial}_{T}^{(2)} &\longrightarrow& \partial_{T} \\
\lbrack\epsilon^1\tilde{\partial}_{T}^{(1)}\rbrack^2 &\longrightarrow& \partial_{T}^{2}
\end{eqnarray}
and using the standard scaling
\begin{eqnarray}
\epsilon k_c \sqrt{A} \hat{\xi}_i &\longrightarrow& \xi_i\,,
\end{eqnarray}
the amplitude equation can be written as
\begin{eqnarray}
\partial_{T}\xi_{1} + \frac{\delta}{2}\partial_{T}^{2}\xi_{1}
&=&
\frac{1}{2}\tilde{\epsilon}\xi_{1}
-
\frac{1}{2\sqrt{A}}\xi_{2}^*\xi_{3}^*
- 
\mid\!\xi_{1}\!\mid^2\xi_{1} - \frac{B_{120}}{A}(\mid\!\xi_{2}\!\mid^2 + \mid\!\xi_{3}\!\mid^2)\xi_{1}\,,\quad
\label{AmplitudeEquationNormalForm}
\end{eqnarray}
where we introduce the dimensionless parameter $\delta=\mu_{2}k_{c}(\rho G +\mu_{2}k_{c})^{-1}$ and where the abbreviations $A$ and $B_{120}$ are given by
\begin{eqnarray}
A = \frac{A'}{32 \mu_2 k_c^3} \approx 5.750 
\quad&\mathrm{and}&\quad
B_{120} = \frac{B'(\theta_{ij}\!=\!2\pi/3)}{64 \mu_2 k_c^3} \approx 3.544\,.
\end{eqnarray}

Starting from Eq.~(\ref{AmplitudeEquationsThirdOrderFredholmSquares}) instead of Eq.~(\ref{AmplitudeEquationsThirdOrderFredholm}) we obtain the corresponding amplitude equation for the square pattern
\begin{eqnarray}
\partial_{T}\xi_{1} + \frac{\delta}{2}\partial_{T}^{2}\xi_{1}
&=&
\frac{1}{2}\tilde{\epsilon}\xi_{1}
-
\mid\!\xi_{1}\!\mid^2\xi_{1} - \frac{B_{90}}{A}\mid\!\xi_{5}\!\mid^2\xi_{1}\,,
\label{AmplitudeEquationNormalFormSquares}
\end{eqnarray}
where the cubic coefficient $B_{90}$ is analogously given as 
\begin{eqnarray}
B_{90} = \frac{B'(\theta_{ij}\!=\!\pi/2)}{64 \mu_2 k_c^3} \approx 4.021 \,.
\end{eqnarray}
The fact that the linear contribution on the right hand side of Eqs.~(\ref{AmplitudeEquationNormalForm}) and (\ref{AmplitudeEquationNormalFormSquares}) is only proportional to the control parameter $\tilde{\epsilon}$ justifies a posteriori our choice of the typical time scale $\tau_{0}$.

\begin{figure}[t]
\includegraphics[width=14cm]{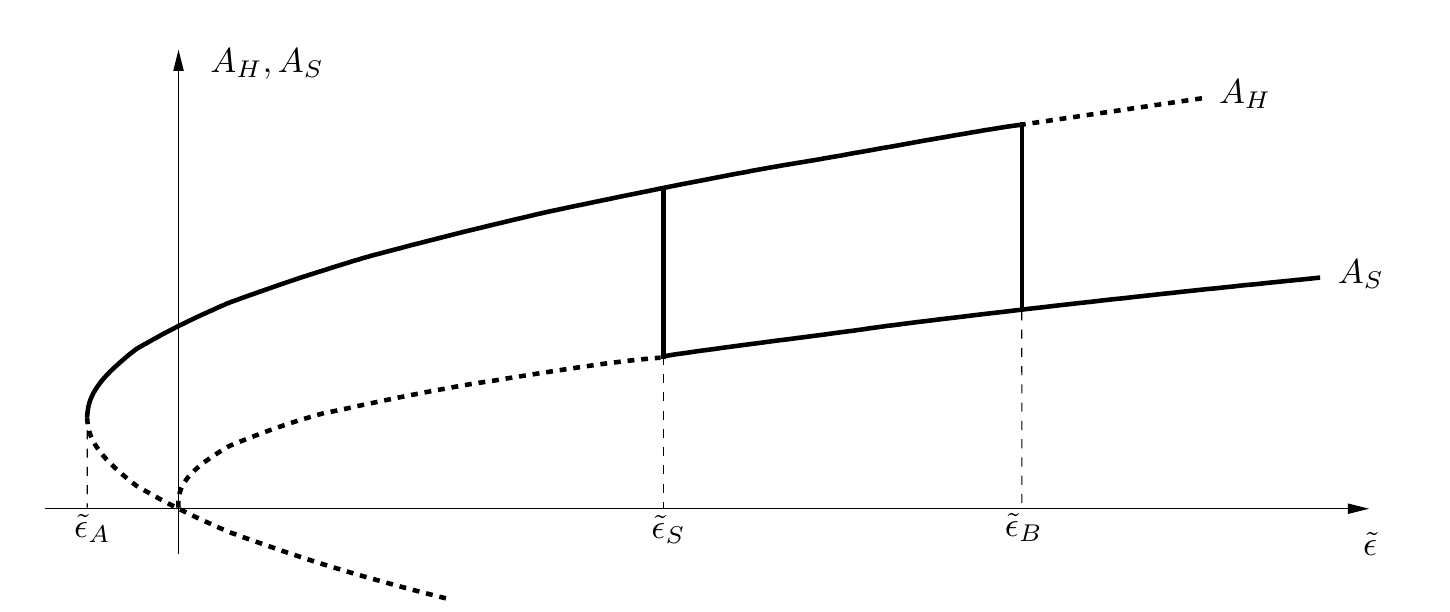}
\caption{
\label{BifDiagr}
Bifurcation diagram: The stationary amplitudes of the hexagonal pattern, $A_H$ (the $\xi_1$ in Eq.~(\ref{AmplitudeEquationNormalForm})) and the square pattern, $A_S$ (the $\xi_1$ in Eq.~(\ref{AmplitudeEquationNormalFormSquares})) as functions of the control parameter; the special values of the latter are explained in the text. The topology of this diagram is similar to that obtained by the energy method \cite{Bohlius2006b}.
}
\end{figure}

Let us first consider the static solutions of Eq.~(\ref{AmplitudeEquationNormalForm})
summarized in Fig.~\ref{BifDiagr}. 
The quadratic contribution gives rise to a transcritical bifurcation from the flat surface to a hexagonal pattern at the linear threshold \cite{Ciliberto1990a}. A bistable regime exists for negative control parameter values $\tilde{\epsilon}$ with its lower boundary given by
\begin{eqnarray}
\tilde{\epsilon}_{A} &=&
- \frac{1}{8 (A + 2B_{120})}\,.
\end{eqnarray}
The solution for the hexagonal pattern takes the form $\xi_{i} = -\!\mid\!\xi_{i}\!\mid e^{i\Phi_{i}}$ for $i\in\{1,2,3\}$, where the magnitude of the amplitudes reads
\begin{eqnarray}
\mid\!\xi_{i}\!\mid
&=&
\frac{1 + \sqrt{1 + 8(A + 2B_{120})\tilde{\epsilon}}}{4\sqrt{A}(1+2B_{120}/A)}
\label{HexagonAmplitudeSolution}
\end{eqnarray}
and where the phases have to fulfill the condition $\sum_{i}\Phi_{i}=0$.

Investigating the values of the cubic coefficients we realize, that $B_{120}/A < 1$ indicating that the hexagon solution is always stable with respect to stripe solutions at the linear threshold. Stripes and squares are mutually exclusive pattern and since $B_{90}+2B_{30}<A+2B_{120}$ and $B_{90}/A < 1$ \cite{Bragard1998a}, the hexagons are losing stability with respect to squares at the critical control parameter $\tilde{\epsilon}_{B}$ given by
\begin{eqnarray}
\tilde{\epsilon}_{B} &=&
\frac{B_{90} + 2 B_{30}}{2(A+2B_{120} - B_{90} - 2B_{30})^2}\,,
\end{eqnarray}
where the cubic coefficient $B_{30}\approx 4.188$ describes the nonlinear interaction between the hexagonal and the square pattern.

The square pattern is stable for control parameters larger than
\begin{eqnarray}
\tilde{\epsilon}_{S}
&=&
\frac{A+B_{90}}{2(A+B_{90}-B_{120}-B_{30})^2}\,.
\end{eqnarray}
Since $\tilde{\epsilon}_{S}<\tilde{\epsilon}_{B}$, also a bistable regime between the hexagons and squares exists.

Let us now focus on the dynamical behavior of the patterns beyond the linear threshold. We assume that the hexagonal pattern with the amplitude $\mid\!\xi_{i}\!\mid$, Eq.~(\ref{HexagonAmplitudeSolution}), has developed and disturb it homogeneously in space by a small excess amplitude $r$, $\mid\!\xi_{i}\!\mid\to\mid\!\xi_{i}\!\mid\! +\, r$. The linearized amplitude equation (\ref{AmplitudeEquationNormalForm}) for the disturbances $r$ then reads
\begin{eqnarray}
\partial_{T} r + \frac{\delta}{2}\partial_{T}^{2} r
&=&
\left\lbrack
\frac{1}{2}\tilde{\epsilon} - \frac{1}{\sqrt{A}} \mid\!\xi_{i}\!\mid - 3\left(1+2\frac{B_{120}}{A}\right)\mid\!\xi_{i}\!\mid^2
\right\rbrack
r\,.
\label{LinearHexagonDynamics}
\end{eqnarray}
Substituting the solution (\ref{HexagonAmplitudeSolution}) in the right hand side of Eq.~(\ref{LinearHexagonDynamics}) it can be simplified to $-(\tilde{\epsilon}/2+\mid\!\xi_{i}\!\mid/(2\sqrt{A}))r$, which is always negative above the linear threshold. This reflects the fact that the exponential growth of the infinitesimal disturbances of the flat surface above the linear threshold gets nonlinearly saturated by the cubic coefficients and a stable pattern develops. Eq.~(\ref{LinearHexagonDynamics}) therefore takes the form of a damped harmonic oscillator which can be solved by using the ansatz $r=\mid\!r\!\mid e^{\lambda T}$ with the eigenvalues
\begin{eqnarray}
\lambda_{1/2}
&=&
-\frac{1}{\delta}\pm\sqrt{\frac{1}{\delta^2}-\frac{\tilde{\epsilon}\sqrt{A}+\mid\!\xi_{i}\!\mid}{\sqrt{A}\delta}}\,,
\end{eqnarray}
where the eigenfrequency $\Omega$ of the oscillator is given by 
\mbox{$\Omega^2 = (\tilde{\epsilon}\sqrt{A}+\mid\!\xi_{i}\!\mid)/{\sqrt{A}\delta}$}.
\begin{figure}
\begin{center}
\includegraphics[width=13.8cm]{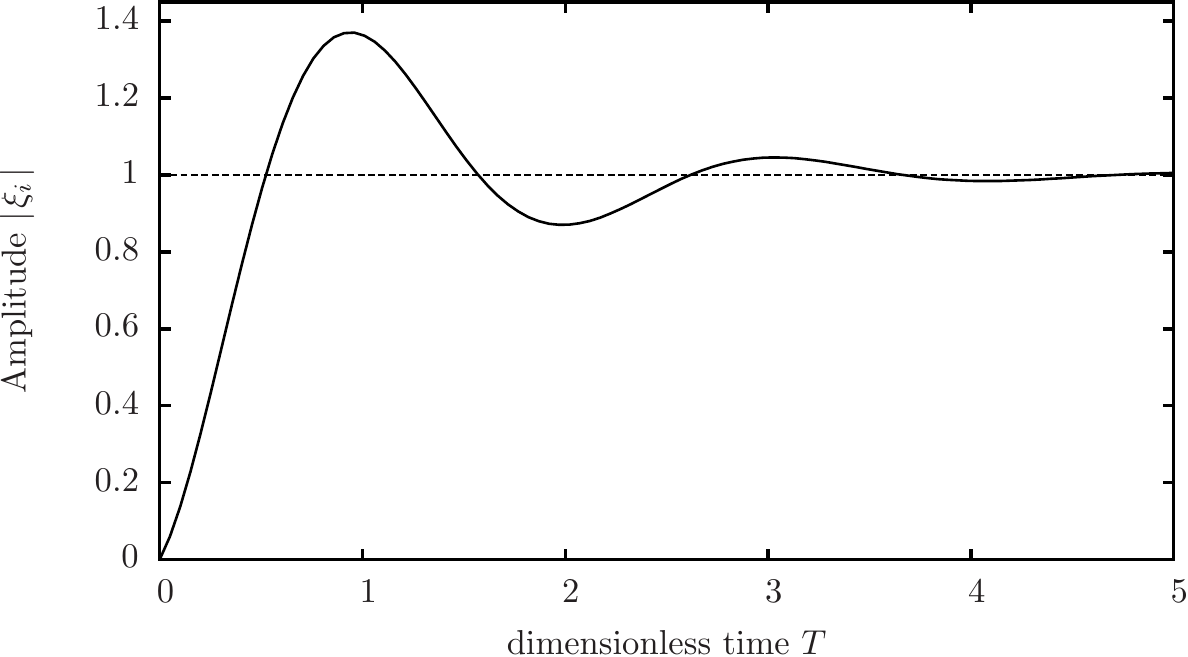}
\caption[Qualitative dynamical growth of the surface spikes in ferrogels]{
\label{SpikeGrowth}
Qualitative time dependent behavior (not to scale) of the surface spikes according to Eq.~(\ref{LinearHexagonDynamics}). The time $T$ as well as the amplitudes $\mid\!\xi_{i}\!\mid$ are dimensionless variables. If the control parameter $\tilde{\epsilon}$ is slightly beyond the critical threshold the plot can be viewed as the qualitative dynamics from the flat surface $\mid\!\xi_{i}\!\mid =0$ to the spiked surface $\mid\!\xi_{i}\!\mid=1$.
}
\end{center}
\end{figure}%

These last results are still in dimensionless units. If we choose the time scale $\nu_{2}/\mu_{2}$ to compare dissipative and oscillatory processes as suggested by Eq.~(\ref{AmplEqAppendixThirdOrderNormalStress}), the eigenvalues read
\begin{eqnarray}
\lambda_{1/2}
&=&
-\frac{(\sqrt{\rho G \sigma_{T}}+\mu_{2})}{\nu_{2}}
\nonumber \\ && \quad
\pm
\sqrt{\frac{(\sqrt{\rho G \sigma_{T}}+\mu_{2})^2}{\nu_{2}^2}
\!-\!
\frac{\mu_{2}(\tilde{\epsilon}\sqrt{A}+\mid\!\xi_{i}\!\mid)\sqrt{\rho G \sigma_{T}+\mu_{2}}}{\sqrt{A}\nu_{2}^2}}\,.
\qquad
\end{eqnarray}
This result is intuitive, since the damping rate is inversely proportional to the dissipative processes, given by $\nu_{2}$, whereas the eigenfrequency increases with increasing shear modulus. We also realize that the relaxation towards the equilibrium pattern becomes faster in a stronger gravitational field as well as for larger surface tensions and elastic higher shear moduli of the medium.

The bifurcation from the flat surface towards hexagons is transcritical and therefore involves a non continuous transition. If the control parameter is slightly above its critical value, the still flat surface (at $T=0$) can be interpreted as a disturbance to the stable stationary solution (\ref{HexagonAmplitudeSolution}). The dynamics towards hexagons from the flat surface is then described by equation (\ref{LinearHexagonDynamics}) giving rise to an overshoot and a damped oscillation towards the equilibrium value (cf. Fig.~\ref{SpikeGrowth}).

\section{Usual Ferrofluids\label{UsualFerrofluis}}

In the case of ferrofluids, the dynamic equation for the strain field (\ref{Elasticitygenerell}) is absent and we only retain the continuity equation (\ref{Continuitygenerell}) and the Navier-Stokes equation (\ref{NSgenerell}). Furthermore, all contributions to the stress tensor (\ref{StressTensorgenerell}) that are proportional to the elastic shear modulus $\mu_{2}$ drop out. The expansion to the nonlinear regime follows the same lines as for magnetic gels. The solvability for the bulk hydrodynamic equations for ferrofluids then reads
\begin{eqnarray}
\langle\bar{v}_{i}\!\mid\!
\partial^{(1)}_{t}(\rho v^{(1)}_{i} )\rangle
&=&
-
\langle\bar{v}_{i}\!\mid\!
\partial_{j}
(\rho v^{(1)}_{i}v^{(1)}_{j})
\rangle\,.
\label{ExplFredholmSecondOrderFF}
\end{eqnarray}
Following the same lines as in \S\ref{FredholmSecondOrder} we obtain for the left hand side of Eq.~(\ref{ExplFredholmSecondOrderFF}), if we retain the lowest order of the expansion in terms of $\partial_{t}^{(0)}$,
\begin{eqnarray}
\langle\bar{v}_{i}\!\mid\!
\partial^{(1)}_{t}(\rho v^{(1)}_{i} )\rangle
&=&
- \frac{12\rho}{k_c} ([\omega^{(0)}]^2-[\sigma^{(0)}]^2)\sigma^{(1)}\sum_{i=1}^{N}
\hat{\xi}_{i}\hat{\xi}_{i}^{*}\,.
\end{eqnarray}
Similarly the right hand side of equation (\ref{ExplFredholmSecondOrderFF}) reads
\begin{eqnarray}
\langle\bar{v}_{i}\!\mid\!
\partial_{j}
(\rho v^{(1)}_{i}v^{(1)}_{j})
\rangle
&=&
-24\rho\sigma^{(0)}([\omega^{(0)}]^2-[\sigma^{(0)}]^2)
(
\hat{\xi}_{1}\hat{\xi}_{2}\hat{\xi}_{3}
+
\hat{\xi}_{1}^{*}\hat{\xi}_{2}^{*}\hat{\xi}_{3}^{*}
)\,.
\end{eqnarray}
We realize, that the lowest order in the expansion of Eq.~(\ref{ExplFredholmSecondOrderFF}) in terms of $\partial_{t}^{(0)}$ is at least proportional to $[\partial_{t}^{(0)}]^3$ due to the deformable surface: The velocity in the original and the adjoint space have to be proportional to the time derivative of the surface deflection and, consequently, the product $\langle\bar{v}_{i}\!\mid\!\partial_{j}(\rho v^{(1)}_{i}v^{(1)}_{j})\rangle$ is proportional to $[\partial_{t}^{(0)}]^3$. This is also the reason why the contributions to the amplitude equation in the case of ferrofluids do not contribute in the case of magnetic gels. The common factor $([\omega^{(0)}]^2-[\sigma^{(0)}]^2)$ in (\ref{ExplFredholmSecondOrderFF}) cancels and, finally, we end up with
\begin{eqnarray}
\frac{\sigma^{(1)}}{\sigma^{(0)}}\hat{\xi}_{1}
=
-\frac{4}{3}k_c\hat{\xi}_{2}^{*}\hat{\xi}_{3}^{*}
\qquad&\mathrm{and}&\qquad
\mid\!\hat{\xi}_{1}\!\mid^2
=
\mid\!\hat{\xi}_{2}\!\mid^2
=
\mid\!\hat{\xi}_{3}\!\mid^2
\label{SecondOrderFredholmExplicitFF}
\end{eqnarray}
and the corresponding conditions for all cyclic permutations $1\to 2 \to 3 \to 1$. As for magnetic gels, Eq.~(\ref{SecondOrderFredholmExplicitFF}) only exists for the hexagonal pattern, whereas for any other surface pattern the amplitudes show no nonlinear interaction in the second order.

The solutions for the second order eigenvectors can be taken from the discussion of magnetic gels in \S\S\ref{HydroSolutSecondOrderMainModes} and \ref{HydroSolutSecondOrderHarmonicModes} by simply substituting $\mu_{2}=0$ and are therefore not shown here. With the solutions of the second order, the third order Fredholm's theorem can be fulfilled. The latter reads in the case of ferrofluids
\begin{eqnarray}
\langle\bar{v}_{i}\!\mid\!\rho\partial_{t}^{(2)}v_{i}^{(1)}\rangle
&+&
\langle\bar{v}_{i}\!\mid\!\rho\partial_{t}^{(1)}v_{i}^{(2,1)}\rangle
\nonumber \\ 
&=&
- \langle\bar{v}_{i}\!\mid\!\rho\partial_{t}^{(1)}v^{(2,2)}\rangle
- \rho \langle\bar{v}_{i}\!\mid\!\partial_{j}(
v_{i}^{(1)}v_{j}^{(2,1)} + v_{i}^{(2,1)}v_{j}^{(1)}
)\rangle
\nonumber \\ &&
- \rho\langle\bar{v}_{i}\!\mid\!\partial_{j}(v_{i}^{(1)}v_{j}^{(2,2)} + v_{i}^{(2,2)}v_{j}^{(1)})\rangle\,.
\label{AmplitudeEquationFredholmsTheoremThirdOrderFF}
\end{eqnarray}
Since the analytical expressions for the eigenvectors $v^{(2,2)}_{i}$ are bulky, the explicit calculation of the cubic coefficients has been performed with {\sl Mathematica} and the results are shown below. We should mention, however, that also in the third order the right hand side of Eq.~(\ref{AmplitudeEquationFredholmsTheoremThirdOrderFF}) is proportional to $[\partial_{t}^{(0)}]^3$ and the global factor $([\omega^{(0)}]^2-[\sigma^{(0)}]^2)$ can be canceled.

The discussion of the normal stress boundary condition in the case of ferrofluids can be taken from \S\S\ref{SecondOrderFredholmBoundary} and \ref{AmplitudeEquationThirdOrderDiscussion}. In the second order, the additional condition to the amplitudes (\ref{PrimAmplEqBoundarySecondOrder}) is valid for ferrogels and ferrofluids, alike, and in the corresponding third order condition (\ref{ThirdOrderNormalStressSolvabilityCondition}) we have to substitute $\mu_{2}\to 0$ with the consequence that there is no second order time derivative. The typical time scale in the case of ferrofluids is then given by $\tau_{0}=\nu_{2}k_{c}/(\rho G)$ which is in accordance with previous theoretical discussions \cite{Friedrichs2003a}. The final amplitude equation is derived in the same way as in \S\ref{FinalAmplitudeEquation} for magnetic gels and finally results for the hexagonal pattern in
\begin{eqnarray}
\partial_{T}\xi_{1}
&=&
\frac{1}{2}\tilde{\epsilon}^{\mathrm{fl}}\xi_{1}
-
\frac{2}{3\sqrt{A^{\mathrm{fl}}}}\xi_{2}^*\xi_{3}^*
-
\mid\!\xi_{1}\!\mid^2\xi_{1} - \frac{B^{\mathrm{fl}}_{120}}{A^{\mathrm{fl}}}(\mid\!\xi_{2}\!\mid^2 \!+\! \mid\!\xi_{3}\!\mid^2)\xi_{1}
\label{AmplitudeEquationNormalFormFF}
\end{eqnarray}
with the control parameter $\tilde{\epsilon}^{fl}$ defined by
\begin{eqnarray}
(M^2 - M_{c,fl}^2) = M_{c,fl}^2 \,\tilde{\epsilon}^{fl}\,,
\end{eqnarray}
where $M_{c,fl}^{2} = 2 \frac{1+\mu}{\mu} \sqrt{\sigma_T\rho\, G}$ is the linear threshold for ferrofluids.\cite{Ferrohydrodynamics} 

For the square pattern the quadratic coefficient is absent and we obtain
%
\begin{eqnarray}
\partial_{T}\xi_{1}
&=&
\tilde{\epsilon}^{\mathrm{fl}}\xi_{1}
-
\mid\!\xi_{1}\!\mid^2\xi_{1} - \frac{B^{\mathrm{fl}}_{90}}{A^{\mathrm{fl}}}\mid\!\xi_{5}\!\mid^2\xi_{1}\,,
\label{AmplitudeEquationNormalFormFFSquares}
\end{eqnarray}
where the cubic coefficients are given by
\begin{eqnarray}
A^{\mathrm{fl}}
&\approx& 8.625
\\
B^{\mathrm{fl}}_{120}
&\approx& 3.150
\\
B^{\mathrm{fl}}_{90} &\approx& 4.266\,.
\end{eqnarray}

The discussion for the different stable patterns follows the same lines as in \S\ref{FinalAmplitudeEquation}. At the linear onset we find hexagons to be the preferred pattern, which remains subcritically stable for control parameters larger than
\begin{eqnarray}
\tilde{\epsilon}_{A} &=&
- \frac{4}{9 (A^{\mathrm{fl}} + 2B^{\mathrm{fl}}_{120})}\,.
\end{eqnarray}
Since $B^{\mathrm{fl}}_{90}+2B^{\mathrm{fl}}_{30}<A^{\mathrm{fl}}+2B_{120}^{\mathrm{fl}}$ and $B_{90}^{\mathrm{fl}}/A^{\mathrm{fl}}<1$, where the cubic coefficient accounting for the nonlinear interaction between hexagons and squares is given by $B^{\mathrm{fl}}_{30}\approx 4.545$, the hexagon pattern transforms into a square pattern for control parameters larger than
\begin{eqnarray}
\tilde{\epsilon}_{B} &=&
\frac{2(B^{\mathrm{fl}}_{90} + 2 B^{\mathrm{fl}}_{30})}{9(A^{\mathrm{fl}}+2B^{\mathrm{fl}}_{120} - B^{\mathrm{fl}}_{90} - 2B^{\mathrm{fl}}_{30})^2}\,.
\end{eqnarray}
The square pattern in turn becomes unstable again for control parameters lower than
\begin{eqnarray}
\tilde{\epsilon}_{S}
&=&
\frac{2(A^{\mathrm{fl}}+B^{\mathrm{fl}}_{90})}{9(A^{\mathrm{fl}}+B^{\mathrm{fl}}_{90}-B^{\mathrm{fl}}_{120}-B^{\mathrm{fl}}_{30})^2}\,.
\end{eqnarray}

In contrast to the nonlinear discussions based on the energy method \cite{Gailitis1977a,Friedrichs2001a} we find stable solutions beyond the linear threshold. In particular, our stability boundaries, except the linear stability, do not depend on the magnetic susceptibility. The reason for this is twofold. First we neglected magnetostriction resulting in a decoupling of the magnetic and hydrodynamic degrees of freedom in the bulk. And second, we assumed a linear magnetization law for the superparamagnetic medium. The bifurcation scenario in realistic ferrofluids, however, depends on the magnetic susceptibility which we 
trace back to the nonlinear magnetizability of ferrofluids and to the 
fact that the macroscopic parameters, as the viscosity, depend on 
the magnetization. To capture these effects, one therefore has to start 
from a macroscopic model for magnetic fluids with nonlinear material 
properties.

\section{Discussion}

In this paper we have derived the amplitude equation for the Rosensweig instability in isotropic magnetic gels based on the fundamental hydrodynamic equations. An important step was to find the adjoint linear system of equations together with its corresponding boundary conditions in the presence of a deformable surface. Two assumptions turned out to be crucial in order to find the adjoint system. Besides the dynamic treatment of the Rosensweig instability, the medium has to be considered compressible for the adjoining process. The reason for the latter assumption is to maintain the symmetry of the stress tensor during the adjoining process. While we can assume an incompressible medium after the adjoining process, the dynamic treatment of the system of equations turns out to be also important in the discussion of the higher perturbative orders.

With the help of the adjoint system we were able to satisfy Fredholm's theorem and to perform a weakly nonlinear analysis. However, due to the decoupling of the magnetic bulk equations from the hydrodynamic ones, Fredholm's theorem does not contain the control parameter, which enters the boundary conditions, only.

We solved this problem by observing that the normal stress boundary condition consists of two parts. One is proportional to the higher harmonics of the characteristic wavelength and merely increases the hydrostatic pressure in the medium. The other one is proportional to the main characteristic wave vector and serves as an additional solvability condition providing the dependence between the scaled growth rate and the control parameter. Both solvability conditions show qualitatively different behavior in the static limit. While the solvability condition obtained from the normal stress boundary remains finite, the bulk contributions scale with the linear growth rate. The latter behavior is mediated by the kinematic boundary condition and has been taken into account while combining both solvability conditions into one. Furthermore it reveals the fact that both states, the initial flat surface and the final spiked one, are motionless states where the velocity field vanishes identically. This does not mean, however, that the Rosensweig instability can be simply treated as a static, energetic one, and the present work has shown why.

While combining the bulk solvability condition with the normal stress boundary one has some freedom to choose the relative weight of the boundary with respect to the bulk via the two differently scaled time derivatives. It seems most natural to weigh these single contributions equally with respect to each other.
We note that this has also been done implicitly, for example, 
in the nonlinear discussions using an extended scalar product \cite{Engel2000a}.

Upon combining both parts of the second and the third order solvability condition, then following the standard procedure, we obtained a set of amplitude equations for the special cases of stripes, squares and hexagons. The latter contains a quadratic coefficient that renders the bifurcation from the flat surface to the hexagonal pattern transcritical. The calculated cubic coefficients additionally reveal that at the linear onset hexagons are the stable surface pattern. For high magnetic field strengths instead, hexagons become unstable and a square pattern develops. Both transitions, from the flat surface to hexagons and from hexagons to squares, involve bistable regions. We obtain qualitatively the same results in the case of ferrofluids, where the derivation of the corresponding amplitude equation and the determination of the nonlinear coefficients has been discussed in \S\ref{UsualFerrofluis}.
We note, however, a qualitative difference regarding the temporal derivatives.
For magnetic fluids the amplitude equation is first order in time,
while for magnetic gels it also contains a second time derivative
reflecting the elasticity of the gel.
Amplitude equations that are second order in time are well established 
in problems related to the buckling of
plates and shells. For example, Lange and Newell \cite{ln}
have analyzed in detail the post-buckling problem for thin
elastic shells.
Along the same lines one has already derived phase equations, the analog
of hydrodynamic equations for large aspect-ratio pattern-forming
systems, containing first and second order time derivatives for elastic
systems under the influence of an external load \cite{wesfreid}.

The results for the static patterns in this article are in qualitative 
agreement with the bifurcation scenario obtained with the energy method \cite{Bohlius2006b}. The cubic coefficients in 
the present paper, however, are independent of the elastic shear modulus and the magnetic susceptibility. This is due to the assumptions of \S\ref{BasicEquations}, where we modeled the magnetic gel as a linear elastic and a linearly magnetizable medium and where we neglected magnetostrictive effects. Within the energy method, where the same approximations have been used, the fourth order coefficients (these coefficients qualitatively correspond to the cubic coefficients in the present treatment) showed an inverse proportionality on the control parameter $\tilde{\epsilon}$. As we analyzed in Ref.~\citen{Bohlius2006b}, this is due to independently minimizing the energy density with respect to the higher harmonics and the main characteristic modes in that method. This dependence has been omitted in the subsequent discussions of the energy method for simplicity rendering this approach valid in the asymptotic limit of a vanishing magnetic susceptibility only. In retrospect this minimization procedure and the simplification afterwards appears to be unsystematic. The results of the present article, however, are valid for a finite magnetic susceptibility and for finite shear moduli.

In our discussion of the nonlinear properties of the Rosensweig instability we assumed spatially homogeneous patterns with no long wavelength variations. By additionally rescaling the spatial coordinates in the same way as the time coordinate (\ref{AmplEqExpansionTimeDerivative}), one could also implement these possible variations in space. As a consequence, the amplitude equation additionally contains derivatives of the amplitudes with respect to the scaled spatial coordinates. For typical nonlinear differential equations these linear contributions to the amplitude equation can be obtained systematically by a standard method exploiting the linear properties of the system \cite{Newell,Brand1986a,Brand1986b}. 

In the case of the Rosensweig instability, however, we additionally have to take into account the deformability of the surface and along with it the kinematic boundary condition. If the surface deforms, also its normal vector ${\bf n}$ changes in the course of time, which we took into account in our previous discussions by explicitly expanding the latter in terms of the surface deflection $\xi$. All the different orders of $\bf n$ involve gradients of the surface deflection $\xi$ as can be seen in Eqs.~(\ref{AnhangMagneticFieldsNormalVector}-\ref{A4}). Upon rescaling the spatial coordinates we also must expand the gradients appearing in ${\bf n}$ in terms of $\epsilon$, which leads to additional contributions to the higher order boundary conditions solely due to the large scale spatial variations of the normal vector. These contributions are not contained in the linear dispersion relation and, of course, cannot be implemented into it by any means, since the dispersion relation only considers the linear properties of the system of equations and therefore assumes a still flat surface. One rather has to expand the set of boundary conditions with the scaled spatial coordinates from the beginning. The contributions to the second spatial derivative in the amplitude equation may then be separated into those due to gradients in the stress tensor (for example $\partial_{j}v_{i}$), which are the ones that follow directly from the dispersion relation, and those solely due to the deformability of the surface. Furthermore we have to evaluate the boundary conditions at the physical boundary, $z=\xi$. In our calculations we accounted for this fact by expanding the eigenvectors in term of $\xi$ around $z=0$. This again involves gradients with respect to $z$, which have to be rescaled as well and which are not contained in the dispersion relation. Additionally, one has to expect contributions to the second order spatial derivatives in the amplitude equation that are due to the bulk equations. In the case of the scaled time derivative we showed that possible contributions due to the bulk scale out in the static limit, but it is  not expected  that this is also the case for the spatial derivatives. 
In conclusion, it is rather obvious that spatial gradients enter the amplitude equation in the form of diffusion terms, although in the present case it is very cumbersome to derive the appropriate coefficient and requires a rather lengthy new calculation.

The amplitude equations have been derived using the critical value $k_c$ for the transverse wavevector. Above threshold, however, a whole range of $k$ values around $k_c$ are allowed for the patterns (Busse balloon) and the most unstable mode may have a transverse wavevector somewhat different from $k_c$. The explicit determination of those properties is again rather unwieldy. There is also the possibility of secondary instabilities, which however we have not looked at.

In addition to the static properties of the surface patterns, the analysis in this article provides us with nonlinear dynamical processes. We obtain the typical first order time derivative that describes the growth of the surface spikes beyond the linear threshold but that also accounts for the dissipative processes in the medium. The typical time scale of the growth (or relaxation) processes increases for increasing viscosities and becomes smaller for increasing surface tension and shear moduli. Additionally, however, we find a second order time derivative in the case of magnetic gels.

Throughout our analysis we treated the system dynamically and allowed for the static limit in the very end. The way we discuss the Rosensweig instability therefore hardly differs from discussions of oscillatory instabilities. When considering amplitude equations for oscillatory instabilities, Coullet et al. \cite{Coullet1985a} 
demonstrated that depending on the relation between the cubic coefficients either propagating or standing patterns are obtained. In our analysis we realized that the imaginary parts of the scaled time derivatives, $\omega^{(1)}$ and $\omega^{(2)}$, vanish and we can therefore exclude such oscillatory soft mode instabilities above the linear threshold.

The analysis in this article elucidated the main aspects of the underlying mechanisms that lead to the Rosensweig instability. But it also unraveled that for a better quantitative understanding additional phenomena have to be taken into account. Two nonlinear properties have been neglected. The nonlinear magnetization behavior, that already affects the linear threshold, and nonlinear elastic properties. Additionally, the magnetostrictive effect might influence the bifurcation behavior in magnetic gels.

\section*{Acknowledgements}
We thank the Deutsche Forschungsgemienschaft for partial 
support of our work, in particular H.R.B. through the Forschergruppe FOR 608
`Nichtlineare Dynamik komplexer Kontinua' and S.B. and H.P. through Schwerpunkt 1104 'Magnetic colloidal fluids'.

\begin{appendix}

\section{Magnetic boundary conditions\label{AppMag}}
\setcounter{equation}{0}
\renewcommand{\theequation}{\ref{AppMag}$\cdot$\arabic{equation}}

In this appendix we derive the magnetic boundary conditions for the second and third order. Since the surface normal ${\bf n}$ is not constant but depends on the surface deflection (as do the distorted field contributions), a higher harmonic coupling to previous orders is possible (in contrast to the system of bulk equations). For the upcoming calculation it is useful to determine first the fields at the boundary $z=\xi$
\begin{eqnarray}
\!\!\!\!\!\!\!\!\!\! {\bf H}
&=&
{\bf H}_{c}
+ \epsilon \! \left({\bf H}^{(1)} - (\bm{\partial}\Phi^{(1)})_{z=0}\right)
+ \epsilon^2 \! \left(
{\bf H}^{(2)} - (\bm{\partial}\Phi^{(2)})_{z=0} - \xi^{(1)}(\partial_{z}\bm{\partial}\Phi^{(1)})_{z=0}
\right)
\nonumber \\ &&
+ \epsilon^3 \Big(
{\bf H}^{(3)} - (\bm{\partial}\Phi^{(3)})_{z=0} - \xi^{(1)}(\partial_{z}\bm{\partial}\Phi^{(2)})_{z=0}
\nonumber \\ && \qquad
-\frac{1}{2}\lbrack(\xi^{(1)2}\partial_{z}^{2} +2\xi^{(2)}\partial_{z})\bm{\partial}\Phi^{(1)}\rbrack_{z=0}
\Big)
\label{AnhangMagneticFieldsHExpansion}
\end{eqnarray}
and accordingly for the magnetic field ${\bf H}^{\mathrm{vac}}$ and the magnetic flux densities ${\bf B}$ and ${\bf B}^{\mathrm{vac}}$. The contributions in (\ref{AnhangMagneticFieldsHExpansion}) that are explicitly proportional to $\xi^{(1)}$ or $\xi^{(2)}$ are due to the deformable surface.

As mentioned, the surface normal ${\bf n}$, initially directed parallel to the $z-$axis, changes its orientation in course of time as the surface perturbation grows (cf. Fig.~\ref{SurfaceModeCartoon}). To give a proper expansion of the boundary conditions, we additionally have to expand the surface normal as a function of the surface deflection $\xi(x,y,t)$
\begin{eqnarray}
{\bf n}
&=&
{\bf n}_0 + \epsilon \, {\bf n}^{(1)} + \epsilon^2 {\bf n}^{(2)} + \epsilon^3 {\bf n}^{(3)}
\label{AnhangMagneticFieldsNormalVector}
\end{eqnarray}
with the different perturbative contributions given by
\begin{eqnarray}
{\bf n}^{(1)}
&=&
\left(
\begin{array}{c}
-\partial_{x}\xi^{(1)} \\
-\partial_{y}\xi^{(2)} \\
0
\end{array}
\right),
\qquad
{\bf n}^{(2)}
\,=\,
\left(
\begin{array}{c}
-\partial_{x}\xi^{(2)} \\
-\partial_{y}\xi^{(2)} \\
\frac{1}{2}
(\partial_{x}\xi^{(1)})^2
+ \frac{1}{2}
(\partial_{y}\xi^{(1)})^2
\end{array}
\right) \label{A3}
\\[1ex]
\mathrm{and} &&
{\bf n}^{(3)}
=
\left(
\begin{array}{c}
-\partial_{x}\xi^{(3)} - \frac{1}{2}(\partial_y \xi^{(1)})^2(\partial_x\xi^{(1)}) - \frac{1}{2}(\partial_x\xi^{(1)})^3
\\
-\partial_{y}\xi^{(3)} - \frac{1}{2}(\partial_x \xi^{(1)})^2(\partial_y\xi^{(1)}) - \frac{1}{2}(\partial_y\xi^{(1)})^3
\\
(\partial_y\xi^{(1)})(\partial_y\xi^{(2)}) + (\partial_x\xi^{(1)})(\partial_x\xi^{(2)})
\end{array}
\right) \label{A4}
\end{eqnarray}

With the previous considerations on hand, we are able to expand the boundary conditions in terms of $\epsilon$. The fact that the normal component of the magnetic flux density is continuous at the boundary gives the following condition
\begin{eqnarray}
{\bf n}\cdot\big({\bf H}^{\mathrm{vac}}-{\bf H}\big)
&=&
{\bf n}\cdot
\big(
{\bf B}^{\mathrm{vac}}-{\bf B} + {\bf M}
\big)
=
{\bf n}\cdot{\bf M}
\end{eqnarray}
Consider the linear perturbative order of the last equation
\begin{eqnarray}
{\bf n}^{(1)}\cdot
\big(
{\bf H}_{c}^{\mathrm{vac}} - {\bf H}_{c}
\big)
+
{\bf n}^{(0)}\cdot
\big(
{\bf H}^{(1)\mathrm{vac}} - {\bf H}^{(1)}
\big)
&=&
{\bf n}^{(1)}\cdot {\bf M}_{c} + {\bf n}^{(0)}\cdot{\bf M}^{(1)}
\end{eqnarray}
For the constant contributions (constant with respect to $x$ and $y$), we find
\begin{eqnarray}
H^{(1)\mathrm{vac}}_{z} - H^{(1)}_{z} &=& M^{(1)}_{z}
\label{AnhangMagnetfelderMagnetisierung}
\end{eqnarray}
while the contributions proportional to ${\bf n}^{(1)}$ cancel identically. The corresponding expression for the second order contribution to the applied field, $H^{(2)\mathrm{vac}}_{z} - H^{(2)}_{z} = M^{(2)}_{z}$, can be obtained straightforwardly.

The boundary condition for the tangential components of the magnetic field (\ref{GeneralBCTangentialHField}) is given in linear order
\begin{eqnarray}
{\bf n}^{(1)}\times\left({\bf H}_c^{\mathrm{vac}} - {\bf H}_c\right)
+
{\bf n}^{(0)}\times\left({\bf H}^{(1)\mathrm{vac}} - {\bf H}^{(1)}\right)
&=&
0
\end{eqnarray}
which can be simplified substituting Eq.\@ (\ref{AnhangMagnetfelderMagnetisierung}) to (with $a\in\lbrace x,y \rbrace$)
\begin{eqnarray}
h_{a}^{(1)\mathrm{vac}} - h_{a}^{(1)}
&=&
-(\partial_{a}\xi^{(1)})M_c
\label{AnhangMagnetfelderTangentialBCFirstOrder}
\end{eqnarray}
In the second perturbative order we find
\begin{eqnarray}
0 &=&
{\bf n}^{(2)}
\times
\big({\bf H}_{c}^{\mathrm{vac}}-{\bf H}_{c}\big)
+
{\bf n}^{(1)}\times
\big(
{\bf H}^{(1)\mathrm{vac}} - {\bf H}^{(1)}
- \bm{\partial}\Phi^{(1)\mathrm{vac}} + \bm{\partial}\Phi^{(1)}
\big)
\nonumber \\
&+&  \,
{\bf n}^{(0)}\times
\big(
{\bf H}^{(2)\mathrm{vac}}
-
\bm{\partial}\Phi^{(2)\mathrm{vac}}
+
k\xi^{(1)}\bm{\partial}\Phi^{(1)\mathrm{vac}}
\nonumber \\ && \qquad\qquad
-
{\bf H}^{(2)}
+
\bm{\partial}\Phi^{(2)}
+
k\xi^{(1)}\bm{\partial}\Phi^{(1)}
\big)
\label{AnhangMagnetifelderTangentialSecondOrderVektor}
\end{eqnarray}
%
%
which is simplified in the same manner (by exploiting the results of the previous order) to
\begin{eqnarray}
&& \partial_{a}\Phi^{(2)\mathrm{vac}} - \partial_{a}\Phi^{(2)} -
M_{c}\, \partial_{a}\xi^{(2)}
- M^{(1)}
\partial_{a}\xi^{(1)}
\qquad\qquad\quad
\nonumber \\
&&+ (\partial_{a}\xi^{(1)})
\big(
\partial_{z}\Phi^{(1)\mathrm{vac}}
-
\partial_{z}\Phi^{(1)}
\big)
-
k\xi^{(1)}
\big(
\partial_{a}\Phi^{(1)\mathrm{vac}}
+
\partial_{a}\Phi^{(1)}
\big)
=
0
\label{AnhangMagnetfelderTangentialBCSecondOrder}
\end{eqnarray}
with $a\in\lbrace x,y \rbrace$. This immediately leads to expression (\ref{TangBoundMag02}) used in the main text. Finally we deduce for the tangential boundary condition in the third perturbative order
\begin{eqnarray}
0 &=& {\bf n}^{(3)} \times
\left({\bf H}_c^{\mathrm{vac}}-{\bf H}_c\right)
+
{\bf n}^{(2)}\times\left(
{\bf H}^{(1)\mathrm{vac}}-{\bf H}^{(1)}
- \bm{\partial}\Phi^{(1)\mathrm{vac}} + \bm{\partial}\Phi^{(1)}
\right)
\nonumber \\
&+& {\bf n}^{(1)}\times\Big(
{\bf H}^{(2)\mathrm{vac}}-{\bf H}^{(2)}
- \bm{\partial}\Phi^{(2)\mathrm{vac}} + \bm{\partial}\Phi^{(2)}
+ k\xi^{(1)}\bm{\partial}\left(\Phi^{(1)\mathrm{vac}}
+ \Phi^{(1)}\right)
\Big)
\nonumber \\
&+& {\bf n}^{(0)}\times\Big(
{\bf H}^{(3)\mathrm{vac}} - {\bf H}^{(3)}
- \bm{\partial}\Phi^{(3)\mathrm{vac}} + \bm{\partial}\Phi^{(3)}
- \xi^{(1)}\partial_{z}\bm{\partial}\left(\Phi^{(2)\mathrm{vac}} - \Phi^{(2)}\right)
\nonumber \\ &&\qquad
-\bigl( \frac{1}{2} k^2\xi^{(1)2} -  k \xi^{(2)}\bigr)\bm{\partial}\Phi^{(1)\mathrm{vac}}
+\bigl( \frac{1}{2} k^2\xi^{(1)2} +  k \xi^{(2)}\bigr)\bm{\partial}\Phi^{(1)}
\Big)
\end{eqnarray}
where it will be sufficient for our discussion to consider only the contributions proportional to the main characteristic modes $\xi^{(1)}$ as discussed in \S\ref{HydroSolutSecondOrderMainModes}.

Along the same lines the boundary condition that guarantees the continuity of the normal component of the magnetic flux density is derived. In first perturbative order we get
\begin{eqnarray}
{\bf n}^{(1)}\cdot\left({\bf B}_c^{\mathrm{vac}} - {\bf B}_c\right)
+
{\bf n}^{(0)}\cdot\left({\bf B}^{(1)\mathrm{vac}} - {\bf B}^{(1)}\right)
&=& 0
\end{eqnarray}
which is straightforwardly simplified to
\begin{eqnarray}
b_{z}^{(1)\mathrm{vac}} - b_{z}^{(1)} &=& 0
\label{AnhangMagnetfelderNormalBCFirstOrder}
\end{eqnarray}
For the corresponding condition in the second perturbative order we obtain
\begin{eqnarray}
0 &=&
{\bf n}^{(2)}
\cdot
\big( {\bf B}_{c}^{\mathrm{vac}} - {\bf B}_{c} \big)
+
{\bf n}^{(1)}\cdot
\big(
{\bf B}^{(1)\mathrm{vac}} - {\bf B}^{(1)}
- \bm{\partial}\Phi^{(1)\mathrm{vac}}
+ \mu \bm{\partial}\Phi^{(1)}
\big)
\nonumber \\
&& + \,
{\bf n}^{(0)}\cdot
\big(
{\bf B}^{(2)\mathrm{vac}} - {\bf B}^{(2)}
- \bm{\partial}\Phi^{(2)\mathrm{vac}}
+ \mu \bm{\partial}\Phi^{(2)}
\nonumber \\ && \qquad\qquad
+ k\xi^{(1)}\bm{\partial}\Phi^{(1)\mathrm{vac}}
+ \mu k\xi^{(1)}\bm{\partial}\Phi^{(1)}
\big)
\end{eqnarray}
%
%
which is simplified by exploiting the previous order to
\begin{eqnarray}
\mu\partial_{z}\Phi^{(2)} - \partial_{z}\Phi^{(2)\mathrm{vac}}
-
(\partial_{x}\xi^{(1)})
\big(
\mu\partial_{x}\Phi^{(1)} - \partial_{x}\Phi^{(1)\mathrm{vac}}
\big)
\qquad\qquad\quad
\nonumber \\
-
(\partial_{y}\xi^{(1)})
\big(
\mu\partial_{y}\Phi^{(1)} - \partial_{y}\Phi^{(1)\mathrm{vac}}
\big)
+
k\xi^{(1)}
\big(
\mu\partial_{z}\Phi^{(1)\mathrm{vac}} + \partial_{z}\Phi^{(1)}
\big)
&=&
0
\end{eqnarray}
Finally, the third order boundary conditions takes the form
\begin{eqnarray}
0 &=& {\bf n}^{(3)}
\cdot
\left({\bf B}_c^{\mathrm{vac}} - {\bf B}_c\right)
+
{\bf n}^{(2)}\cdot\left(
{\bf B}^{(1)\mathrm{vac}} - {\bf B}^{(1)}
-\bm{\partial}\Phi^{(1)\mathrm{vac}} + \mu \bm{\partial}\Phi^{(1)}
\right)
\nonumber \\
&+& {\bf n}^{(1)}\cdot
\Big(
{\bf B}^{(2)\mathrm{vac}} - {\bf B}^{(2)}
- \bm{\partial}\Phi^{(2)\mathrm{vac}} + \mu \bm{\partial}\Phi^{(2)} 
+ k\xi^{(1)} \bm{\partial} \left(\Phi^{(1)\mathrm{vac}} + \mu \Phi^{(1)} \right)
\Big)
\nonumber \\
&+& {\bf n}^{(0)}\cdot
\Big(
{\bf B}^{(3)\mathrm{vac}}-{\bf B}^{(3)}
-\bm{\partial}\Phi^{(3)\mathrm{vac}} + \mu\bm{\partial}\Phi^{(3)}
- \xi^{(1)} \partial_{z}\bm{\partial} \left(\Phi^{(2)\mathrm{vac}} + \mu \Phi^{(2)} \right)
\nonumber \\ && \qquad
-\bigl(\frac{1}{2} k^2\xi^{(1)2} -  k \xi^{(2)}\bigr)\bm{\partial}\Phi^{(1)\mathrm{vac}}
+\bigl(\frac{1}{2} k^2\xi^{(1)2} +  k \xi^{(2)}\bigr)\bm{\partial}\Phi^{(1)}
\Big)
\end{eqnarray}
where again it will be sufficient for our discussion to focus on the contributions proportional to the main characteristic modes $\xi^{(1)}$.

\section{Hydrodynamic boundary conditions\label{AnhangHydroBC}}
\setcounter{equation}{0}
\renewcommand{\theequation}{\ref{AnhangHydroBC}$\cdot$\arabic{equation}}

\subsection{Expansion of the boundary conditions}

In this section we discuss the expansion of the hydrodynamic boundary conditions to the second and third order in terms of $\epsilon$. Recall first, that we require the tangential stress at the free surface to vanish whereas the normal stress is balanced by surface tension (\ref{GeneralBCTangentialStress}, \ref{GeneralBCNormalStress}). The contributions of the stress tensor to the different perturbative orders are defined by the expansions of the macroscopic variables, Eqs.\@ (\ref{ExpansionFields01}, \ref{ExpansionFields02}), and by the expansion of the surface normal ${\bf n}$, Eq.\@ (\ref{AnhangMagneticFieldsNormalVector}).
The linear eigenvectors of the hydrodynamic set of equations are either proportional to $e^{kz}$ or $e^{qz}$ \cite{Bohlius2006b}. For the boundary conditions one has to evaluate them at $z=\xi$ and therefore an expansion similar to Eq.~(\ref{AnhangMagneticFieldsHExpansion}) is needed that explicitly accounts for the deformability of the surface.
The linear order of the boundary conditions is discussed extensively in Refs.~\citen{Bohlius2006a,Bohlius2007a} and is therefore skipped here.

\subsection{The second perturbative order \label{AppSubSecSecondOrderBC}}

In the second order we find as the tangential boundary conditions involving the hydrodynamic fields
\begin{eqnarray}
\Omega^{(2)}_{xz} &\equiv& 2\mu_{2}\epsilon^{(2)}_{xz}
+
\nu_{2}
\big( \partial_{z}v^{(2)}_{x} \!+\! \partial_{x}v^{(2)}_{z} \big)
\label{AnhangHydroBCTangentialBC02SecondOrder} \nonumber \\
&=&
-\xi^{(1)}\partial_{z}
\big\lbrack2\mu_{2}\epsilon_{xz}^{(1)} + \nu_2 ( \partial_{x}v^{(1)}_{z}
                     \!+\! \partial_{z}v^{(1)}_{x} ) \big\rbrack
\nonumber \\ &&
+ (\partial_{y}\xi^{(1)})
  \big\lbrack 2\mu_{2}\epsilon_{yz}^{(1)} + \nu_{2}(\partial_{y}v^{(1)}_{x} \!+\! \partial_{x}v^{(1)}_{y})\big\rbrack
\nonumber \\
&&
- 2 (\partial_{x}\xi^{(1)})\big\lbrack
  \mu_{2}(\epsilon_{zz}^{(1)} + \epsilon_{xx}^{(1)}) + \nu_{2}
  ( \partial_{z}v^{(1)}_{z} + \partial_{x}v^{(1)}_{x} ) \big\rbrack
+ \rho v^{(1)}_{x} v^{(1)}_{z}
\\[1.5ex]
\Omega^{(2)}_{yz} &\equiv& 2\mu_{2}\epsilon^{(2)}_{yz} +
\nu_{2}
\big( \partial_{z}v^{(2)}_{y} \!+\! \partial_{y}v^{(2)}_{z} \big)
\label{AnhangHydroBCTangentialBC01SecondOrder} \nonumber \\
&=&
-\xi^{(1)}\partial_{z}
\big\lbrack 2\mu_{2}\epsilon_{yz}^{(1)} + \nu_2 ( \partial_{y}v^{(1)}_{z}
                     \!+\! \partial_{z}v^{(1)}_{y} ) \big\rbrack
\nonumber \\ &&
+ (\partial_{x}\xi^{(1)})\big\lbrack
  2\mu_{2}\epsilon_{xy}^{(1)} + \nu_{2}
  (\partial_{x}v^{(1)}_{y} \!+\! \partial_{y}v^{(1)}_{x})\big\rbrack
\nonumber \\
&&
- 2(\partial_{y}\xi^{(1)})\big\lbrack
  \mu_{2}(\epsilon_{zz}^{(1)} + \epsilon_{yy}^{(1)})
  + \nu_{2}( \partial_{z}v^{(1)}_{z} + \partial_{y}v^{(1)}_{y} )\big\rbrack
+ \rho v^{(1)}_{y} v^{(1)}_{z}
\end{eqnarray}
In Eqs.\@ (\ref{AnhangHydroBCTangentialBC01SecondOrder}) and (\ref{AnhangHydroBCTangentialBC02SecondOrder}) the inhomogeneities on the right hand side have been abbreviated by $\Omega^{(2)}_{xz}$ and $\Omega^{(2)}_{yz}$, respectively. In particular these inhomogeneities are proportional to $\xi^{(1)2}$.

The normal stress boundary condition (\ref{GeneralBCNormalStress}) in the second order reads
\begin{eqnarray}
2\mu_{2}\epsilon^{(2)}_{zz}
&+&
2\nu_{2}\partial_{z}v^{(2)}_{z}
- p^{(2)}
+ G\rho\xi^{(2)}
- 
       \mu H_{c}\partial_{z}\Phi^{(2)}
       + H_{c}^{\mathrm{vac}}\partial_{z}\Phi^{(2)\mathrm{vac}}
\nonumber \\
&=&
- 2\mu_{2}[\epsilon^{(1)}_{zz}]^2
+ \rho [v^{(1)}_{z}]^2
- M_{c}B_{c} \big\lbrack
                 (\partial_{y}\xi^{(1)})^2
                 + (\partial_{x}\xi^{(1)})^2
             \big\rbrack
\nonumber \\ &&
- \frac{1}{2}\mu\big(\partial_{z}\Phi^{(1)}\big)^2
+ \frac{1}{2}\big(\partial_{z}\Phi^{(1)\mathrm{vac}}\big)^2
\nonumber \\ &&
- 2\mu H_{c}(\partial_{y}\xi^{(1)})(\partial_{y}\Phi^{(1)})
+ 2 H^{\mathrm{vac}}_{c}
  (\partial_{y}\xi^{(1)})
  (\partial_{y}\Phi^{(1)\mathrm{vac}})
\nonumber \\ &&
+ \frac{1}{2}\mu(\partial_{y}\Phi^{(1)})^2
- \frac{1}{2}(\partial_{y}\Phi^{(1)\mathrm{vac}})^2
+ \frac{1}{2}\mu(\partial_{x}\Phi^{(1)})^2
- \frac{1}{2}(\partial_{x}\Phi^{(1)\mathrm{vac}})^2
\nonumber \\ &&
- 2\mu H_{c}(\partial_{x}\xi^{(1)})(\partial_{x}\Phi^{(1)})
+ 2 H^{\mathrm{vac}}_{c}
  (\partial_{x}\xi^{(1)})
  (\partial_{x}\Phi^{(1)\mathrm{vac}})
\nonumber \\ &&
+ \xi^{(1)}\partial_{z}
\Big(
2\mu_{2}\epsilon^{(1)}_{zz} + 2\nu_{2}\partial_{z}v^{(1)}_{z}
- p^{(1)} - \frac{\mu}{1+\mu}M_{c}^2k_{c}\xi^{(1)}
\Big)
\nonumber \\ &&
+ \frac{\mu}{1 + \mu} M^{(1)}M_{c}k_c \xi^{(1)}
- \sigma_T \Delta \xi^{(2)}
\label{AnhangHydroBCNormalBCSecondOrder}
\end{eqnarray}
Furthermore we obtain for the kinematic boundary condition in second order
\begin{eqnarray}
\partial_{t}^{(0)}\xi^{(2)} + \partial_{t}^{(1)}\xi^{(1)}
+ ({\bf v}^{(1)}\cdot\bm{\partial})\xi^{(1)}
&=&
v_{z}^{(2)} + \xi^{(1)}\partial_{z}v_{z}^{(1)}
\label{AnhangHydroBCKinBCSecondOrder}
\end{eqnarray}
The physical boundary is at $z=\xi$, giving rise to an additional dependence on $\xi$. In Eqs.~(\ref{AnhangHydroBCTangentialBC02SecondOrder})-(\ref{AnhangHydroBCKinBCSecondOrder}) such terms have been made explicit (e.g. the last one of (\ref{AnhangHydroBCKinBCSecondOrder})). Thus these boundary conditions are effective ones that have to be taken at $z=0$.

What can be realized immediately in the expressions (\ref{AnhangHydroBCNormalBCSecondOrder}) and (\ref{AnhangHydroBCKinBCSecondOrder}) is, that two qualitatively different contributions arise. On the one hand we obtain the expected contributions proportional to the higher harmonic coupling $\xi^{(1)2}$ of the main characteristic mode. On the other hand, there are still contributions proportional to the main characteristic mode $\xi^{(1)}$ itself. The latter will allow us to find the linear contributions in an amplitude equation even though no explicit control parameter is present in the bulk equations.

To solve the corresponding hydrodynamic bulk equations, we introduced a scalar $\Phi^{(2)}$ and a vector potential ${\bf\Psi}^{(2)}$ in \S\ref{HydroContribSecondOrder}, to discuss potential and rotational flow contributions separately. Following the same lines as done in the linear order \cite{Bohlius2006a,Bohlius2007a}, we can translate the boundary conditions into a corresponding set of equations for the amplitudes of the second order potentials $\Phi^{(2)}$ and ${\bf\Psi}^{(2)}$. We obtain for the tangential contributions (\ref{AnhangHydroBCTangentialBC01SecondOrder}) and (\ref{AnhangHydroBCTangentialBC02SecondOrder})
\begin{equation}
\tilde{\mu}_{2}
(\partial_{z}^{2} \!-\! \partial_{y}^{2})\Psi^{(2)}_{x}
\!+\!
\tilde{\mu}_{2}
(\partial_{y}\partial_{x})\Psi^{(2)}_{y}
\!+\!
2\tilde{\mu}_{2}\partial_{y}\partial_{z}\varphi^{(2)}
=
2\mu_{2}v^{(1)}_{k}\partial_{k}\epsilon^{(1)}_{yz}
\!+\!
\partial^{(0)}_{t}\Omega_{yz}^{(2)}
\label{AnhangHydroBCTangentialBCPotentialsSecondOrder}
\end{equation}
\begin{equation}
-\tilde{\mu}_{2}
(\partial_{x}\partial_{y})\Psi^{(2)}_{x}
\!-\!\tilde{\mu}_{2}
(\partial^{2}_{z} \!-\! \partial^{2}_{x})\Psi^{(2)}_{y}
\!+\!2\tilde{\mu}_{2}
\partial_{x}\partial_{z}\varphi^{(2)}
=
2\mu_{2}v^{(1)}_{k}\partial_{k}\epsilon^{(1)}_{xz}
\!+\!
\partial^{(0)}_{t}\Omega_{xz}^{(2)}
\label{AnhangHydroBCTangentialBCPotentialsSecondOrder2}
\end{equation}
using $\epsilon_{xz} = \epsilon_{yz} \equiv0$ at the boundary. The normal stress boundary condition (\ref{AnhangHydroBCNormalBCSecondOrder}) translates into
\begin{eqnarray}
-(2\tilde{\mu}_{2}
&\partial_{y}& \partial_{z}
+ \rho G\partial_{y})\Psi^{(2)}_{x}
+(2\tilde{\mu}_{2}\partial_{z}\partial_{x}
+ \rho G\partial_{x})\Psi^{(2)}_{y}
+(2\tilde{\mu}_{2}\partial^{2}_{z} + \rho G\partial_{z})\varphi^{(2)}
\nonumber \\
&=&
\partial^{(0)}_{t}
\big(
H_{c}\mu\partial_{z}\Phi^{(2)}
- H_{c}^{\mathrm{vac}}\partial_{z}\Phi^{(2)\mathrm{vac}}
\big)
+M^{(1)}M_{c}k_{c}\frac{\mu}{1+\mu}\partial^{(0)}_{t}\xi^{(1)}
\nonumber \\ && 
+ 2\mu_{2}v^{(1)}_{k}\partial_{k}\epsilon^{(1)}_{zz}
-2\rho G\xi^{(1)}\partial_{z}v^{(1)}_{z}
+ \rho G\partial_{t}^{(1)}\xi^{(1)}
+2\mu_{2}\partial^{(1)}_{t}\epsilon^{(1)}_{zz}
\nonumber \\ &&
- \sigma_T  \partial_{t}^{(0)}\Delta \xi^{(2)}
+ \partial^{(0)}_{t}p^{(2)}
+ \partial^{(0)}_{t}\Omega_{zz}^{(2)}
\quad
\label{AnhangHydroBCNormalBCPotentialsSecondOrder}
\end{eqnarray}
Eqs.\@ (\ref{AnhangHydroBCTangentialBCPotentialsSecondOrder}-\ref{AnhangHydroBCNormalBCPotentialsSecondOrder}) follow from (\ref{AnhangHydroBCTangentialBC01SecondOrder}-\ref{AnhangHydroBCNormalBCSecondOrder}) by taking the time derivative with respect to the fast time scale $t^{(0)}$ without loss of generality. This is why in Eq.\@ (\ref{AnhangHydroBCNormalBCPotentialsSecondOrder}) only the contribution $\partial_{t}^{(0)} p^{(2)}$ and no contribution $\partial_{t}^{(1)} p^{(1)}$ arises, while $\partial_{t}^{(0)}\epsilon_{ij}^{(2)}$ gives rise to contributions $\sim\! v_{i}^{(2)}$ and $\sim\! \partial_{t}^{(1)}\epsilon_{ij}^{(1)}$ (cf. Eq.\@ (\ref{SecondOrderElasticity})).

\subsection{The third perturbative order}

We take over the procedure of the previous section to the third order. If we use the solutions (\ref{SolutionsSecondOrderMainModesInhomo},\ref{SolutionsSecondOrderMainModesVectorPotential},\ref{SolutionsSecondOrderMainModesScalarPotential}) of the hydrodynamic bulk equations in second order, the kinematic boundary condition reads
\begin{eqnarray}
\partial_{t}^{(0)}\xi^{(3)}
&+&
\partial_{t}^{(1)}\xi^{(2)} + \partial_{t}^{(2)}\xi^{(1)} + ({\bf v}^{(1)}\cdot\bm{\partial})\xi^{(2)} + ({\bf v}^{(2)}\cdot\bm{\partial})\xi^{(1)}
\nonumber \\
&=&
v_{z}^{(3)} + \xi^{(1)}\partial_{z}v_{z}^{(2,2)} + \xi^{(1)}\partial_{z}v_{z}^{(2,1)\mathrm{hom}} \!-\! \frac{\mu_2 + \tilde{\mu}_{2}}{q\tilde{\mu}_{2}}k_{c}^{2}\xi^{(1)}\partial_{t}^{(1)}\xi^{(1)}
\nonumber \\ &&
+ \xi^{(2)}\partial_{z}v_{z}^{(1)} + \frac{1}{2}\xi^{(1)2}\partial_{z}^{2}v_{z}^{(1)}
\label{AnhangHydroBCKinBCThirdOrder}
\end{eqnarray}
The tangential boundary conditions are of the usual structure and can be written as
\begin{eqnarray}
2\mu_{2}\epsilon^{(3)}_{yz} +
\nu_{2}
\big( \partial_{z}v^{(3)}_{y} + \partial_{y}v^{(3)}_{z} \big)
&=&
\Omega^{(3)}_{yx}
\label{AnhangHydroBCTangentialBC01ThirdOrder}
\\[1.5ex]
2\mu_{2}\epsilon^{(3)}_{yz} + 
\nu_{2}
\big( \partial_{z}v^{(3)}_{y} + \partial_{y}v^{(3)}_{z} \big)
&=&
\Omega^{(3)}_{xz}
\label{AnhangHydroBCTangentialBC02ThirdOrder}
\end{eqnarray}
where the contributions to the inhomogeneities that are at least proportional to the higher harmonic couplings are collected in the abbreviation $\Omega^{(3)}_{ij}$, similarly as done in second order. Taking the time derivative of Eqs.\@ (\ref{AnhangHydroBCTangentialBC01ThirdOrder}, \ref{AnhangHydroBCTangentialBC02ThirdOrder}) together with (\ref{AnhangHydroBCKinBCThirdOrder}) we find
\begin{eqnarray}
&& \tilde{\mu}_{2}
(\partial_{z}^{2} - \partial_{y}^{2})\Psi^{(3)}_{x}
+
\tilde{\mu}_{2}
\partial_{y}\partial_{x} \Psi^{(3)}_{y}
+
2\tilde{\mu}_{2}\partial_{y}\partial_{z}\varphi^{(3)}
\nonumber \\
&& =
\partial_{t}^{(0)}\Omega^{(3)}_{yx}
+2\mu_2(\partial_{t}^{(1)}\epsilon_{yz}^{(2)} + \partial_{t}^{(2)}\epsilon_{yz}^{(1)})
+2\mu_2(v_{k}^{(1)}\partial_{k}\epsilon_{yz}^{(2)}
+v_{k}^{(2)}\partial_{k}\epsilon_{yz}^{(1)})
\\[1.5ex]
&-&\tilde{\mu}_{2}
\partial_{x}\partial_{y}\Psi^{(3)}_{x}
- \tilde{\mu}_{2}
(\partial^{2}_{z} - \partial^{2}_{x})\Psi^{(3)}_{y}
+2\tilde{\mu}_{2}
\partial_{x}\partial_{z}\varphi^{(3)}
\nonumber \\
&& =
\partial_{t}^{(0)}\Omega^{(3)}_{xz}
+2\mu_2(\partial_{t}^{(1)}\epsilon_{xz}^{(2)} + \partial_{t}^{(2)}\epsilon_{xz}^{(1)})
+2\mu_2(v_{k}^{(1)}\partial_{k}\epsilon_{xz}^{(2)}
+v_{k}^{(2)}\partial_{k}\epsilon_{xz}^{(1)})
\end{eqnarray}

For the normal stress boundary condition we obtain from Eq.\@ (\ref{GeneralBCNormalStress})
\begin{eqnarray} \label{ThirdOrderBCNormalStress}
2 \! && \mu_{2}\epsilon_{zz}^{(3)}
+
2\nu_2\partial_{z}v_{z}^{(3)} - p^{(3)} + \rho G \xi^{(3)} - \mu H_{c}\partial_{z}\Phi^{(3)} + H_{c}^{\mathrm{vac}}\partial\Phi^{(3)\mathrm{vac}}
\nonumber \\ &&=
\mu H^{(2)}\partial_{z}\Phi^{(1)} - H^{(2)\mathrm{vac}}\partial_{z}\Phi^{(1)\mathrm{vac}}
+
\mu H^{(1)}\partial_{z}\Phi^{(2)} - H^{(1)\mathrm{vac}}\partial_{z}\Phi^{(2)\mathrm{vac}}
\nonumber \\ && \,\,\,\,
+ \Omega_{zz}^{(3)} + \sigma_T \bm{\partial}\cdot{\bf n}^{(3)}
\end{eqnarray}
which by a similar procedure can be written as
\begin{eqnarray}
-(2\tilde{\mu}_{2}&\partial_{z}&
\partial_{y}
+ \rho G \partial_{y})\Psi^{(3)}_{x}
+(2\tilde{\mu}_{2}\partial_{z}\partial_{x} + \rho G \partial_{x})\Psi^{(3)}_{y}
+(2\tilde{\mu}_{2}\partial_{z}^2 + \rho G \partial_{z})\varphi^{(3)}
\nonumber \\ &=&
\mu H_{c}\partial_{z}\Phi^{(3)} - H_{c}^{\mathrm{vac}}\partial\Phi^{(3)\mathrm{vac}}
+
\frac{\mu}{1+\mu}(
M_c M^{(2)} + M^{(1)2}
)k_{c}\partial_{t}^{(0)}\xi^{(1)}
\nonumber \\ &&
+ \rho G
\big(\partial_{t}^{(2)}\xi^{(1)} + \partial_{t}^{(1)}\xi^{(2)} + v_{k}^{(2)}\partial_{k}\xi^{(1)} - \xi^{(1)}\partial_{z}v_{z}^{(2)}
- \xi^{(2)}\partial_{z}v_{z}^{(1)}
\nonumber \\ &&
- \frac{1}{2}\xi^{(1)2}\partial_{z}^{2}v_{z}^{(1)}\big)
+ 2 \mu_{2}\bigl( \partial^{(2)}_{t}\epsilon_{zz}^{(1)}
+ \partial^{(1)}_{t}\epsilon_{zz}^{(2)} + v_{k}^{(1)}\partial_{k} \epsilon_{zz}^{(2)}
\nonumber \\ &&
+ v_{k}^{(2)}\partial_{k} \epsilon_{zz}^{(1)}\bigr)
+ \sigma_T \partial_{t}^{(0)}\bm{\partial}\cdot{\bf n}^{(3)}
+ \partial_{t}^{(0)}p^{(3)}
+ \partial_{t}^{(0)}\Omega_{zz}^{(3)}
\end{eqnarray}
%
%

\section{Eigenvectors in second order\label{AppendixEigenvectors}}
\setcounter{equation}{0}
\renewcommand{\theequation}{\ref{AppendixEigenvectors}$\cdot$\arabic{equation}}

In this appendix we give the contributions to the eigenvectors in the second perturbative order that are proportional to the higher harmonic couplings $\xi^{(2)}$. Due to the fact that we have to treat the system dynamically throughout all orders, the expressions become tedious and have therefore been calculated with {\sl Mathematica}. In the following the solutions for the hydrodynamic potentials are represented for the patterns under consideration, hexagons ($\theta_{ij}=2\pi/3$), squares ($\theta_{ij}=\pi/2$) and stripes ($\theta_{ij}=0$) as well as for the interaction between hexagons and squares ($\theta_{ij}=\pi/6$).

The inhomogeneous contributions to the vector potential, cf. Eq.~(\ref{InhomSolSecOrderPsi}), separate into a contribution $\sim\!e^{(k_c+q)z}$ and $\sim\!e^{2qz}$. For the hexagonal case ($ij = ji =12 = 23 = 31$) we obtain
\begin{eqnarray} \label{PsiInhomohex}
\Psi^{\mathrm{inhom}}_{NMij}(z)
&=&
\frac{(k_c^2+q^2)(2\mu_{2}q^2-5\mu_2qk_c+\rho[D_{t}^{(0)}]^2)e^{(k_c+q)z}D_{t}^{(0)}}{2q(k_c^2-q^2)(2k_cq\tilde{\mu}_{2}+q^2\tilde{\mu}_{2}-\rho[D_{t}^{(0)}]^2)}
\nonumber \\ && \quad
+\frac{3k_c^2q(\mu_{2}k_c^2+2\mu_{2}q^2-\rho[D_{t}^{(0)}]^2)e^{2qz}D_{t}^{(0)}}{(k_c^4-5k_c^2q^2+4q^4)(\tilde{\mu}_{2}k_c^2-4\tilde{\mu}_{2}q^2+\rho[D_{t}^{(0)}]^2)}
\end{eqnarray}
with $\{N,M\}\in\{R,L\}$. The abbreviation $D^{(0)}_t$ stands for $2i \omega^{(0)} + 2 \sigma^{(0)}$, $2 \sigma^{(0)}$, and $-2i \omega^{(0)} + 2 \sigma^{(0)}$ for $N=M=R$, $N \neq M$, and $N=M=L$, respectively. The second coefficient $\tilde{\Psi}^{\mathrm{inhom}}_{NMij}$  reads
\begin{eqnarray}
&\tilde{\Psi}&^{\mathrm{inhom}}_{NMij}(z)
=
\frac{- k_c^2q(k_c^2\mu_{2} - 2 q^2\mu_{2} + \rho[D_{t}^{(0)}]^2)e^{2qz}D_{t}^{(0)}}{(3k_c^4 - 7 k_c^2q^2 + 4q^4)(3k_c^2\tilde{\mu}_{2} -4q^2\tilde{\mu}_{2} + 2\rho[D_{t}^{(0)}]^2)}
\\ &&\quad +
\frac{(k_c^2 \!+\! q^2)(6k_c^3\mu_{2} \!-\! 10 k_c^2 q \mu_{2} \!+\! k_c q^2 \mu_{2} \!+\! 2q^3 \mu_{2} \!+\! q\rho[D_{t}^{(0)}]^2)e^{(k_c+q)z}D_{t}^{(0)}}{2(2k_c^4 \!-\! 2k_c^3q \!-\! 3k_c^2q^2 \!+\! 2k_cq^3 \!+\! q^4)(2k_c^2\tilde{\mu}_{2} \!-\! 2k_cq\tilde{\mu}_{2} \!-\! q^2\tilde{\mu}_{2} \!+\! \rho[D_{t}^{(0)}]^2)}
\nonumber
\end{eqnarray}

For the square pattern we get ($ij=ji=15$)
\begin{eqnarray} \label{PsiInhomosquare}
\Psi^{\mathrm{inhom}}_{NMij}
&=&
\frac{(k_c^2\!+\!q^2)(4\mu_{2}k_c^3 \!-\! 10\mu_{2}qk_c^2 \!+\! 2\mu_{2}q^3 \!+\! (k_c\!+\!q)\rho[D_{t}^{(0)}]^2)e^{(q+k_c)z}D_{t}^{(0)}}{2(k_c^4 \!-\! 2qk_c^3 \!-\! 2k_c^2q^2 \!+\! 2k_cq^3\!+\!q^4)(\tilde{\mu}_{2}k_c^2 \!-\! 2\tilde{\mu}_{2}qk_c \!-\! \tilde{\mu}_{2}q^2 \!+\! \rho[D_{t}^{(0)}]^2)}
\nonumber \\ && \quad
+ \frac{qk_c^2(2\mu_{2}q^2 \!-\! \rho[D_{t}^{(0)}]^2)e^{2qz}D_{t}^{(0)}}{4(k_c^4 \!-\! 3q^2k_c^2 \!+\! 2q^4)(2\tilde{\mu}_{2}k_c^2 \!-\! 4\tilde{\mu}_{2}q^2 \!+\! \rho[D_{t}^{(0)}]^2)}
\\ &=&
\tilde{\Psi}^{\mathrm{inhom}}_{NMij}(z)
\end{eqnarray}

For the stripe geometry, $i=j$, we obtain
\begin{eqnarray}
\Psi^{\mathrm{inhom}}_{NMij}(z)
&=&
\frac{(k_c^2\!+\!q^2)\bigl(6k_c^2\mu_{2}\!-\!4 k_cq\mu_{2} \!-\! 2 q^2\mu_{2} \!-\! \rho[D_{t}^{(0)}]^2\bigr)e^{(k_c+q)z}D_{t}^{(0)}}{2(k_c\!-\!q)(3k_c^2 \!+\! 4k_cq \!+\! q^2)\bigl(3k_c^2\tilde{\mu}_{2} \!-\! 2k_cq\tilde{\mu}_{2} \!-\! q^2\tilde{\mu}_{2} \!+\! 2\rho[D_{t}^{(0)}]^2\bigr)}
\nonumber \\
\end{eqnarray}
and

\begin{eqnarray}
&\tilde{\Psi}&^{\mathrm{inhom}}_{NMij}(z)
=
\frac{4 k_c^4\mu_{2} \!-\! 4 q^2 (2 q^2\mu_{2} \!-\! \rho[D_{t}^{(0)}]^2) \!-\! k_c^2(20 q^2\mu_{2} \!+\! 3\rho[D_{t}^{(0)}]^2)}{4(k_c^2-q^2)^2(4q^3\tilde{\mu}_{2} + q \rho[D_{t}^{(0)}]^2)} \,k_c^2 e^{2qz}D_{t}^{(0)}
\nonumber \\ && +
Z_{1}
\Bigl[
4 (k_c\!-\!q)^2(k_c\!+\!q)^3\bigl(k_c^2\tilde{\mu}_{2} \!+\! 2 k_c q \tilde{\mu}_{2} \!+\! q^2 \tilde{\mu}_{2} \!-\! 2 \rho[D_{t}^{(0)}]^2\bigr)
\Bigr]^{-1}
e^{(k_c+q)z}D_{t}^{(0)}
\end{eqnarray}
with the numerator $Z_{1}$ being given by
\begin{eqnarray}
Z_{1}
&=&
(k_c^2\!+\!q^2)\left(24 k_c^3q\mu_{2} \!-\! 4 q^4\mu_{2} \!-\! 2 q^2\rho[D_{t}^{(0)}]^2 \!+\! 20 k_c^2q^2\mu_{2} \!+\! 3k_c^2\rho[D_{t}^{(0)}]^2 \right) \nonumber \\ 
&&+ 8k_cq^3\mu_2 \!-\!4k_cq\rho[D_{t}^{(0)}]^2
\end{eqnarray}

In addition, to describe the interaction between the stripe and the hexagonal pattern we need to consider also the case $\theta_{ij}=\pi/6$, ($ij = ji = 14 = 36 = 25$)
\begin{eqnarray}
&\Psi&^{\mathrm{inhom}}_{NMij}
=
N_{1}^{-1}
\left\lbrace
(k_c^2 \!+\! q^2)(2q^3\mu_{2}\!-\!10k_c^2q\mu_{2} \!+\! k_c^2\mu_{2}(1\!+\!3\sqrt{3})\!+\!q\rho[D_{t}^{(0)}]^2
\right.
\\ &&\qquad\qquad\qquad
\left.
+ \sqrt{3}k_cq^2\mu_{2} \!+\! k_c(1\!-\!\sqrt{3})\rho[D_{t}^{(0)}]^2)
\right\rbrace
e^{(k_c+q)z}D_{t}^{(0)}
\nonumber \\ &&
-\frac{k_c^2q(k_c^2\mu_{2}(2\sqrt{3}-3) - 2 (2-\sqrt{3})(q^2\mu_{2} - \rho[D_{t}^{(0)}]^2))e^{2qz}D_{t}^{(0)}}
{([2+\sqrt{3}]k_c^2 - 4q^2)(k_c^2-q^2)([2+\sqrt{3}]k_c^2\tilde{\mu}_{2} - 4 q^2\tilde{\mu}_{2} + 2 \rho[D_{t}^{(0)}]^2)}
\nonumber
\end{eqnarray}
where the denominator $N_{1}$ is given by
\begin{eqnarray}
N_{1} &=&
2(k_c^2\!-\!q^2)\Bigl\{2(2\!+\!\sqrt{3})k_c^4\tilde{\mu}_{2} \!-\! 4 (1\!+\!\sqrt{3})k_c^3q\tilde{\mu}_{2} \!+\! q^4\tilde{\mu}_{2} \!-\! q^2\rho(D_{t}^{(0)})^2
\nonumber \\ &+&
 k_c^2\bigl[(1\!+\!\sqrt{3})\rho[D_{t}^{(0)}]^2\!+\!2(1\!-\!\sqrt{3})q^2\tilde{\mu}_{2}\bigr]\!+\!4k_cq^3\tilde{\mu}_{2}\! -\! 2k_cq\rho[D_{t}^{(0)}]^2\Bigr\}
\end{eqnarray}
and
\begin{eqnarray}
&\tilde{\Psi}&^{\mathrm{inhom}}_{NMij}
=
\nonumber \\ && +
\frac{\big((3+2\sqrt{3})k_c^2\mu_{2} + (2+\sqrt{3})(2 q^2\mu_{2}-\rho[D_{t}^{(0)}]^2)\big)qk_c^2e^{2qz}
D_{t}^{(0)}}{(k_c^2-q^2)\bigl[(\sqrt{3}-2)k_c^2 + 4 q^2\bigr]\bigl[(\sqrt{3}-2)k_c^2\tilde{\mu}_{2}+4q^2\tilde{\mu}_{2}-\rho[D_{t}^{(0)}]^2\bigr]}
\nonumber \\ &&
-N_{2}^{-1}
\left\lbrace
(k_c^2\!+\!q^2)
\bigl[
(3\sqrt{3}\!-\!1)k_c^3\mu_{2}\!+\!10k_c^2q\mu_{2}\!-\!2 q^3\mu_{2}
\right.
\nonumber \\ && \qquad
\left.
-2q\rho[D_{t}^{(0)}]^2\!+\!k_c\sqrt{3}q^2\mu_{2}\!-\!k_c[1\!+\!\sqrt{3}]\rho[D_{t}^{(0)}]^2
\bigr]
\right\rbrace
e^{(k_c+q)z}D_{t}^{(0)}
\qquad
\end{eqnarray}
with
\begin{eqnarray}
N_{2} &=&
2(k_c^2\!-\!q^2)
\Big[
2(2\!-\!\sqrt{3})k_c^4\tilde{\mu}_{2} \!+\! 4 (\sqrt{3}\!-\!1) k_c^3q\tilde{\mu}_{2} \!+\! q^4 \tilde{\mu}_{2} \!-\! q^2\rho[D_{t}^{(0)}]^2
\\ && \qquad
+ 2k_c^2 [1\!+\!\sqrt{3}]q^2\tilde{\mu}_{2} \!+\! k_c^2[1\!-\!\sqrt{3}]\rho[D_{t}^{(0)}]^2 \!+\! 4k_c q^3\tilde{\mu}_{2} \!-\! qk_c\rho[D_{t}^{(0)}]^2
\Big]
\nonumber
\end{eqnarray}

For the scalar potential we obtain from Eq.~(\ref{ScalarPotentialSecondOrderHigherHarmonics}) in the geometry of hexagons ($ij=ji=12=23=31$)
\begin{eqnarray} \label{Phihex}
\hat{\varphi}_{NMij}
&=&
\left\lbrack 8\rho D_{t}^{(0)} q (q\!+\!k_c)(k_c^2\!-\!4q^2) \right\rbrack^{-1}
\nonumber \\  &\times& \left\lbrace
2 k_c^5 q (12\tilde{\mu}_{2} \!-\! 14\mu_{2} \!-\! 3\nu_{2}D_{t}^{(0)})
\!-\! 12q^2\rho[D_{t}^{(0)}]^2
\right.
\nonumber \\ &&
\left.
+ 2 k_c^4 \big(4\tilde{\mu}_{2} \!-\! 16\mu_{2}q^2 \!+\! \nu_{2}q^2D_{t}^{(0)} \!+\! 2\rho[D_{t}^{(0)}]^2\big)
\right.
\nonumber \\ &&
\left.
+ k_c^2 \big(19q^2\rho[D_{t}^{(0)}]^2 \!+\! 8 q^4[4\tilde{\mu}_{2} \!+\! 4\mu_{2} \!+\!\nu_{2}]D_{t}^{(0)}\big)
\right.
\nonumber \\ &&
\left.
- 4k_c \big(9q^3\rho[D_{t}^{(0)}]^2 \!-\! 4q^5[3\mu_{2} \!+\! \nu_{2}D_{t}^{(0)}] \big)
\right.
\nonumber \\ &&
\left.
+ k_c^3 \big(17q\rho[D_{t}^{(0)}]^2 \!-\! 96\tilde{\mu}_{2}q^3 \!+\! 52\mu_{2}q^3 \!+\! 20 q^3\nu_{2}D_{t}^{(0)}\big)
\right\rbrace 
\end{eqnarray}
and for the case of squares ($ij=ji=15$)
\begin{eqnarray} \label{Phisquare}
\hat{\varphi}_{NMij}
&=&
\left\lbrack
4\sqrt{2}\rho D_{t}^{(0)} (k_c \!+\! q) (k_c^2 \!-\! 2q^2) (k_c^2\!-\!2k_cq\!-\!q^2)
\right\rbrack^{-1}
\nonumber \\  &\times& \left\lbrace
4k_c^7(6\tilde{\mu}_{2} \!-\! 7\mu_{2} \!-\! 2\nu_{2}D_{t}^{(0)})
+ 2 q^5\rho[D_{t}^{(0)}]^2
\right. \nonumber \\ && \left.
- 4k_c^6q (10\tilde{\mu}_{2} \!-\! 14\mu_{2} \!-\! 5\nu_{2}D_{t}^{(0)})
\right.
\nonumber \\ &&
\left.
+ k_c^2 \big(16q^5\tilde{\mu}_{2} \!-\! 16q^5\mu_{2} \!-\! 8q^5\nu_{2}D_{t}^{(0)} \!+\! 9q^3\rho[D_{t}^{(0)}]^2\big)
\right.
\nonumber \\ &&
\left.
+ 2k_c \bigl(5q^4\rho[D_{t}^{(0)}]^2 \!-\! 4q^6[3\mu_{2} \!+\! \nu_{2}D_{t}^{(0)}]\bigr)
\right.  \nonumber \\ &&  \left. + k_c^5 \bigl(3\rho[D_{t}^{(0)}]^2
- 4q^2 [22\tilde{\mu}_{2} \!-\! 14\mu_{2} \!-\! 3\nu_{2}D_{t}^{(0)}]\bigr)
\right.
\nonumber \\ &&
\left.
- k_c^3 \big(35q^2\rho[D_{t}^{(0)}]^2 \!-\! 4 q^4 [20\tilde{\mu}_{2} \!+\! 15\mu_{2} \!+\! 3\nu_{2}D_{t}^{(0)}]\big)
\right.
\nonumber \\ &&
\left.
- k_c^4 \big(13q\rho[D_{t}^{(0)}]^2 \!-\! 4q^3 [18\tilde{\mu}_{2} \!-\! 22\mu_{2} \!-\! 9\nu_{2}D_{t}^{(0)}]\big)
\right\rbrace 
\end{eqnarray}
For stripes ($i=j$) we obtain
\begin{eqnarray}
\hat{\varphi}_{NMij}
&=&
k_c
\left\lbrack
(k_c\!-\!q)(k_c\!+\!q)(3k_c\!+\!q)\rho D_{t}^{(0)}
\right\rbrack^{-1}
 \\  &\times& \left\lbrace
2 k_c(2k_c\!-\!q)\rho[D_{t}^{(0)}]^2 - \nu_{2}(k_c\!-\!q)(3k_c\!+\!q)(3k_c^2\!-\!2k_cq \!+\! q^2)D_{t}^{(0)}
\right.
\nonumber \\ &&
\left.
+ (k_c\!-\!q)(3k_c\!+\!q)
\big[k_c^2(6\tilde{\mu}_{2}\!-\!7\mu_{2}) \!-\! 3q^2\mu_{2} \!+\! 2k_cq(\tilde{\mu}_{2}\!+\!2\mu_{2})\big]
\right\rbrace  \nonumber
\end{eqnarray}

The homogeneous contributions to the vector potential follow from Eq.~(\ref{VectorPotentialHomSecondOrderHigherHarmonics}). The solutions for hexagons ($ij=ji=12=23=31$) read
\begin{eqnarray} \label{PsiHomHex}
&\hat{\Psi}&^{\mathrm{hom}}_{NMij}
= \nonumber
\\ &&
\left\lbrack
8k_c^2(k_c \!+\! q)\rho D_{t}^{(0)}
\big(
\rho [D_{t}^{(0)}]^2 \!-\!q(2k_c\!+\!q)
\big)
\big(
\tilde{\mu}_{2} k_c^2 \!-\! 4\tilde{\mu}_{2} q^2 \!+\! \rho[D_{t}^{(0)}]^2
\big)
\right\rbrack^{-1}
\nonumber \\ &\times& 
\Big\lbrace
4 k_c q (2k_c\!+\!q)\tilde{\mu}_{2}^2
\big[
k_c^4(6\tilde{\mu}_{2}\!-\!7\mu_{2}) \!+\! 2k_c^3q(\tilde{\mu}_{2} \!-\! 4\mu_{2}) \!-\! 8k_cq^3(\tilde{\mu}_{2}\!+\!2\mu_{2})
\nonumber \\ && \qquad\qquad\qquad
+ k_c^2q^2(13\mu_{2}\!-\!24\tilde{\mu}_{2})
\big]
\nonumber \\ &&
-\nu_{2}\tilde{\mu}_{2}k_c(k_c\!-\!2q)q(2k_c\!+\!q)(k_c\!+\!2q)(2k_c^2\!-\!k_cq \!+\! 2q^2)D_{t}^{(0)}
\nonumber \\ &&
+\tilde{\mu}_{2}\rho[D_{t}^{(0)}]^2
\big[
4q^5\tilde{\mu}_{2} \!+\! 3k_c^3q^2(\mu_{2}\!-\!2\tilde{\mu}_{2})
\!+\! k_c^2q^3(79\tilde{\mu}_{2}\!+\!8\mu_{2})
\big]
\nonumber \\ &&
+ 2k_c\rho[D_{t}^{(0)}]^3\tilde{\mu}_{2}\nu_{2} (k_c^2\!-\!2k_cq\!-\!5q^2)(3k_c^2 \!-\! k_cq \!+\!2q^2)
\nonumber \\ &&
-\rho^2[D_{t}^{(0)}]^4
\big[
5\tilde{\mu}_{2}q^3 \!+\! k_c^3(17\tilde{\mu}_{2}\!-\!8\mu_{2}) \!+\!k_cq^2(\tilde{\mu}_{2}\!-\!4\mu_{2})
+ k_c^2q(17\tilde{\mu}_{2}\!+\!4\mu_{2})
\big]
\nonumber \\ &&
+ 2k_c\rho^2\nu_{2} [D_{t}^{(0)}]^5(3k_c^2\!-\!k_cq\!+\!2q^2)
\!-\!\rho^3[D_{t}^{(0)}]^6(k_c\!-\!q)
\Big\rbrace 
\end{eqnarray}
and for squares ($ij=ji=15$) we obtain
\begin{eqnarray} \label{PsiHomSquare}
&\hat{\Psi}&^{{\mathrm{hom}}}_{NMij}
=
\nonumber \\ &&
\left\lbrack
8\rho D_{t}^{(0)} k_c^2(k_c\!+\!q)
\left(
(k_c^2\!-\!2k_c q \!-\! q^2)\tilde{\mu}_{2} \!-\! \rho[D_{t}^{(0)}]^2
\right)\right\rbrack^{-1}
\nonumber \\ &\times& \left(
4\tilde{\mu}_{2}q^2 \!-\! \rho[D_{t}^{(0)}]^2 \!-\! 2\tilde{\mu_{2}}k_c^2
\right)^{-1}
\nonumber \\ &\times&
\Big\lbrace
8 k_c\tilde{\mu}_{2}(k_c^2\!-\!2k_cq\!-\!q^2)
\big[
2k_c\tilde{\mu}_{2}(3k_c\!+\!q)(k_c^2\!-\!2q^2)
\nonumber \\ && \qquad\qquad
-\mu_{2} (7k_c^4\!-\!7k_c^2q^2\!+\!8k_cq^3\!-\!6q^4)
\big]
\nonumber \\ &&
-8\nu_{2}\tilde{\mu}_{2}k_c(k_c^2\!-\!2q^2)(k_c^2\!-\!2k_cq\!-\!q^2)(2k_c^2\!-\!k_cq\!+\!q^2)D_{t}^{(0)}
\nonumber \\ &&
+ 2 \tilde{\mu}_{2}\rho[D_{t}^{(0)}]^2
\big[
k_c^5(33\tilde{\mu}_{2}\!-\!26\mu_{2})\!-\!2q^5\tilde{\mu}_{2} \!-\! 2k_cq^4(5\tilde{\mu}_{2}-7\mu_{2})
\nonumber \\ && \qquad\qquad
-k_c^2q^3(29\tilde{\mu}_{2}\!+\!12\mu_{2})
\!+\!k_c^3q^2(20\mu_{2}\!-\!81\tilde{\mu}_{2})+k_c^4q(36\mu_{2}\!-\!15\tilde{\mu}_{2})
\big]
\nonumber \\ &&
+4\tilde{\mu}_{2}k_c\rho\nu_{2}(3k_c\!-\!5q)(k_c\!+\!q)(2k_c^2\!-\!k_c q \!+\! q^2)[D_{t}^{(0)}]^3
\nonumber \\ &&
+\rho^2[D_{t}^{(0)}]^4
\big[
5q^3\tilde{\mu}_{2}\!+\!k_c^3(23\tilde{\mu}_{2}\!-\!12\mu_{2})\!+\!k_cq^2(\tilde{\mu}_{2}\!-\!4\mu_{2})\!+\!k_c^2q(11\tilde{\mu}_{2}\!+\!8\mu_{2})
\big]
\nonumber \\ &&
-4k_c\rho^2\nu_{2}[D_{t}^{(0)}]^5(2k_c^2\!-\!k_cq\!+\!q^2)\!+\!(k_c\!-\!q)\rho^3[D_{t}^{(0)}]^6
\Big\rbrace 
\end{eqnarray}
And for stripes
\begin{eqnarray}
\hat{\Psi}^{\mathrm{hom}}_{NMij}
&=&
-\left\lbrack
8k_c^2\rho D_{t}^{(0)} (k_c\!+\!q)\left(
(k_c\!-\!q)(3k_c\!+\!q)\tilde{\mu}_{2}\!+\!\rho[D_{t}^{(0)}]^2
\right)
\right\rbrack^{-1}
\nonumber \\ && \times
\Big\lbrace
4 \tilde{\mu}_{2}k_c(k_c\!-\!q)(3k_c\!+\!q)
\big[
k_c^2(6\tilde{\mu}_{2}-7\mu_{2})\!-\!q^2\mu_{2}\!+\!2k_cq(\tilde{\mu}_{2}+2\mu_{2})
\big]
\nonumber \\ && \,\,\,
- 4 \tilde{\mu}_{2}k_c\nu_{2}D_{t}^{(0)}(k_c\!-\!q)(3k_c\!+\!q)(3k_c^2\!-\!2k_cq\!+\!q^2)
\nonumber \\ && \,\,\,
+\rho[D_{t}^{(0)}]^2
\big[
\tilde{\mu}_{2}(31k_c^3\!+\!3k_c^2q\!+\!5k_cq^2\!+\!q^3) \!-\! 4k_c(5k_c^2\!-\!4k_cq\!+\!q^2)
\big]
\nonumber \\ && \,\,\,
- 4 k_c\rho\nu_{2}[D_{t}^{(0)}]^3(3k_c^2\!-\!2k_cq\!+\!q^2)
\!+\! \rho^2[D_{t}^{(0)}]^4(k_c\!-\!q)
\Big\rbrace 
\end{eqnarray}

With these potentials the components of the velocity field can be calculated straightforwardly. To determine the components of the strain field via Eq.\@ (\ref{SecondOrderElasticity}), however, the inhomogeneous contributions $-v_{k}^{(1)}\partial_{k}\epsilon_{ij}^{(1)}$  have to be calculated additionally
\begin{eqnarray} \label{eps1zz}
v_{k}^{(1)}\partial_{k}\epsilon_{zz}^{(1)}
&=&
\frac{D_{t}^{(0)}}{(k_c^2-q^2)^2}
\Big\lbrace
(1-\cos\theta_{ij})\bigl[ 4k_c^4q^2 e^{2qz}
+
(k_c^2+q^2)k_c^2e^{2k_cz} \bigr]
\nonumber \\ && \quad
-2k_c^2(k_c^2+q^2)(k_c^2-2qk_c\cos\theta_{ij}+q^2)e^{(k_c+q)z}
\Big\rbrace
\xi_{iN}\xi_{jM}
\\[0.6cm] 
v_{k}^{(1)}\partial_{k}\epsilon_{xy}^{(1)}
&=&
\frac{-D_{t}^{(0)}\cos\theta_{ij}}{2(k_c^2-q^2)^2}
\Big\lbrace
(1-\cos\theta_{ij})\bigl[ 4k_{c}^4q^2 e^{2qz}
+
k_c^2(k_c^2+q^2)e^{2k_cz}\bigr]
\nonumber \\ && \quad
-2k_c^2(k_c^2+q^2)(k_c^2-2qk_c\cos\theta_{ij}+q^2)e^{(k_c+q)z}
\Big\rbrace
\xi_{iN}\xi_{jM}
 \\
v_{k}^{(1)}\partial_{k}\epsilon_{xx}^{(1)}
&=&
\frac{-D_{t}^{(0)}(k_{i,x}^2+k_{j,x}^2)}{2(k_c^2-q^2)^2}
\Big\lbrace
(1-\cos\theta_{ij})\bigl[ 4k_c^2q^2e^{2qz} + (k_c^2+q^2)^2e^{2k_cz} \bigr]
\nonumber \\ && \quad
-2(k_c^2+q^2)(k_c^2-2qk_c\cos\theta_{ij}+q^2 )e^{(q+k_c)z}
\Big\rbrace
\xi_{iN}\xi_{jM}
\\ 
v_{k}^{(1)}\partial_{k}\epsilon_{yy}^{(1)}
&=&
\frac{-D_{t}^{(0)}(k_{i,y}^2+k_{j,y}^2)}{2(k_c^2-q^2)^2}
\Big\lbrace
(1-\cos\theta_{ij})\bigl[ 4k_c^2q^2e^{2qz} + (k_c^2+q^2)^2)e^{2k_cz} \bigr]
\nonumber \\ && \quad
-2(k_c^2+q^2)(k_c^2-2qk_c\cos\theta_{ij}+q^2)e^{(q+k_c)z}
\Big\rbrace
\xi_{iN}\xi_{jM} \\ 
v_{k}^{(1)}\partial_{k}\epsilon_{xz}^{(1)}
&=&
-\frac{iD_{t}^{(0)}(k_{i,x}\!+\!k_{j,x})(k_c^2\!+\!q^2)}{2(k_c^2\!-\!q^2)^2}
\Big\lbrace
(1\!-\!\cos\theta_{ij})
 \\ && \times
\bigl[2k_c^2q e^{2qz}
+
k_c(k_c^2\!+\!q^2)e^{2k_cz}\bigr]
\nonumber \\ && 
-
(k_c\!+\!q)\bigl[2k_c^2\!-\!qk_c(1\!+\!\cos\theta_{ij})\!+\!q^2\!-\!k_c^2\cos\theta_{ij}\bigr]e^{(k_c+q)z}
\Big\rbrace
\xi_{iN}\xi_{jM} \nonumber
\end{eqnarray}

\begin{eqnarray} 
v_{k}^{(1)}\partial_{k}\epsilon_{yz}^{(1)}
&=&
-\frac{iD_{t}^{(0)}(k_{i,y}\!+\!k_{j,y})(k_c^2\!+\!q^2)}{2(k_c^2\!-\!q^2)^2}
\Big\lbrace
(1\!-\!\cos\theta_{ij}) \\ && \quad\quad\times
\bigl[ 2k_c^2qe^{2qz}
+
k_c(k_c^2\!+\!q^2)e^{2k_cz} \bigr]
\nonumber \\ && 
-
(k_c\!+\!q)\bigl[2k_c^2\!-\!qk_c(1\!+\!\cos\theta_{ij})\!+\!q^2\!-\!k_c^2\cos\theta_{ij}\bigr]e^{(k_c+q)z}
\Big\rbrace
\xi_{iN}\xi_{jM} \label{eps1yz}  \nonumber
\end{eqnarray}
where we have displayed only the $\xi_{iN}\xi_{jM}$ contributions to $-v_{k}^{(1)}\partial_{k}\epsilon_{ij}^{(1)}$. The contributions $\sim \xi_{iN}\xi_{jM}^*$ can be derived from Eqs.~(\ref{eps1yz}) - (\ref{eps1zz}) by the replacements
\begin{eqnarray}
\xi_{iN}\xi_{jM} \longrightarrow \xi_{iN}\xi_{jM}^*
\quad&\mathrm{and}&\quad
{\bf k}_{j} \longrightarrow -{\bf k}_{j}
,\,
\cos\theta_{ij} \longrightarrow -\cos\theta_{ij}
\end{eqnarray}
%
%

\section{Satisfying the normal stress boundary condition \label{AppendixNormalBC}}
\setcounter{equation}{0}
\renewcommand{\theequation}{\ref{AppendixNormalBC}$\cdot$\arabic{equation}}

\subsection{The contributions in the second order\label{AppendixNormalBC2nd}}

In this section we present the explicit derivation of the surface contributions to the amplitude equation that are due to the second order surface boundary conditions, in particular in the case of the Rosensweig instability due to the normal stress boundary condition. We restrict ourselves to the contributions in the normal stress boundary condition that are proportional to the main modes $\xi^{(1)}$. This is exactly the part the gives the relevant condition, while the contributions proportional to $\xi^{(2)}$ are compensated by a pressure offset and do not give rise to additional restrictions (cf. \S\ref{SecondOrderFredholmBoundary}).

Starting from the general second order normal stress boundary condition Eq.~(\ref{AnhangHydroBCNormalBCPotentialsSecondOrder}), its $\xi^{(1)}$ part has been given in Eq.~(\ref{BCnormXi1}). 
Using the expression for the linear eigenvector \cite{Bohlius2006a} $\epsilon_{zz}^{(1)}$ and the expressions for the solutions of the second perturbative order (Eqs.~(\ref{SolutionsSecondOrderMainModesInhomo}), (\ref{SolutionsSecondOrderMainModesHomoVecPot})-(\ref{SolutionsSecondOrderMainModesScalarPotential})) we can rewrite this equation as
\begin{eqnarray}
2k_c^2\nu_{2}(q &-&  k_c)^2\tilde{\mu}_{2}\frac{\partial_{t}^{(1)}}{[\partial_{t}^{(0)}]^2}\hat{\xi}^{(1)}
+
2\rho(q^2 \tilde{\mu}_{2} - k_c^2 \mu_{2})\frac{\partial_{t}^{(1)}}{\partial_{t}^{(0)}}\hat{\xi}^{(1)}
\nonumber \\
&-& 4\frac{k_c^4 q (q-k_c)^2 - 2 k_c^3(q^2-k_c^2)^2}{q(q^2+k_c^2)}\tilde{\mu}_{2}(\mu_2+\tilde{\mu}_2)\frac{\partial_{t}^{(1)}}{[\partial_{t}^{(0)}]^3}\hat{\xi}^{(1)}
\qquad \nonumber \\
&=& 2\rho k_c^2 \frac{\mu}{1+\mu}M^{(1)}M_{c}\hat{\xi}^{(1)} \quad
\end{eqnarray}
If we expand the last expression in terms of $\partial_{t}^{(0)}$ and keep the lowest order, we find
\begin{eqnarray}
(\sigma^{(1)} \pm i\omega^{(1)})\hat{\xi}^{(1)}
&=&
\frac{\mu M^{(1)}M_{c}}{\nu_{2}(1+\mu)}\hat{\xi}^{(1)}
\end{eqnarray}

The real and the imaginary part have to be satisfied separately and provide the scaled growth rate $\sigma^{(1)}$ and the scaled frequency $\omega^{(1)}$ as a function of the control parameter
\begin{eqnarray}
\sigma^{(1)}\hat{\xi}^{(1)} &=& \frac{\mu M^{(1)}M_{c}}{\nu_{2}(1+\mu)}\hat{\xi}^{(1)}
\label{AppendixNormalStressSecondOrderGrowthRate}
\\
\omega^{(1)} &=& 0
\end{eqnarray}
Using the scaled dimensionless growth rate $\tilde{\partial}_{T}^{(1)} \equiv \tau_{0}\sigma^{(1)}$ with the typical time scale $\tau_{0}=\nu_{2}k_{c}(\rho G+\mu_{2}k_c)^{-1}$, we can rewrite the growth as
\begin{eqnarray}
\tilde{\partial}_{T}^{(1)}\hat{\xi}^{(1)}
&=&
\frac{k_c \mu M^{(1)}M_{c}}{(1+\mu)(\rho G+\mu_{2}k_c)}\hat{\xi}^{(1)}
\label{AppendixNormalStressSecondOrderPrimitiveAmplEq}
\end{eqnarray}
The choice of the typical time scale $\tau$ seems arbitrary at this stage, but in \S\ref{FinalAmplitudeEquation} where we combine the second and the third order, this particular choice is a posteriori justified.

\subsection{The contributions in the third order \label{AppE2}}

In this section we apply the same arguments as in \S\ref{AppendixNormalBC2nd} to the normal stress boundary condition for the third perturbative order. Taking into account only contributions proportional to $\xi^{(1)}$, Eq.\@ (\ref{ThirdOrderBCNormalStress}) reduces to 
\begin{eqnarray}
2\mu_{2}\epsilon_{zz}^{(3,1)} &+& 2 \nu_{2}\partial_{z}v_{z}^{(3,1)}
-
p^{(3,1)}
\!-\! \mu H_{c}\partial_{z}\Phi^{(3,1)} \!+\! H_{c}^{\mathrm{vac}}\partial_{z}\Phi^{(3,1)\mathrm{vac}}
\label{ThirdOrderNormalBCLinear}
\\
&=&
\mu H^{(2)}\partial_{z}\Phi^{(1)} \!-\! H^{(2)\mathrm{vac}}\partial_{z}\Phi^{(1)\mathrm{vac}}
\!+\! \mu H^{(1)}\partial_{z}\Phi^{(2,1)} \!-\! H^{(1)\mathrm{vac}}\partial_{z}\Phi^{(2,1)\mathrm{vac}}
\nonumber
\end{eqnarray}
With the help of the explicit expressions of the eigenfunctions, Eq.\@ (\ref{ThirdOrderNormalBCLinear}) can be written as
\begin{eqnarray}
2&k_c^2&\nu_{2}(q - k_c)^2\tilde{\mu}_{2}\frac{\partial_{t}^{(2)}}{[\partial_{t}^{(0)}]^2}\hat{\xi}^{(1)}
+
\rho(q^2+k_c^2)\tilde{\mu}_{2}
\left(
2\frac{\partial_{t}^{(2)}}{\partial_{t}^{(0)}}
+
\frac{[\partial_{t}^{(1)}]^2}{[\partial_{t}^{(0)}]^2}
\right)\hat{\xi}^{(1)}
\nonumber \\
&-&
4k_c^4\frac{(q\!-\!k_c)^2}{(q^2\!+\!k_c^2)}\tilde{\mu}_{2}(\mu_{2}\!+\!\tilde{\mu}_{2})\left(
\frac{\partial_{t}^{(2)}}{[\partial_{t}^{(0)}]^3}\!-\!\frac{[\partial_{t}^{(1)}]^2}{[\partial_{t}^{(0)}]^4}
\right)\hat{\xi}^{(1)}
\nonumber \\
&+&
4 k_c^3q\frac{\mu_{2}\!+\!\tilde{\mu}_{2}}{q^2\!+\!k_c^2}\rho
\left(
\frac{\partial_{t}^{(2)}}{\partial_{t}^{(0)}}\!-\!\frac{[\partial_{t}^{(1)}]^2}{[\partial_{t}^{(0)}]^2}
\right)\hat{\xi}^{(1)}
- 2k_c^3\rho\frac{\mu_{2}\!+\!\tilde{\mu}_{2}}{q}\frac{\partial_{t}^{(2)}}{\partial_{t}^{(0)}}\hat{\xi}^{(1)}
\nonumber \\
&+&
2k_c^3\rho
\left(
\frac{\mu_{2}\!+\!\tilde{\mu}_{2}}{q}
\!+\!
\rho[\partial_{t}^{(0)}]^2\frac{(\mu_{2}\!+\!\tilde{\mu}_{2})^2}{4q^3\tilde{\mu}_{2}^2}
\right)
\frac{[\partial_{t}^{(1)}]^2}{[\partial_{t}^{(0)}]^2}\hat{\xi}^{(1)}
-
2 k_c^2\rho(\mu_{2}\!+\!\tilde{\mu}_{2})\frac{\partial_{t}^{(2)}}{\partial_{t}^{(0)}}\hat{\xi}^{(1)}
\nonumber \\
&+&
2\mu_{2}k_c\frac{\mu_{2}\!+\!\tilde{\mu}_{2}}{\tilde{\mu}_{2}}
\left(
2k_c^2(k_c\!-\!q)\tilde{\mu}_{2}+\frac{k_c^2}{q}\rho[\partial_{t}^{(0)}]^2
\right)
\frac{[\partial_{t}^{(1)}]^2}{[\partial_{t}^{(0)}]^3}\hat{\xi}^{(1)}
\nonumber \\
&=&
\rho k_c^2\frac{\mu}{1\!+\!\mu}
\left(
2M^{(2)}M_{c} + [M^{(1)}]^2
\right) \hat{\xi}^{(1)}
\end{eqnarray}
If expanded in terms of $\partial_{t}^{(0)}$ we find
\begin{eqnarray}
(\sigma^{(2)} \pm i\omega^{(2)})\hat{\xi}^{(1)}
+
\frac{\mu_2}{\nu_{2}}\frac{[\sigma^{(1)}]^2}{[\sigma^{(0)}]^2}\hat{\xi}^{(1)}
&=&
\frac{\mu(2M^{(2)}M_{c} + [M^{(1)}]^2)}{2\nu_{2}(1+\mu)}\hat{\xi}^{(1)}
\end{eqnarray}
from which $\omega^{(2)}=0$ follows. In the last expression we made use of the results of the previous order, namely that $\omega^{(1)}=0$ in the static limit. For the second contribution on the left hand side we can substitute the scaled time derivative $\tilde{\partial}_{T}^{(1)}$ of the second perturbative order and we obtain
\begin{eqnarray}
\nu_{2}\sigma^{(2)}\hat{\xi}^{(1)}
+
\mu_2[\tilde{\partial}_{T}^{(1)}]^2\hat{\xi}^{(1)}
&=&
\frac{\mu(2M^{(2)}M_{c} + [M^{(1)}]^2)}{2(1+\mu)}\hat{\xi}^{(1)}
\label{AmplEqAppendixThirdOrderNormalStress}
\end{eqnarray}
which results in a second order time derivative of the pattern amplitudes. This contribution is proportional to the elastic shear modulus $\mu_{2}$ and therefore accounts for the reversible bulk processes in the medium whereas the first order time derivative represents a purely dissipative process. Equation (\ref{AmplEqAppendixThirdOrderNormalStress}) additionally suggests the time scale $\nu_{2}/\mu_{2}$ as the typical time scale to compare oscillatory processes with dissipative ones.

To combine the surface condition with the solvability condition from the bulk equations, we multiply by the typical time scale $\tau_{0}$ and finally obtain ($\tilde{\partial}_{T}^{(2)} \equiv  \tau_0 \sigma^{(2)}$)
\begin{eqnarray}
\tilde{\partial}_{T}^{(2)}\hat{\xi}^{(1)}
+
\frac{\mu_{2} k_{c}}{\rho G+\mu_{2}k_c}[\tilde{\partial}_{T}^{(1)}]^2\hat{\xi}^{(1)}
&=&
\frac{k_c \mu(2 M^{(2)}M_{c} + [M^{(1)}]^2)}{2(1+\mu)(\rho G+\mu_{2}k_c)}\hat{\xi}^{(1)}
\label{AppendixNormalStressThirdOrderPrimitiveAmplEq}
\end{eqnarray}
The second time derivative in Eq.~(\ref{AppendixNormalStressThirdOrderPrimitiveAmplEq}) is unique for magnetic gels and vanishes in the limit of pure ferrofluids.

\end{appendix}


\end{document}